\documentclass[]{aastex}

\shorttitle{Inversions of Stellar Spectra}
\shortauthors{Allende Prieto et al.}

\begin{document}

\title{Chemical Abundances  from Inversions of  Stellar Spectra: Analysis of Solar-Type Stars with Homogeneous  and Static Model Atmospheres}
\author{Carlos Allende Prieto}
\affil{McDonald Observatory and Department of Astronomy, The University of Texas, Austin, TX 78712-1083, USA}
\email{callende@astro.as.utexas.edu}
\author{Paul S. Barklem and Martin Asplund}
\affil{Uppsala Astronomical Observatory, Box 515, 751-20 Uppsala, SWEDEN}
\email{barklem,martin@astro.uu.se}
\and
\author{Basilio Ruiz Cobo}
\affil{Instituto de Astrof\'{\i}sica de Canarias, La Laguna E-38200, SPAIN}
\email{brc@ll.iac.es}

\begin{abstract}

Spectra of late-type stars are usually analyzed with static model atmospheres
in local thermodynamic  equilibrium  (LTE) and a homogeneous plane-parallel
or spherically symmetric  geometry. The energy balance requires
particular attention, as two  elements which are particularly difficult 
to model  play an important role:  line blanketing and convection. 
{\it Inversion} techniques are able to 
 bypass the difficulties of a detailed description of
the energy balance. Assuming that the atmosphere is in 
 hydrostatic equilibrium and LTE, it is possible to constrain its 
 structure from  spectroscopic observations. Among the 
most serious approximations still implicit in the method is a 
static and homogeneous geometry. In this paper, we take advantage of a 
realistic three-dimensional radiative hydrodynamical simulation of the
solar surface to check the systematic errors incurred by an inversion 
assuming a plane-parallel horizontally-homogeneous atmosphere. The  
thermal structure recovered resembles the spatial and time average of 
the three-dimensional atmosphere. Furthermore,  
the abundances retrieved are typically within {\it 10 \% (0.04 dex)} of the  
abundances used to construct the simulation. The application to a 
fairly complete dataset from the solar spectrum provides  further 
confidence
in  previous analyses of the solar composition. There is only a 
narrow range
of one-dimensional thermal structures able to fit the absorption lines 
in the spectrum of the Sun. With our carefully selected dataset, 
 random errors are about a factor of two smaller than  
systematic errors. A small number of strong metal lines can provide very reliable results. We foresee no major difficulty in 
applying the technique to other similar stars, and obtaining
 similar accuracies,  
using spectra with $\lambda/\delta\lambda \sim 5 \times 10^4$ and 
a signal-to-noise ratio as low as 30. 

\end{abstract}

\section{Introduction}

Because of the nature of astronomical observations, observed stellar
spectral line profiles are the complex result of a large number of
averages. Our telescopes gather light that once emerged from half of the
stellar surface, after being scattered, absorbed, and re-emitted inside a
 rotating inhomogeneous dynamic atmosphere. The limited resolution
of our spectrographs smears  the flux detected at a given frequency
into its neighbors, partly blurring the messages in the spectral
features. In addition, our detectors add up the light during a given 
time interval,
losing information on the time variability of the stellar flux.  
As a result, a limited amount of information is contained in the observed 
spectrum. This information is definitely insufficient to recover in full 
the physical conditions in the complex atmosphere in which the spectrum 
was shaped: the general 
inverse problem of finding the atmospheric structure 
 
 \begin{equation}
 {\bf N} \equiv [T({\bf r}),N_{\rm e}({\bf r}), {\bf v}({\bf r}),{\bf A}({\bf r})],
 \label{n}
 \end{equation}
 
  \noindent where $T$ represents temperature, $N_{\rm e}$ electron density, 
  {\bf v} velocity, 
 {\bf A} chemical abundances, and 
 ${\bf r} \equiv (r,\theta,\phi,t)$, 
 from the flux reaching Earth

\begin{equation}
f_{\nu} =  \left(\frac{R}{d}\right)^2  
 \int_{\mu=0}^{1} \int_{\phi=0}^{2\pi} I_{\nu} ({\bf R},{\bf n}) \mu d\mu d\phi
\label{f}
\end{equation}

\noindent where ${\bf R} \equiv (R,\theta,\phi,t)$, $R$ is the stellar
radius, ${\bf n} \equiv (\theta,\phi)$ a unit vector that points towards Earth, 
$d$ the distance to Earth, $\nu$ the frequency, 
$\mu = \cos \theta$, and $I_{\nu} ({\bf r},{\bf n})$, the intensity 
 at a position ${\bf r}$ in 
the direction ${\bf n}$,  satisfies the radiative transfer equation 
(RTE)

\begin{equation}
\left[\frac{1}{c} \frac{\partial}{\partial t} + ({\bf n} \cdot \nabla)\right] 
I({\bf r},{\bf n}) = \eta_{\nu} ({\bf r},{\bf n}) - \kappa_{\nu} 
({\bf r},{\bf n}) I ({\bf r},{\bf n})
\label{i}
\end{equation}

\noindent with $\kappa_{\nu}$ and $\eta_{\nu}$, the absorption and 
emission coefficient, functions of ${\bf N}$ and the intensity itself 
(Mihalas 1978). 

 Multiple scenarios, defined by  ${\bf N}$, will be able to produce 
similar spectra, and so  the general inverse problem  has no unique 
solution. In fact, some constraints are needed to recover any information 
at all. This problem  affects all stellar  (including solar)
observations, but the difficulty is not the same in all instances. An
example of a very extreme case could be a non-spherical
rapidly-rotating active and distant star. The best conditions are met for
the
 solar case, as the star's   proximity  makes it possible to obtain
 very fast  observations of the spatially resolved surface, yet having
 enough light to afford very high dispersion. It is precisely in the
latter scenario where inversion methods emerged. Largely avoiding the
lack of spatial, time, and spectral  resolution typically found in
stellar spectra, the best solar observations of intensity line
profiles  still depend upon the physical conditions in a fairly complex
piece of atmosphere.  Moreover, new variables enter the game, and it is
no longer possible to ignore the existence of magnetic fields. The
first inversions adopted simplified Milne-Eddington atmospheres (Auer,
House \& Heasley 1977), but the business has developed to see  more
sophisticated scenarios, such as
  model atmospheres in Local Thermodynamical Equilibrium (LTE) with one
  or more components, as well as two-dimensional atmospheres (Keller et
al. 1990, Bellot Rubio et al. 2000), or Microstructured Magnetic
Atmospheres (MISMAs; S\'anchez Almeida 1997).   Recent noteworthy
advances include the work by Socas-Navarro, Ruiz Cobo \& Trujillo Bueno
(1998) allowing for departures from LTE.

Classical abundance analyses attempt to derive only one or very few
parameters from an observed stellar spectrum. Thus, a model atmosphere
is constructed upon physical principles and assumptions, and only the 
chemical abundance(s) and probably a small set of parameters defining the
theoretical atmosphere are fitted. While it is true that we cannot determine
all the physically relevant information of an stellar atmosphere from the
observed spectrum, high-quality observations contain far more data than 
classical analyses are able to extract. Furthermore, the common practice of 
analyzing equivalent widths, rather than spectral line profiles, implies 
neglecting (usually intentionally) information readily available in  
the observations. Logically,  an observer  would like to obtain 
as much information as possible from a stellar spectrum.  That is 
exactly the purpose that inversions serve. In general, solving 
simplified versions of the inverse problem in Eqs. \ref{n}$-$\ref{i} 
to find out the atmospheric thermal structure 
from  spectroscopic observations goes by the name 
of semi-empirical  modeling. The perennial solar model built by Holweger 
and M\"uller (Holweger 1966, Holweger \& M\"uller 1974) is a prime example.

Inversions of spatially unresolved stellar spectra 
adopting an LTE single-component model atmosphere in hydrostatic 
equilibrium targeted the Sun (Allende Prieto et al. 1998), 
$\epsilon$ Eridani, and Groombridge 1830 (Allende Prieto et al. 2000). 
Frutiger et al. (2000) added some complexity, and showed that a two-component 
model including depth-dependent velocities is indeed a powerful tool 
to study stellar granulation, and  $\alpha$ Cen A and B have already 
been the subject of one such study (Frutiger, Solanki \& Mathys 2001).

Evidently, the mentioned inversion schemes   rely largely on the
adopted scenarios and the adequacy of the underlying approximations.
Dealing with  simplified model atmospheres inevitably introduces
systematic errors in every parameter that we try to quantify: chemical
abundances, magnetic fields, projected  rotational velocity,
etc. The key question is how realistic our models are and how large are
the systematic errors introduced by the differences from reality. The 
favorite approximation to calculate the radiation field, LTE, is also
a matter of  concern in most cases.

Recent work has  shown that the state-of-the-art
radiative hydrodynamical simulations have achieved an extremely high
degree of realism. Models of the outer solar envelope and
photosphere  now  sharply match the granulation topology, flow fields, 
and time-scales  (Stein \& Nordlund 1998), fit the helioseismic measurements
(Rosenthal et al. 1999), and produce spectral line profiles that
closely resemble the most detailed observations 
(Asplund et al. 2000a, b; Asplund 2000). The intricate velocity, 
temperature, and radiation fields predicted by the simulations 
have to be very close to reality, otherwise such diverse tests would 
never have been passed. 

In this situation, it seems possible to use a 
detailed time-dependent three-dimensional model of the Sun to check 
for systematics introduced by adopting simplified model atmospheres
 as part of an inversion method. For example, it is of interest to 
 learn how far  the mean atmospheric 
temperature at a given depth is 
from the value recovered by a single-component inversion. 
Even more importantly, we wonder how different are the chemical 
abundances used in the numerical simulation from those recovered by 
 inversions. As argued by Rutten (1998) for the case of the 
 Holweger \& M\"uller (1974) model, a positive answer 
 may be anticipated by the agreement between the photospheric 
 solar abundances derived by Allende Prieto et al. (1998) 
 and those in  meteorites (Grevesse \& Sauval 1998). 
 Nonetheless, we think that the  independent assessment we attempt 
 in this paper is in order.

Another issue that causes concern to practitioners of inversions 
 is uniqueness. If our simplified model is too crude, degeneracy 
 will emerge. In fact,  it will emerge in any case, if the observations 
 are incomplete, or if the inverse problem is ill-defined. 
  High quality observations are expensive. 
 The typical sizes of  CCD detectors still do not   provide large spectral
  coverage at the highest dispersion, 
and except for the nearest stars, high signal-to-noise ratios 
require large telescopes and long exposures. It is, therefore, 
a must to find out what spectral information is required 
to achieve meaningful inversions: number of lines, species, 
strength -- that will dictate the wavelength coverage--, but 
also resolution and signal-to-noise ratio (SNR). 

 In Section 2, we present a careful selection of atomic spectral lines
for inversions of solar-type spectra. The selected lines have
accurately determined transition probabilities and newly computed
collisional damping cross-sections. In \S 3 we have used a 
detailed radiative hydrodynamical simulation of the solar surface as a
numerical laboratory to test our one-dimensional single-component
inversion method, {\tt MISS}\footnote{Multiline Inversion of Stellar Spectra}. 
In doing so, we  quantify directly the systematic errors
incurred by assuming that the atmosphere is plane-parallel and
homogeneous. Section 4 is devoted  to exploring to what extent a
single-component LTE inversion of the solar spectrum is unique, and
what kind of model structure and abundances it provides. In \S 5 we 
investigate which spectral features contain most of the information. 
In \S 6, we  discuss prospective applications of this type  of 
inversion, and a short summary  is provided in \S 7.

\section{Atomic  Data}

Allende Prieto et al. (1998) employed 40 spectral lines in their inversion 
of the solar flux spectrum. 
 Strict selection criteria were imposed for quality of the lines and data, 
 and as a result, all lines were of neutral species and weak in the solar 
 spectrum, except for one reasonably strong  Ca I line.   With a view to 
 improving the inversion, we have relaxed a number of the criteria while
  attempting to maintain the quality of the atomic data.

We identified three key desired additions to the line list.  The first 
is the inclusion of ionized lines, as they will even more strongly constrain 
the inversion.  The second is the inclusion of the elements magnesium and 
silicon, due to their importance as electron contributors and therefore in 
determining the H$^-$ fraction.  Thirdly, more strong lines are required to 
obtain better depth coverage.  The strongest lines, 
being independent of microturbulence, also provide strong 
constraint on abundances and, in concert with weak lines, microturbulence.  
The first two additions force us to look beyond the transition probabilities 
measured by the Oxford group (for example Blackwell \& Shallis 1979).  
The addition of strong lines is made possible by recent advances in the 
understanding of collisional broadening by hydrogen (Barklem et al. 2000a
and references therein).   We also looked outside the list of lines 
identified as clean by 
Meylan et al. (1993).  There are very few clean strong lines in the 
solar spectrum, however we are able to employ a number of lines by 
removing sections with blends. 

The inversion requires that atomic data are of high quality 
(see Allende Prieto et al. 1998), and so thorough critical evaluation 
was required.  We surveyed available oscillator strengths for clean lines 
of desired species, and made critical appraisal
 of their suitability.  In all cases we favored experimental data, 
 except in 
the case of the Ca II infrared triplet lines, where theoretical 
calculations are considered more reliable based on agreement 
with high accuracy lifetime measurements.  Lines which have been 
measured in good agreement by more than one group were strongly 
favored.  For example, available $f$-values of suitable Cr~II lines at 
present do not show such agreement and thus were omitted.  We 
deemed 25 new lines suitable for addition to the list, which is 
presented in Table  \ref{tab:atomic_data}.  
We removed two Ca I and two Ti I lines 
used in  Allende Prieto et al. (1998) due to possible continuum problems 
or suspected blending.   Line excitations were collected from 
the literature
 and radiative damping widths were all taken directly from the 
Vienna Atomic Line Database (VALD, Kupka et al. 1999), except for in 
the Ca~II triplet where we computed the radiative damping widths
 from the Theodosiou (1989) transition probabilities, due 
this mechanism's importance in these lines.

Oscillator strengths for Fe II lines were adopted from the compilation
of Allende Prieto et al. (2001), which combines
experimental branching ratios from Kroll \& Kock (1987), Whaling (as
included in the compilation by Fuhr, Martin \& Wiese 1988 -- also
private communication), Heise \& Kock (1990), and Pauls, Grevesse \&
Huber (1990), with lifetimes measured by Biemont et al. (1991), Guo et
al. (1992), and Schnabel, Kock \& Holweger (1999).

The collisional broadening by neutral hydrogen for all lines is treated 
using the Anstee, Barklem \& O'Mara (ABO) theories (see Barklem et al. 
2000a and references therein).   For most lines the data is used directly 
from Barklem et al. (2000a).  However, the lines of neutral Si are 
not covered by ABO calculations.  The reason for this is that in more 
excited states exchange effects may start to become important in the 
broadening.  
We made specific calculations in any case, as these will certainly 
provide a reasonable estimate and as the lines are weak this should be 
sufficient.  
For the weak ionized lines, we made special calculations following 
Barklem \& O'Mara (1998),  though always assuming $E_p=-4/9$ 
atomic units.  As the energy levels in these transitions are not low 
lying,  they are not isolated and so this should be acceptable.  Again, 
as the lines are weak,  a good estimate should suffice. 

The atomic data relevant to the calculation of the LTE level populations and
the continuum opacity depend on the code used for the spectral 
synthesis. {\tt MISS} is based on Wittmann's code (Wittmann 1974), and 
the 3D calculations are based on the Uppsala's stellar atmosphere 
package (Gustafsson et al. 1975 and updates). 

\section{Inversion of the time-averaged spectrum from a three-dimensional 
solar model atmosphere}

Hydrodynamical model atmospheres  have achieved an impressive
degree of realism, as evident from detailed comparisons 
with various observational constraints, such as granulation topology
and statistics (Stein \& Nordlund 1998), helioseismological properties
(Rosenthal et al. 1998; Stein \& Nordlund 2001) and spectral line
shapes and asymmetries (Asplund et al. 2000a). It thus appears that
the state-of-the-art time-dependent three-dimensional radiative
hydrodynamical simulations of the solar surface convection indeed
include the relevant physics to properly represent granulation in
the quiet Sun in spite of the still limited numerical resolution
(Asplund et al. 2000c).
These models combine basic physical laws (conservation of mass,
momentum and energy) and a treatment of the radiative transfer
with the  atomic and molecular data
relevant to the equation of state and the opacity. LTE
is still an assumption, and  spectral lines are taken into account
 in the energy balance.

We use the model as a numerical laboratory to check the behavior of our
plane-parallel homogeneous  atmosphere inversion method. The
temperature, velocity, and density  fields predicted by the simulation
evolve with time, giving shape to 
 the   spectrum emerging from the photosphere,
so it is a function of time, direction, and location on the surface. To
simulate spatially unresolved stellar observations, we average the
emergent intensity over those quantities to obtain the average flux, 
displacing the intensity
profiles to reproduce the Doppler shifts induced by rotation with
$v_{\rm rot} \sin i = 1.88$ km s$^{-1}$. The energy input at the lower
boundary of the simulation box was carefully adjusted prior to the 
starting of the simulation sequence, until the temporal average 
of the energy emerging
at the top matched the solar constant or, in other words, until the
bolometric flux was that of a black body with an effective temperature
close to 5777 K. The gravity is  well-known, roughly $\log g = 4.44 $ (c.g.s).
We chose the chemical composition to be  the  {\it photospheric} 
abundances given by Grevesse \& Sauval (1998; hereafter GS98). 
Note that these have some differences to those of Anders 
\& Grevesse (1989) used below as the default initial values 
in the inversion.

The inversion procedure, {\tt MISS},  follows Allende Prieto et al.  
(1998) and
references therein. It assumes LTE, hydrostatic-equilibrium, and a
plane-parallel homogeneous  atmosphere. Compared to classical model
atmospheres, we do not impose 
energy conservation, so the system of ordinary differential equations
that defines the classical homogeneous model atmosphere problem is
no longer closed. The algorithm starts by adopting a guess atmospheric
structure, micro- and macro-turbulence,  and chemical composition. We
solve the transfer equation, obtain the  emergent intensity  at
different angles, and compute the  spectrum from the integrated disk
using $v_{\rm rot} \sin i = 1.88$ km s$^{-1}$.  The difference between
synthetic and observed spectra is quantified by the reduced $\chi^2$

\begin{equation}
\chi^2 = \frac{1}{N-M} \sum_{i=1}^{N} \left( \frac{\bar{f_i} - F_i}{\sigma_i} 
\right)^2, 
\end{equation}

\noindent where $N$ is the number of frequencies (wavelengths), 
$M$ the number of {\it free} parameters 
(usually 6 temperature nodes + 6 abundances + micro + macro = 14), 
$\bar{f_i}$ the synthetic (normalized) flux,  
$F_i$ the observed (normalized) flux, and  $\sigma_i = 1/{\rm SNR}$, 
and we try to minimize it by changing the 
temperature structure, the micro- and   macro-turbulence, and 
the chemical abundances.
The Marquardt algorithm (Marquardt 1963, Press et al. 1988) 
transforms the fitting of a non-linear model
 into solving a succession of linear systems that relate  
 recommended changes in the vector of unknowns to the first and
 second order  derivatives of the $\chi^2$. Singular
Value Decomposition (SVD) is employed to eliminate  singularities 
(null dependency of the spectrum on some parameters).  We calculate 
analytically the  derivatives of the  $\chi^2$ with respect to the 
temperature  from the response function to that parameter.  A small number 
 of optical depths (called nodes), which are equally spaced in the 
 logarithm of the optical depth at 5000 \AA, are selected. The
temperature is only modified at those nodes and returned back to the
original grid by cubic-spline interpolation (or a polynomial 
when the number of nodes is four or less).  We  compute the
corresponding gas and electron pressure by solving the hydrostatic
equilibrium  (with $\log g = 4.44$) and Saha equations, and finally, 
we calculate the emergent flux, starting a new iteration.

The inversion process is normally initialized with an isothermal model,
reasonable or null macro- and micro-turbulent velocities, and the {\it
photospheric} solar abundances given by Anders \& Grevesse (1989).
Continuum opacities are simple but believed complete for the optical
spectrum of solar-type stars (see \S 4), including the following
species: H, He, H$^{-}$, He$^{-}$, H$_2^{-}$, H$_2^{+}$, C I,
Cl$^{-}$,  Mg I, Na I, and electron and Rayleigh scattering terms. The
temperature is then changed at the nodes, whose number is progressively
increased until no significant improvement in the $\chi^2$ is reached,
or until the model looses smoothness.  The appearance of wiggles is
typical for numbers of nodes larger than six, and are interpreted as
the reaching of the depth resolution limit. Higher order polynomials,
although they tend to improve slightly the comparison with
observations, are regarded as unrealistic, and discarded. Sharp
variations of the temperature structure are also found when the input
line profiles are not sufficient to constrain  the model properly.

Multiparametric non-linear minimization problems are  complex, and it is
difficult to guarantee that the absolute minimum has been found. In our
particular case, the adopted chemical abundances affect  the emergent
spectrum directly, if we input lines produced by that particular
element, and indirectly, through the  continuum absorption coefficient.
It is essential to include spectral lines of all
the metals that contribute electrons (mainly iron, magnesium, and 
silicon), and which, ultimately, control
the continuous opacity in a solar-type star through the formation of
H$^{-}$ ions. It is important to assess the independence of
the final answer from the initial  abundances of
 the main electron donors. Not less important is to make sure our 
 results are not conditioned by the initial micro- and macro-turbulence 
 velocities. 
Moreover, we should be able to retrieve the same {\it  preferred} model 
from different sets of lines. Redundancy is indeed desirable and thus we
 should include input lines that provide  a similar depth coverage
but  opposite sensitivity to the searched parameters, so that unavoidable 
errors, such as those in the radiative transition probabilities, cancel out.

The inversion of the solar simulation allows us to get rid of many  
uncertainties  inherent to the analysis of 
real observations, such as  errors in the atomic data, 
continuum opacities, or the treatment of collisional damping. This is
very useful, as it allows us to isolate and quantify 
the systematic errors involved  in the assumption of a static 
horizontally-homogeneous  medium to fit the spectrum of a 
time-dependent inhomogeneous atmosphere. We should note that we forced 
consistency in the atomic line data (excitation potentials, radiative damping,
hydrogen collisional damping, and transition probabilities) between the 
 synthesis routines used with the 3D simulation and the 1D inversion, but 
 we did not so between other atomic data considered to have a much smaller
 impact on the abundances,  such as partition functions or 
 continuum opacities. Later tests confirmed these suspicions (see below).

We synthesized the spectral line profiles selected in \S 2 using the
solar granulation simulation as a model atmosphere. The full simulation
covers two solar hours, but we selected a shorter sequence of 30
minutes with snapshots stored every minute (the  life-time of a granule
is typically less than 15 minutes). More details are given in  Asplund
et al. (2000a), but it should be noted that the  snapshots used here
are a subsample of the 100 snapshots used for the calculations of Fe I
and Fe II lines in Asplund et al. (2000a, b). This restriction was made
to save computing time, since here flux ($4 \phi$- and $4 \mu$-angles)
instead of intensity profiles have been utilized, but test calculations have 
shown that this introduces negligible differences. A wavelength grid 
and a mask was chosen for each spectral line.
 The mask selected the parts of the line profile that 
appeared  free from overlapping with other spectral features in 
the observed spectrum of the Sun, thus limiting  the information to
 that potentially available in the observed spectrum. Departures from LTE are
removed from this experiment, as the synthesis of the input profiles from the
simulation and the inversion assume LTE. Still, to  
minimize departures from LTE in the application to real observations, and
keep consistency in the inversion of the simulation, we excluded 
the part of the line profiles produced by neutral species where 
the flux was lower 
than 50 \% of the continuum level. The wavelength 
grid limited the sampling, also to resemble real observations, 
to 5 m\AA\ for  lines with  equivalent widths ${\rm EW} < 400$ m\AA, 
and to 20 m\AA\ for the three lines stronger than that. 

The line database was divided into three groups: {\it weak-neutral}, 
{\it weak-ionized}, and {\it strong} lines. The weak/strong 
separation was arbitrarily placed at 100 m\AA\ equivalent width. 
Each of these groups 
was itself split into two subgroups containing  the same number of 
lines and, in some cases, a line was common to the two subgroups. 
The subgroups were finally combined to produce eight datasets, each a combination of a {\it weak-neutral}, a {\it weak-ionized}  and a 
{\it strong} subgroup. 
Each of the  datasets contained 34 spectral lines sampled with 
a number of frequencies in the range  2724 -- 2882.

\begin{figure}[!ht]
\centering
\includegraphics[width=15cm,angle=0]{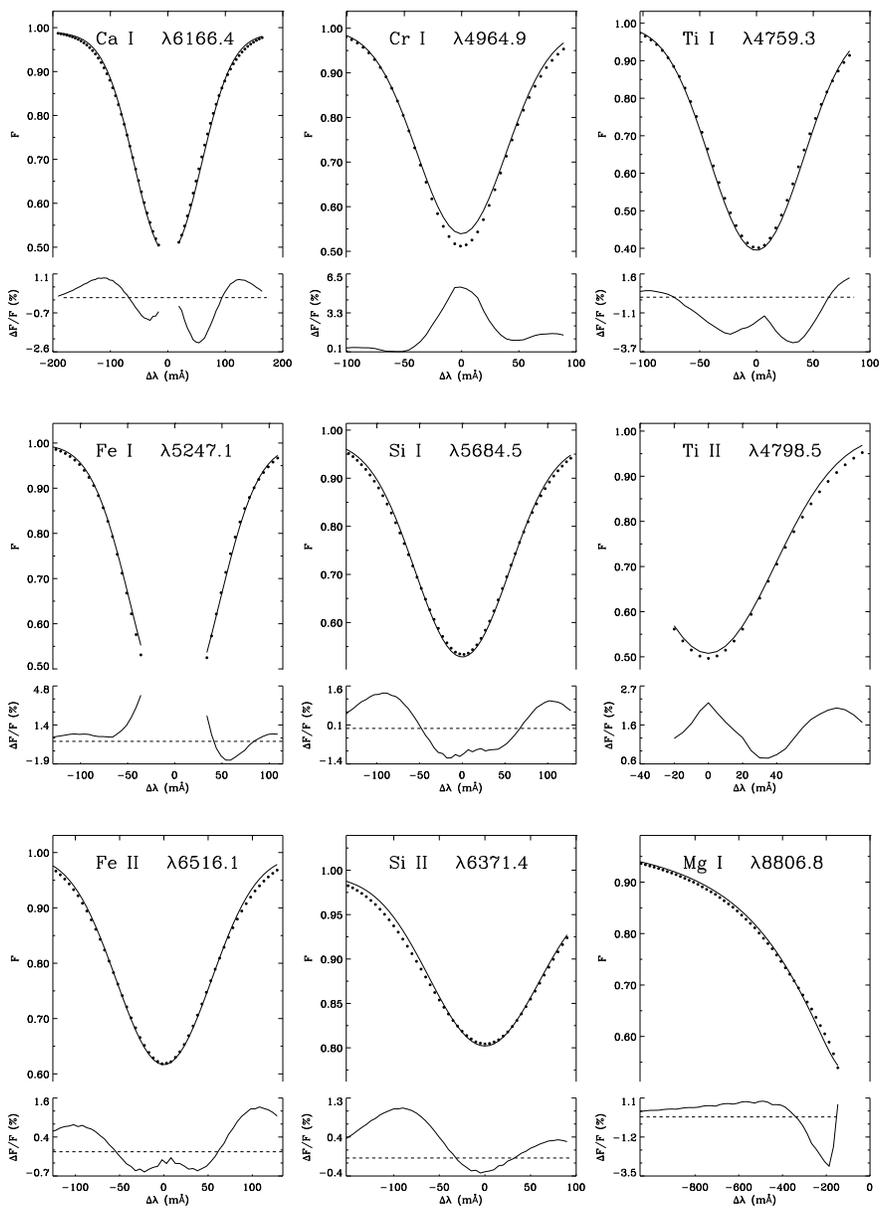}  
\protect\caption[ ]{
Comparison between the time- and spatially-averaged line profiles from the 
hydrodynamical simulation (filled circles) and from the homogeneous 
static model obtained from the inversion. 
\label{f1}}
\end{figure}

We performed inversions for each of the eight datasets with an initial
micro-turbulence of $\xi_0 = 0.1, 1.0, 1.5 $ and 2.0 km s$^{-1}$, and
with the abundances of iron, magnesium and silicon simultaneously
changed by $\Delta$ A $= -0.2, -0.1, 0.0, +0.1$ and $+0.2$ dex, for a
total of $8 \times 4 \times 5 = 160 $ cases. Some of the
models initialized with $\xi_0 = 2.0$ km s$^{-1}$ did not converge to
a solution, as at some point the temperature dropped below 500 K, or the
optical depth for some lines became larger than 0.1 at the top of the
atmosphere. Therefore,  we discarded the inversions starting with
$\xi_0 = 2.0$ km s$^{-1}$, and we dealt with 120 test cases. We also 
performed an inversion of all the lines simultaneously. Figure \ref{f1} 
shows several examples of the closeness between the final synthetic 
profiles produced by the recovered homogeneous atmosphere and the 
input data obtained from the three-dimensional simulation. The comparison is
poor, as expected: the hydrostatic homogeneous model cannot reproduce the convective
line asymmetries.

\begin{figure}[!ht]
\centering
\includegraphics[width=9cm,angle=90]{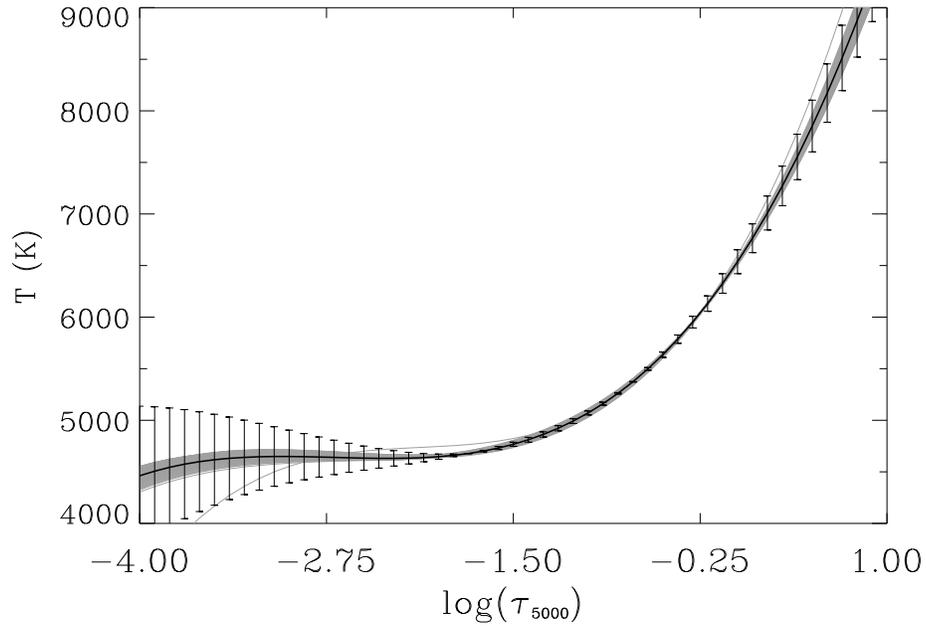}  
\protect\caption[ ]{
Temperature structures derived from the inversion of different datasets 
with different initial conditions (dotted gray lines). The discrepant
case corresponds to one of the inversions with most extreme initial 
parameters
($\xi_0=0.5$ km s$^{-1}$ and $\Delta$ A = $-0.2$). 
The solid black curve with error bars corresponds to the inversion 
of all the available line profiles simultaneously.
\label{f2}}
\end{figure}

As Figure \ref{f2} shows, the 120 atmospheric structures
retrieved (dotted gray lines) are very similar to one another,  and the
scatter around the inversion of the whole database (solid black),  is
small. There is only a single renegade that corresponds to one of the eight
data sets for $\xi_0 = 0.5$ km s$^{-1}$ and $\Delta$ A $= -0.2$. The
error bars  provided by the inversion code  (see Appendix B in 
S\'anchez Almeida 1997)  seem to be slightly
optimistic in the line forming region ($-2.0 \le \log \tau \le -0.5$),
whereas the scatter among the individual cases cannot be realistically
associated to  errors at  depths where the inverted spectrum is not
sensitive -- at optical depths smaller than $\log \tau \sim -2.75$ and
larger than $\log \tau \sim + 0.5$ the thermal structure derived is
just  a extrapolation of neighbor depths that  still affect the
inverted spectrum, and in fact, some tests have shown that our formal
error bars at extreme depths are unrealistically small. The main
conclusion is that, as long as we provide an input spectrum able
 to constrain properly the thermal structure of the atmosphere,  no
 matter which lines are used, we find a single solution.

\begin{figure}[!ht]
\centering
\includegraphics[width=9cm,angle=90]{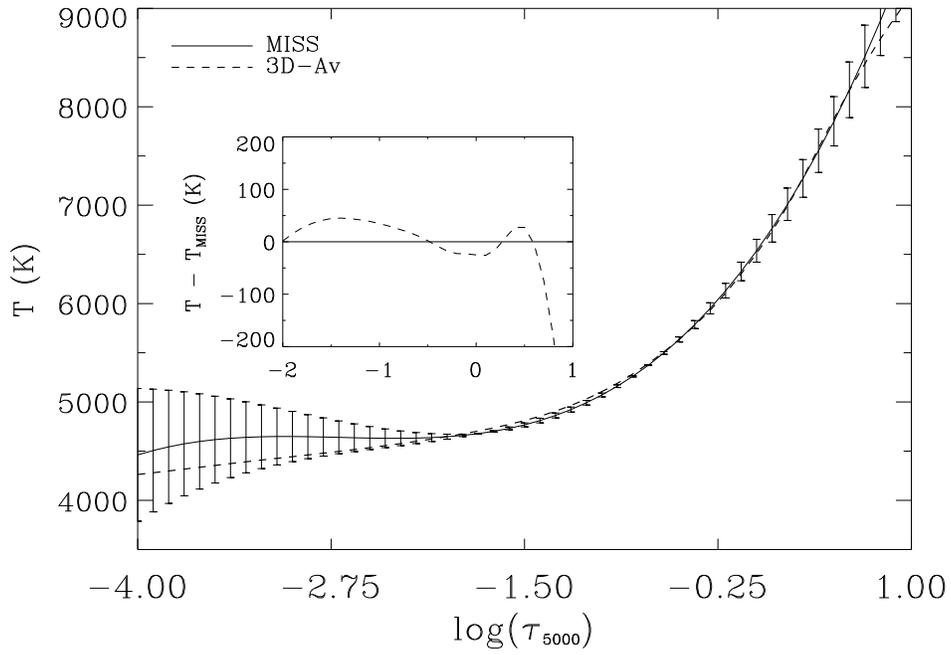}  
\protect\caption[ ]{
Thermal structure derived from the inversion (solid line with error bars) 
and the  horizontal- and time-averaged mean thermal stratification of the 
three-dimensional simulation (dashed curve). 
\label{f3}}
\end{figure}

\begin{figure}[!ht]
\centering
\includegraphics[width=9cm,angle=90]{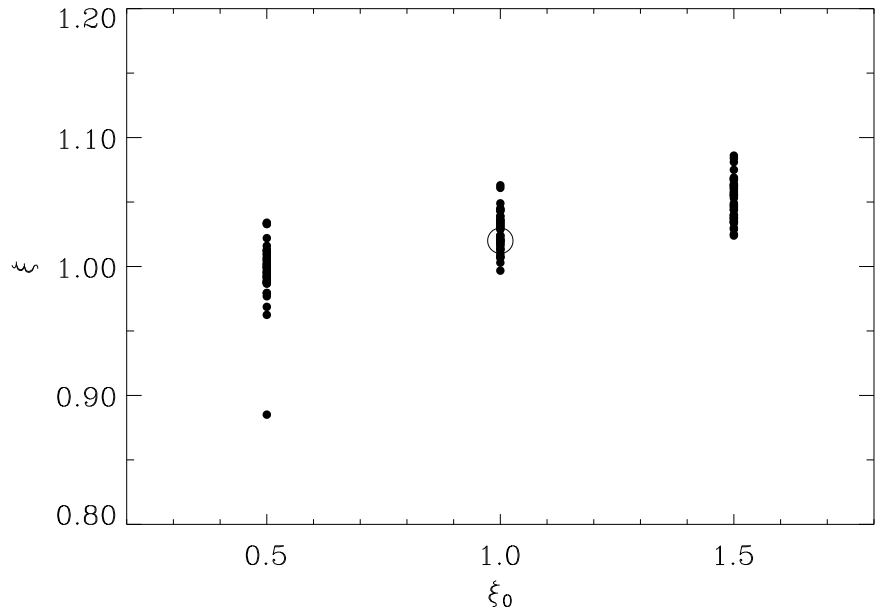}  
\protect\caption[ ]{
Final ($\xi$) versus initial ($\xi_0$) values for the micro-turbulence in 
all the 120 test runs. The open circle shows the value retrieved from the 
inversion of all the lines simultaneously.
\label{f4}}
\end{figure}

Figure \ref{f3} answers the question of whether the recovered
homogeneous atmosphere is some kind of {\it average} of the real
three-dimensional structure. The recovered temperatures (solid line) 
show great resemblance to a temporal average of the simulation
over layers of equal optical depth (dashed curve). This situation is totally
different from the chromospheric case explored by Carlsson \& Stein (1995), 
in which large fluctuations induced by waves on a cool-chromosphere solar model 
produced the same spectrum as a hydrostatic model with a hot-chromosphere. 
Even though at the temperatures of the solar photosphere the source function
is typically on the non-linear (Wien) side of the Planck function, in this case
there are no large fluctuations, and as a consequence, the thermal structure
obtained from inversion of a time-
and spatially-averaged spectrum is very close to the (linear) average of the
3D time-dependent atmosphere. The  micro-turbulence retrieved by the 
inversions has  a weak dependence on the initial value adopted ($\xi_0$), 
as can be seen  in Figure \ref{f4}. We recover a mean value 
$\xi =1.03 \pm 0.02$  (1$\sigma$) km s$^{-1}$. The final macroturbulence is 
$1.58 \pm 0.02$  km s$^{-1}$. This parameter  
was always initialized at 2.0 km s$^{-1}$, as it is much 
less relevant for the atmospheric structure. 

\begin{figure}[!ht]
\centering
\includegraphics[width=5cm,angle=90]{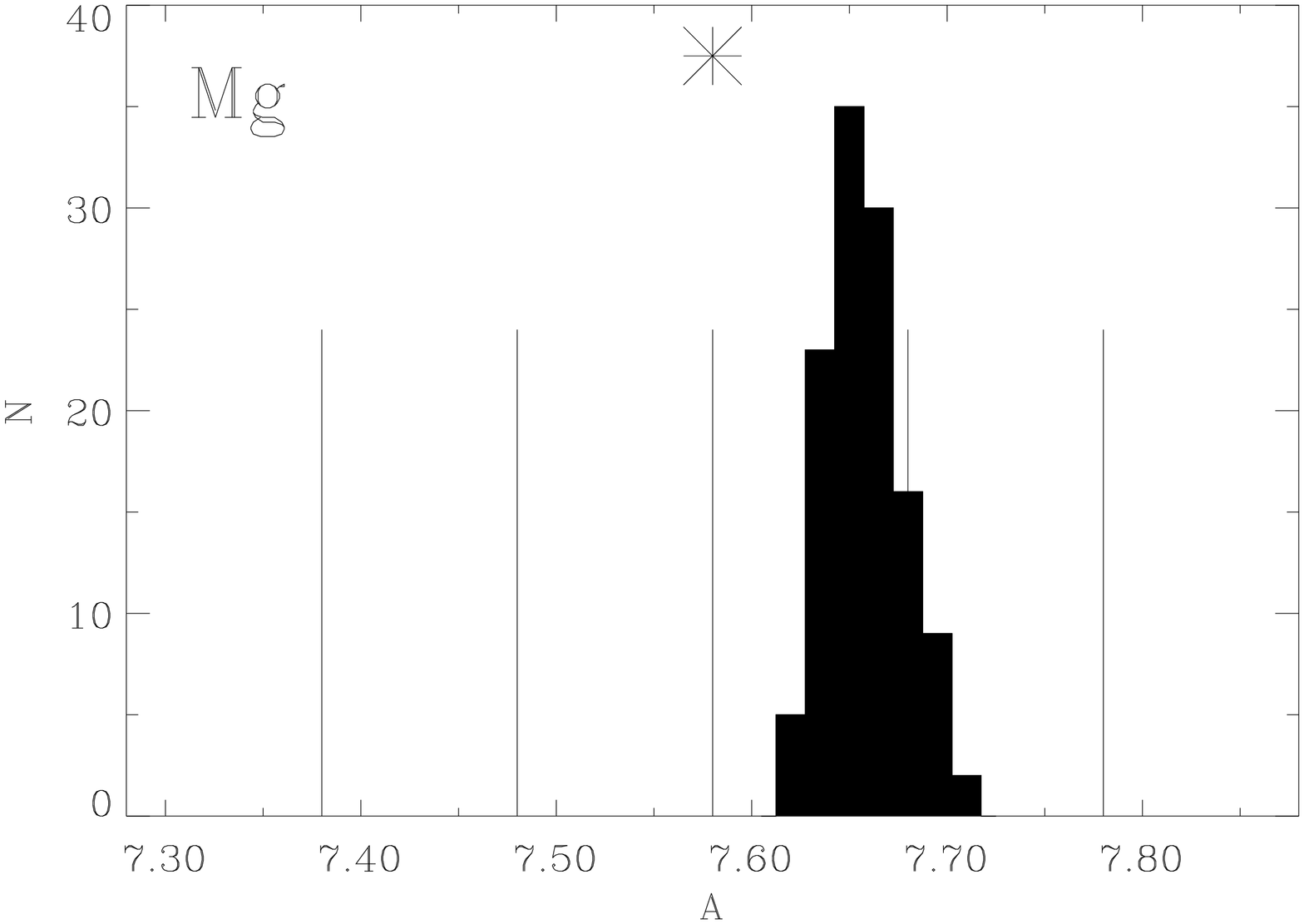}  
\includegraphics[width=5cm,angle=90]{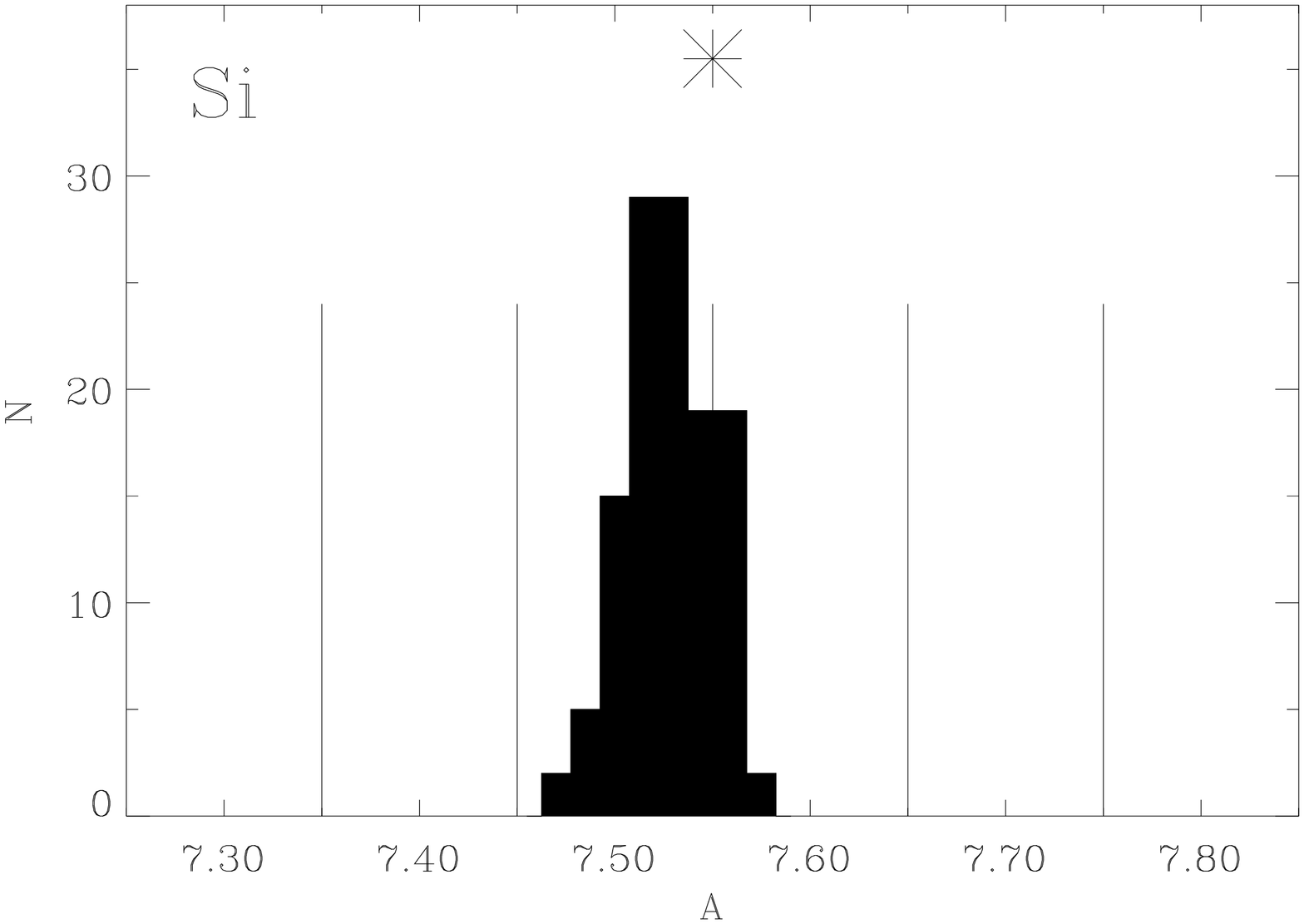}  
\includegraphics[width=5cm,angle=90]{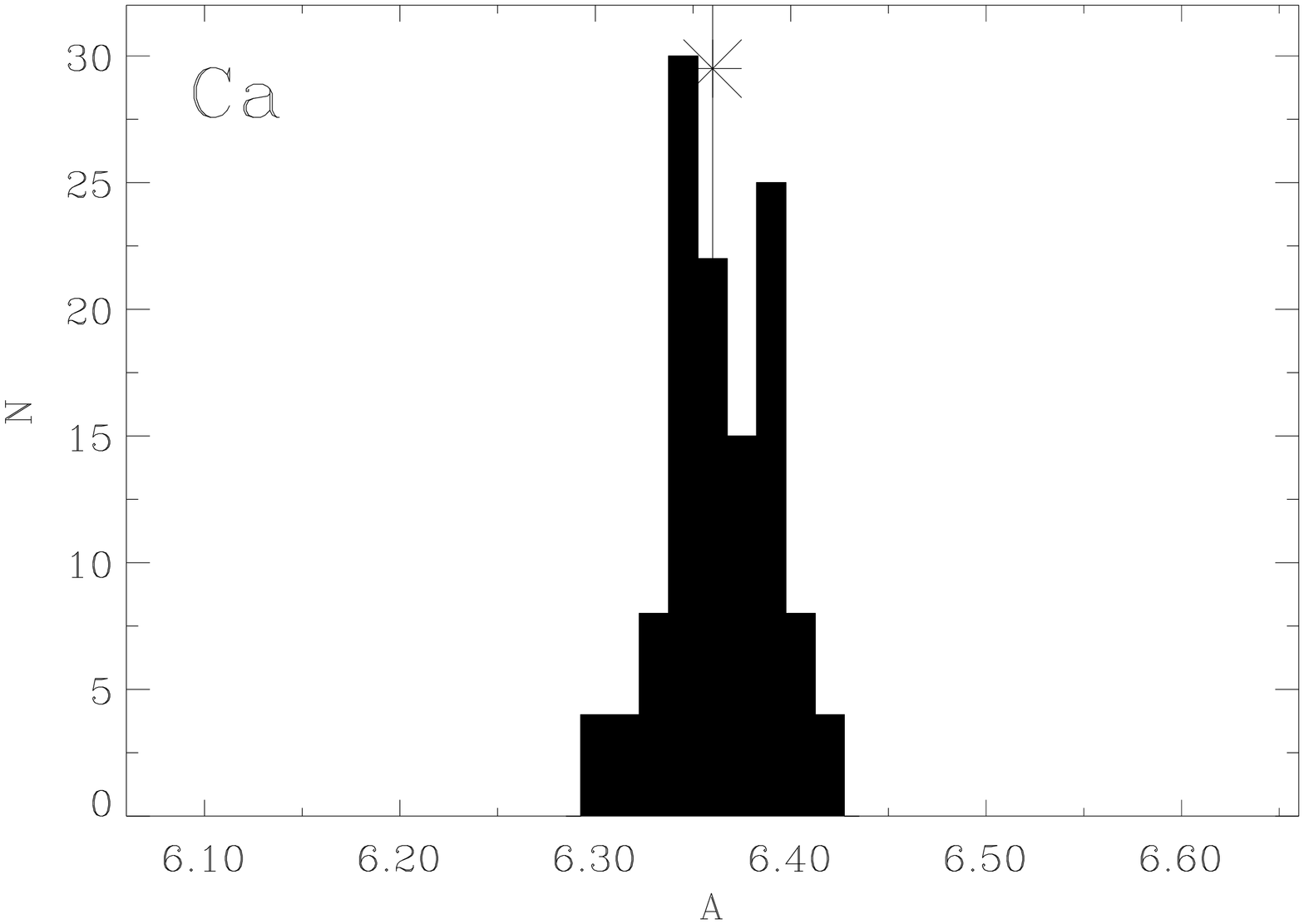}  
\includegraphics[width=5cm,angle=90]{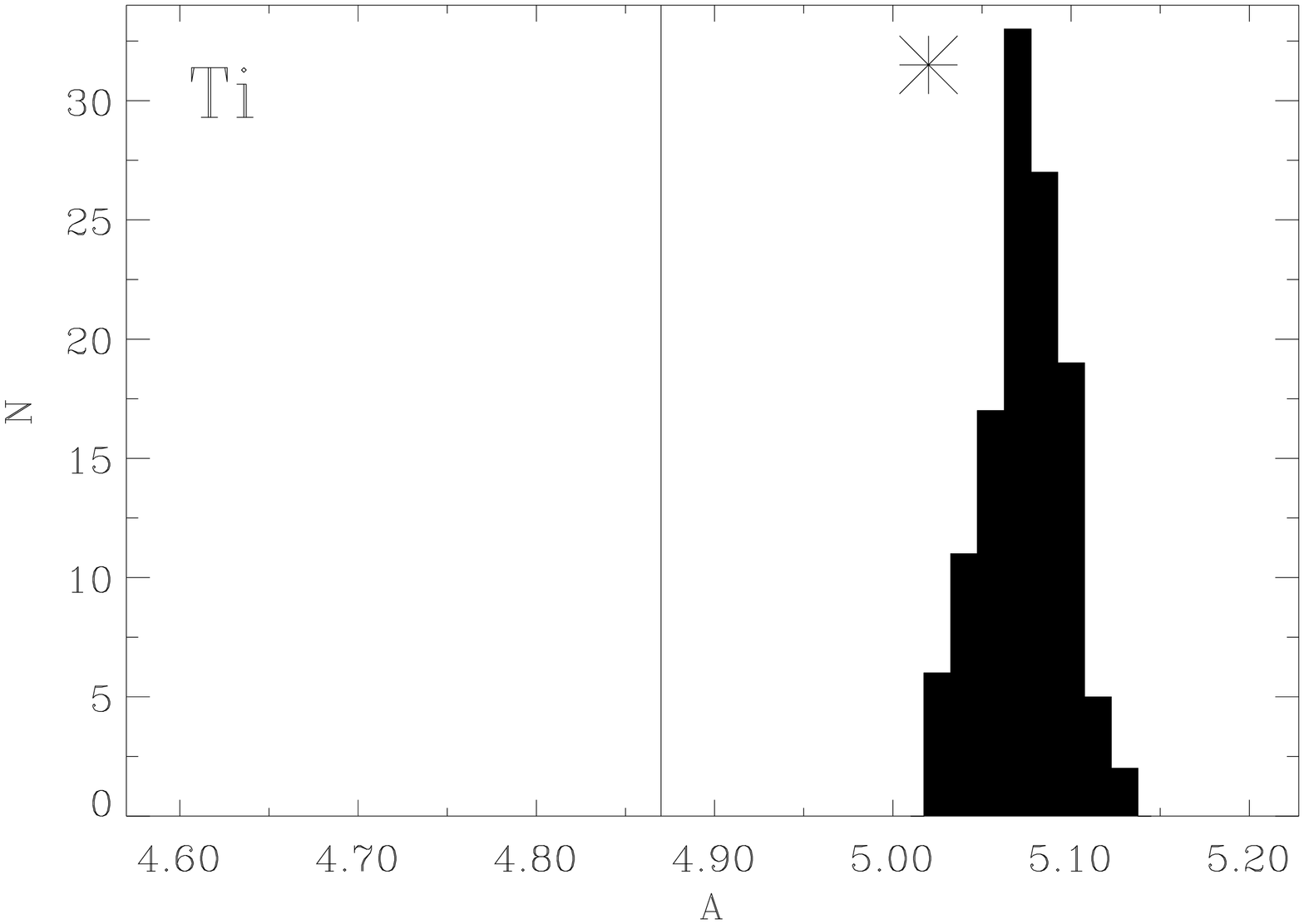}  
\includegraphics[width=5cm,angle=90]{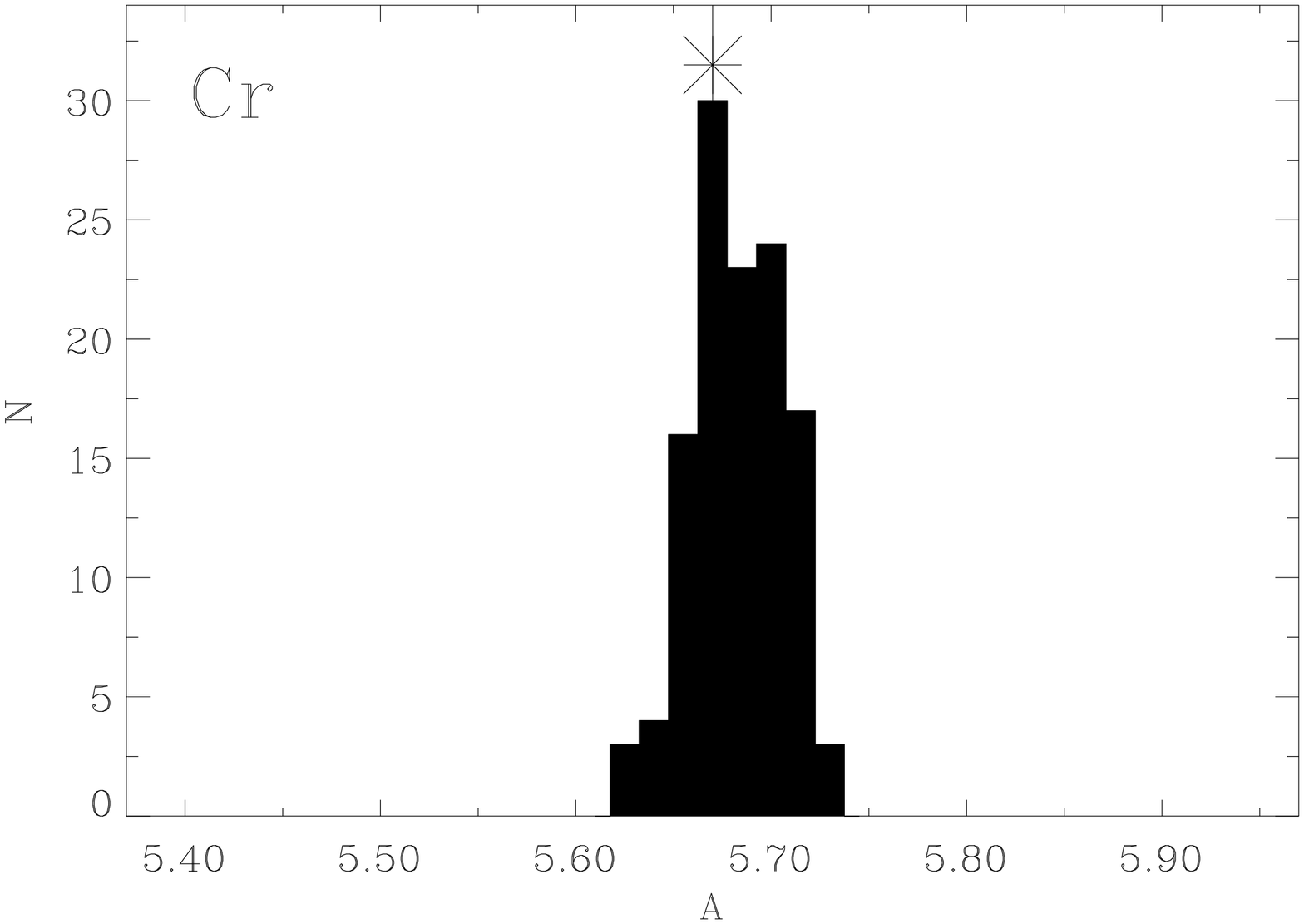}  
\includegraphics[width=5cm,angle=90]{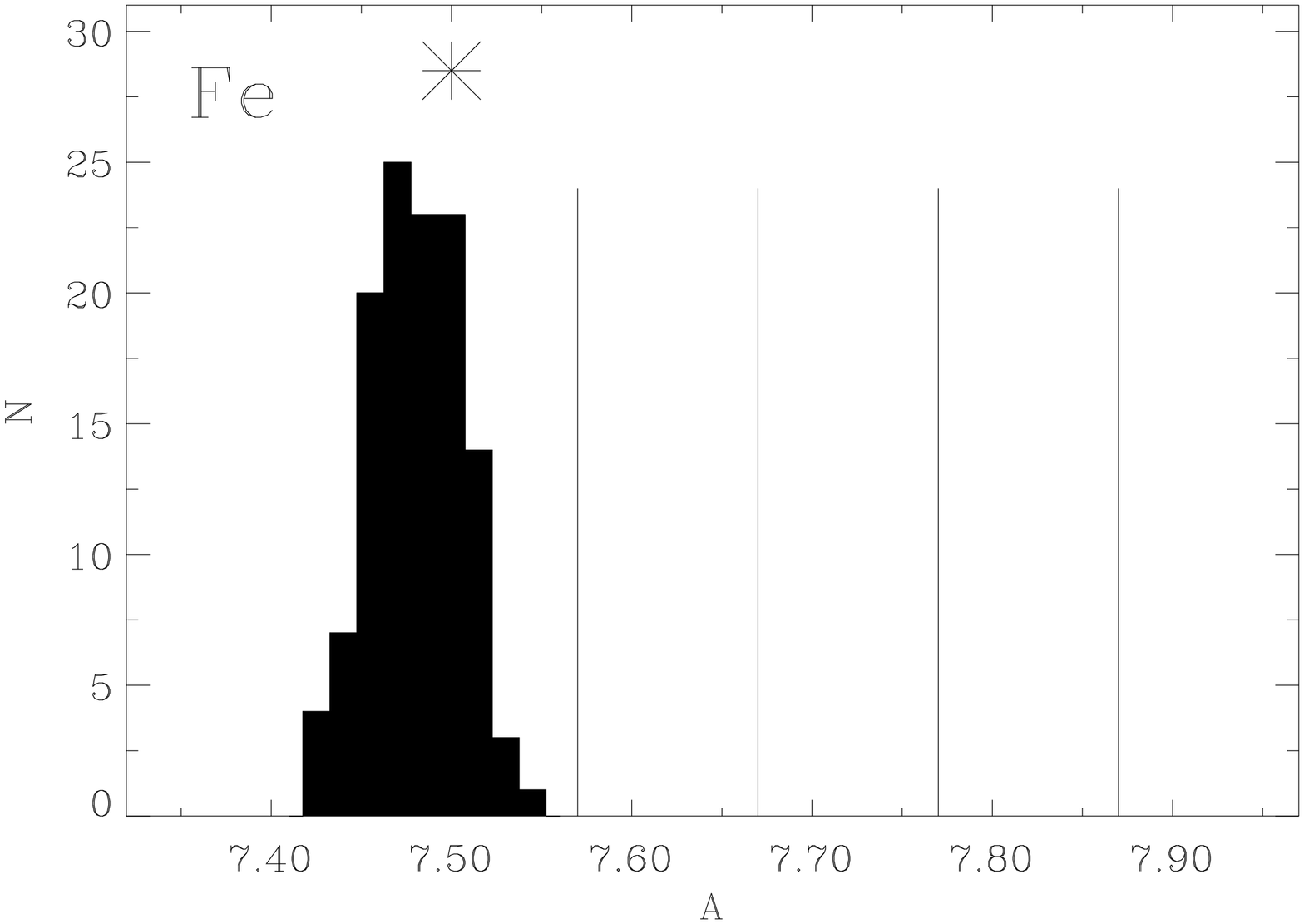}
\protect\caption[ ]{
Histograms of the initial [spikes at $\Delta$ A $= -0.2, -0.1, 0.0, +0.1$,  and $+0.2$ dex or simply $\Delta$ A $= 0.0$ 
from the Anders \& Grevesse (1989) photospheric abundances] and
final abundances for the 120 test runs. The asterisks mark the {\it true} 
abundances used to build the hydrodynamical simulation.
\label{f5}}
\end{figure}

\begin{figure}[!ht]
\centering
\includegraphics[width=15cm,angle=0]{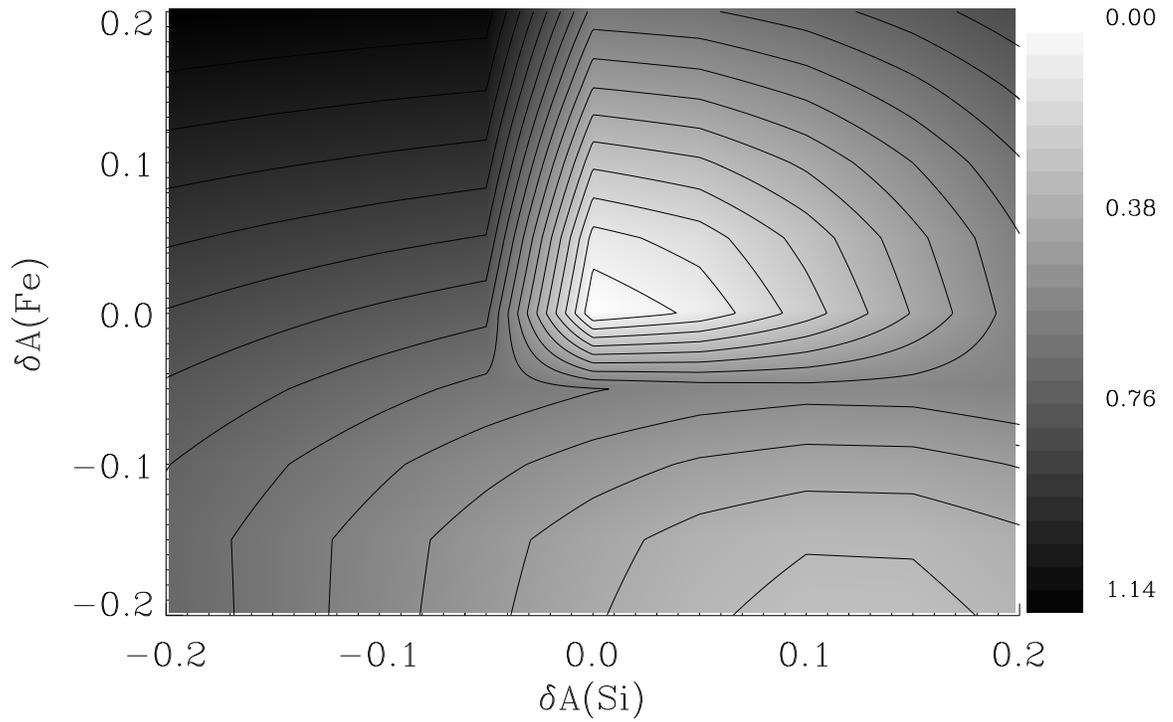}    
\protect\caption[ ]{
Response of the reduced $\chi^2$, normalized to its minimum value, 
to changes
in the optimal abundances of iron and silicon. Both the gray scale and 
the contours
use a logarithmic scale, and thus the units in the color-code bar are dex.
\label{f5_b}}
\end{figure}

More interesting, although closely connected with the thermal
structure, is the scatter among the retrieved abundances and its
resemblance to the actual abundances used in the simulation. Figure
\ref{f5} compares the histograms of the initial and final abundances
for the 120 testing runs. The initial abundances correspond to the
photospheric abundances of Anders \& Grevesse (1989), with the abundances of Mg, Si,  and Fe altered by $\Delta$
A $= 0.0, \pm 0.1$ or $\pm 0.2$ dex, and their histogram corresponds to
the  spikes in each panel of the figure. The histogram of the final abundances is identified by the
filled bars. The {\it true} abundances used in the simulation
 and three-dimensional radiative transfer are marked by asterisks in
 the figure. Table 3 gives the exact values used (GS98), as well as the weighted mean (and standard  deviation) of the abundances recovered,
labeled as "A(Sim)".  The mean difference between  {\it true} and
recovered abundances is  $-0.01 \pm 0.02$ (SEM\footnote{Standard error of
the mean: $\sigma/\sqrt{N}$}) dex, while the rms
scatter is just 0.04 dex. Figure \ref{f5_b} displays the logarithm of the
reduced $\chi^2$, normalized to its minimum, as a function 
of the variation in the abundances of iron and silicon from their optimal 
values [$\delta$A(Fe) and $\delta$A(Si)]. Changes of $\pm 0.2$ dex 
bring with them a degradation in the reduced $\chi^2$ by more than 
an order of magnitude.

The agreement is remarkable, given that the discrepancy between the
{\it true} abundances and those used to initialize the  inversion code
reaches up to 0.4 dex, and that the inversion code tries to reproduce
the input spectrum using a 1D representation, while the line formation
is a highly non-linear process occurring in a 3D environment. The small
differences of about 0.04 dex  could well partly be attributed to the fact
that  continuum opacities,  equation of state,   partition functions, and
other details are not exactly identical for  the numerical simulations 
and  the inversion code.  This result gives confidence in the abundances
derived from inversions working with  homogeneous atmospheres. In the
next section we attempt the inversion of the solar spectrum. Given this
case of success, our prospects to derive reliable abundances are
optimistic, although the practical application involves several error sources
 not present in the previous exercise, such as possible departures from LTE.

\section{The solar spectrum}

Inversions of the solar  spectrum have already been presented
 by Allende Prieto et al. (1998) and Frutiger et al. (2000). Still, neither of them 
can be considered definitive. Both works were aimed at testing  
inversion techniques, rather than producing a serious analysis of 
the chemical abundances in the solar photosphere. Nonetheless, 
in view of the results from the previous section, this is indeed an 
 interesting issue that deserves  consideration.

We address the inversion of the solar flux spectrum  armed
 with a carefully selected 
and comprehensive line list, which includes lines of the
 most important electron donors:  Mg, Si, and Fe.  Similarly to Allende 
Prieto et al. (1998), we extracted the observations from the solar atlas of 
Kurucz et al. (1984). Given that the
inversion adopts a static model, the synthetic profiles are not able to reproduce the observed asymmetric line shapes, and thus 
it is unlikely that noise in the
observed profiles is important, at least down to very low
signal-to-noise ratios. But uncertainties
 in the atomic data  could be an important source of error. We have
 selected radiative transition probabilities from the best sources in
the literature, searching mainly for laboratory measurements, and we  expect
errors to be roughly symmetric. Redundancy, that is, including several
lines that are formed over the same depth range, will then prevent
organized errors. A symmetric error distribution  might not be true for
the adopted values for the natural damping, collected from VALD and
likely coming from a single source (Kurucz's theoretical
calculations), but that parameter's contribution is typically small,
and definitely unimportant within a large sample.

As explained in \S 2, we have included   wings of strong metal lines,
which are expected to be weakly affected by departures from LTE and 
allow us to improve the
depth coverage. An adequate treatment of collisional interactions with
neutral hydrogen is  crucial,  because this is truly a 
potential source of systematic
errors. The traditional Van der Waals approximation is by no means
useful for this task, as it substantially 
underestimates the observed collisional
damping to varying degrees. To the contrary, the studies in 
the literature that employed
the Anstee, Barklem \& O'Mara  theories have proved this approach to be 
very promising (Asplund et al. 2000b; Barklem et al. 2000a).
Fortunately, we do not need to take any risk. Our line database is
sufficiently large and diverse that  without the wings of strong metal lines we should
have an adequate  depth coverage.  As a matter of fact, we can further
test the collisional broadening theories
 by comparing   inversions with, without, and only with, the wings of
strong lines as input data.

\begin{figure}[!ht]
\centering
\includegraphics[width=15cm,angle=0]{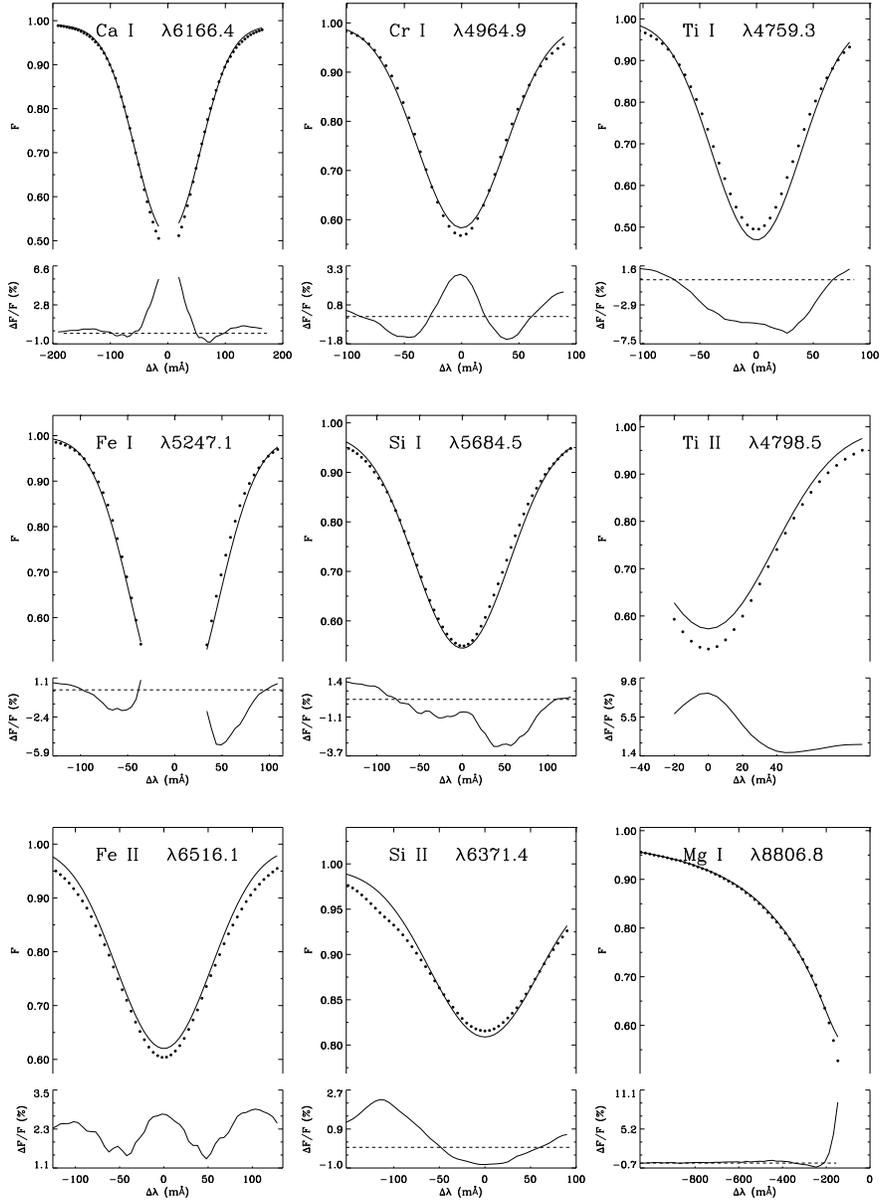}  
\protect\caption[ ]{
Comparison between the solar observed flux spectrum (filled circles) 
and the same lines synthesized with the homogeneous 
static model obtained from the inversion. 
\label{f6}}
\end{figure}

We proceed with the inversion  of the solar spectrum following
 the same scheme 
we applied for the 3D simulation. We carried out runs for each 
of the eight datasets with an initial
micro-turbulence of $\xi_0 = 0.1, 1.0, 1.5 $ and 2.0 km s$^{-1}$, always
 with $v_{\rm rot} \sin i = 1.88$ km s$^{-1}$, and
with the abundances of iron, magnesium and silicon simultaneously
changed by $\Delta$ A $= -0.2, -0.1, 0.0, +0.1$ and $+0.2$ dex, for a
total of $8 \times 4 \times 5 = 160 $ cases. All of them converged. Identically
 to the case of the simulation, we started out from an isothermal
model with a macroturbulence of 2.0 km s$^{-1}$, progressively  
augmenting the complexity of the temperature structure. When attempting 
the inversion with seven nodes in temperature, wiggles started to appear, so
the process was stopped after reaching six nodes. At the same time,
the chemical abundances of the elements with lines in the input spectrum
were allowed to change (taking into account their effect on the continuum opacities), and so was the depth-independent micro-turbulence. For a 
single run, the numbers of frequencies and lines were the same as in the 
inversion of the hydrodynamical simulation, because in \S 3 we used  a 
mask to select the  part of the profiles that would be  useful in a
real application. We also performed an inversion with all the lines 
simultaneously. Figure \ref{f6} shows some examples of 
the degree of agreement achieved between observed and synthetic 
profiles. It is interesting to mention that the global agreement 
between the input profiles and those produced by the atmosphere 
recovered by the inversion was slightly better for the simulation 
than for actual observations. 
Noticeably however, the wings of strong lines produced by the simulations 
could not be matched as well as those observed.

\begin{figure}[!ht]
\centering
\includegraphics[width=9cm,angle=90]{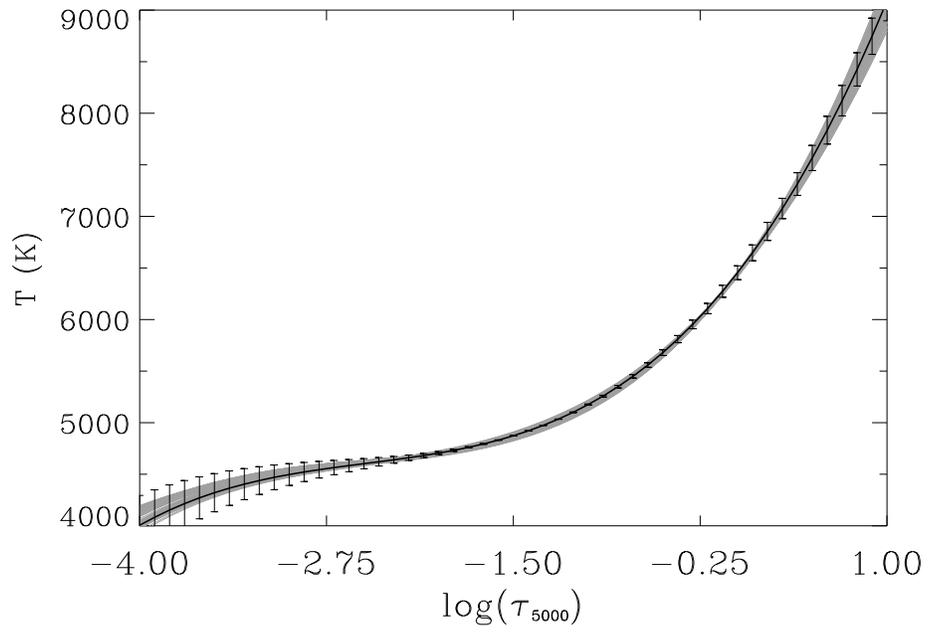}  
\protect\caption[ ]{
Temperature structures derived from the inversion of different datasets from the solar flux spectrum, 
with different initial conditions (dotted gray lines). The solid black  
with error bars corresponds to the inversion of all the available data simultaneously. 
\label{f7}}
\end{figure}

The temperature stratification of the 160 retrieved models is shown in Figure \ref{f7} (gray dotted lines). The black solid 
line with error bars corresponds to the run with all the lines simultaneously. Within the explored ranges, no matter the initial abundances or  micro-turbulence, or which dataset of lines
was involved,  we arrive at a very narrow range of solutions.  The model 
derived from the simultaneous inversion of all lines appears in Table 2.

\begin{figure}[!ht]
\centering
\includegraphics[width=9cm,angle=90]{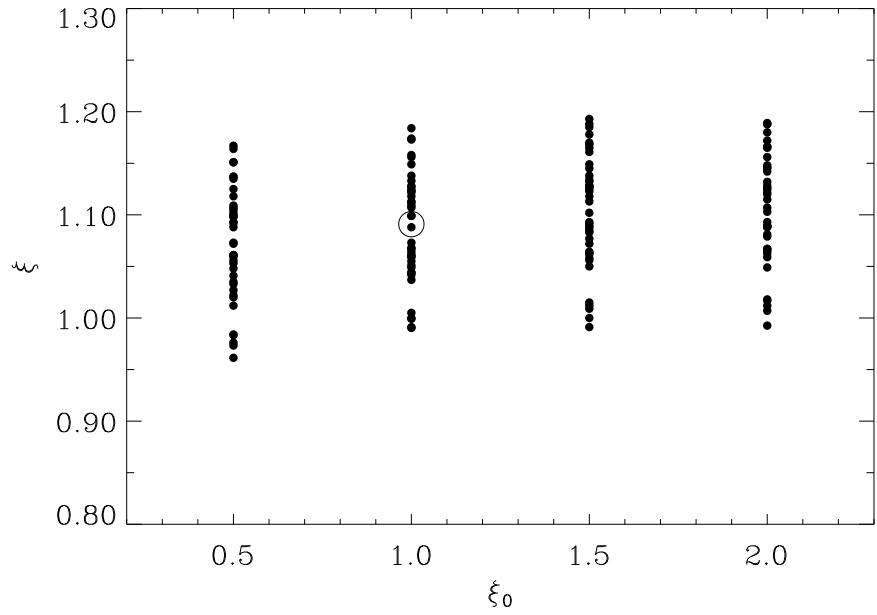}  
\protect\caption[ ]{
Final ($\xi$) versus initial ($\xi_0$) values for the micro-turbulence 
in all the 160  runs with solar data. The mean value $\xi = 1.1 \pm 0.1$ 
km s$^{-1}$
is essentially independent from the initial guess. The open circle shows 
the value retrieved from the inversion of all the lines simultaneously.
\label{f8}}
\end{figure}

The retrieved  micro-turbulence is   $1.1 \pm 0.1$ km s$^{-1}$,  quite 
independently from the value  adopted as a starting guess ($\xi_0)$, 
as shown in Figure \ref{f8}. We find $1.54 \pm 0.02$ km 
s$^{-1}$ for the macro-turbulence.  We convolved the synthetic profiles with 
a Gaussian instrumental profile equivalent to  the spectral  resolution of
the solar spectrum (R$\equiv \lambda/\delta\lambda \simeq 5 \times 
10^5$). We note that the value for the micro-turbulence we refer to 
corresponds to $\sqrt{2} \sigma$  of a Gaussian, (as
traditionally used in the literature, e.g.,  Struve \& Elvey 1934), 
 the quoted value for the macro-turbulence  corresponds to the 
$\sigma$ of a Gaussian, and the $\delta\lambda$ used
 to calculate the spectral resolution is the FWHM of the   
Gaussian instrumental profile ($\simeq 2.355 \sigma$).

\begin{figure}[!ht]
\centering
\includegraphics[width=5cm,angle=90]{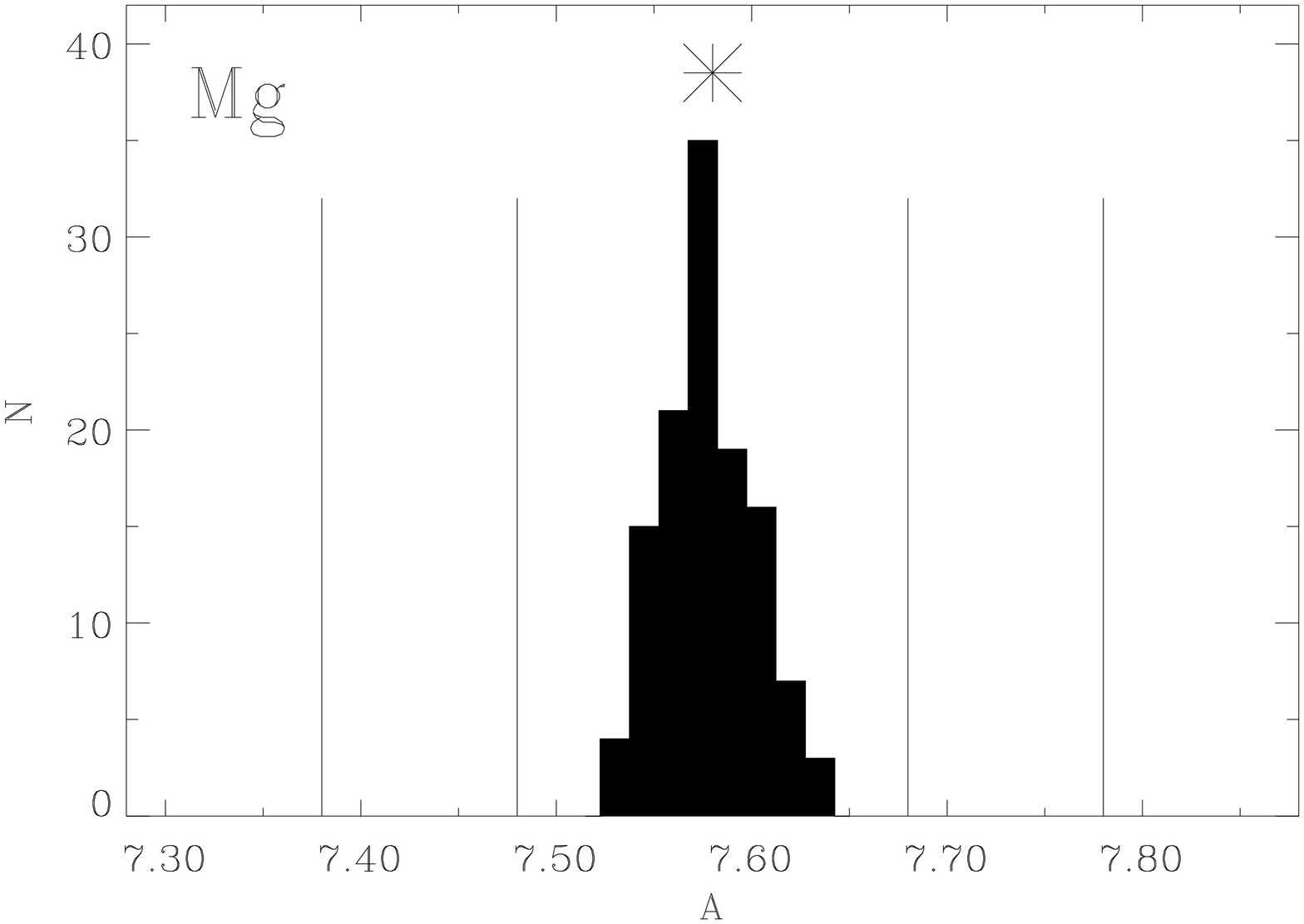}  
\includegraphics[width=5cm,angle=90]{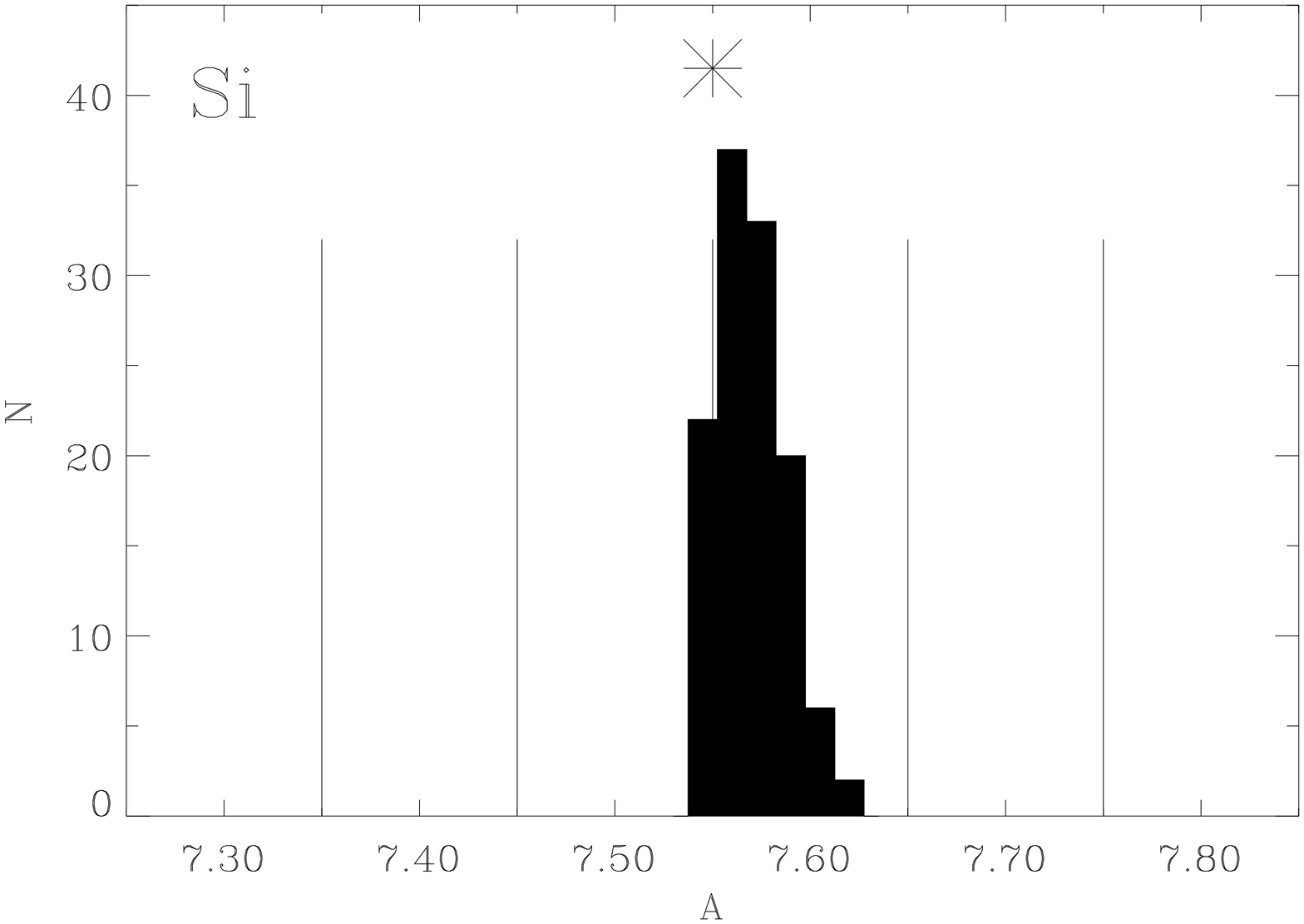}  
\includegraphics[width=5cm,angle=90]{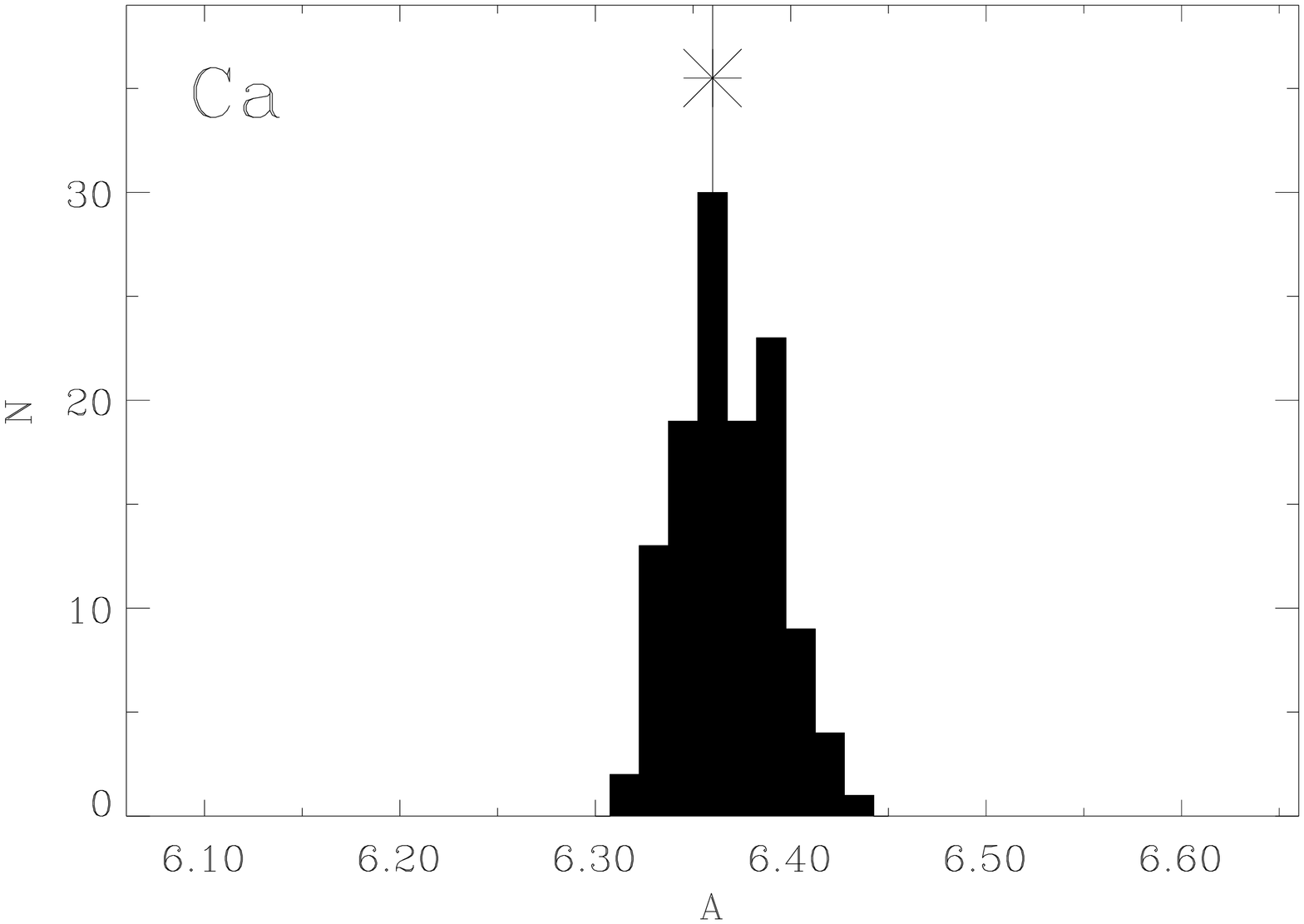}  
\includegraphics[width=5cm,angle=90]{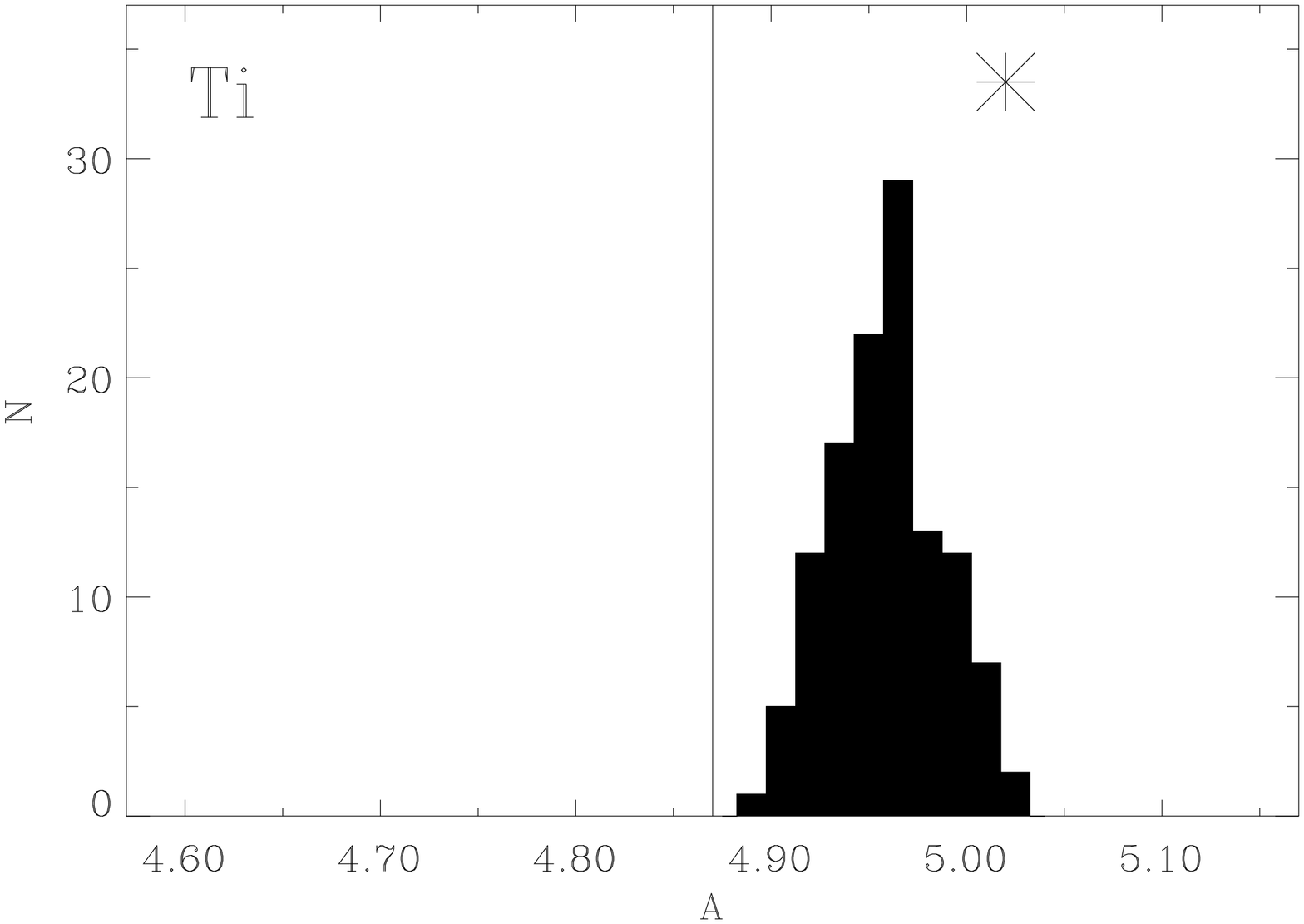}  
\includegraphics[width=5cm,angle=90]{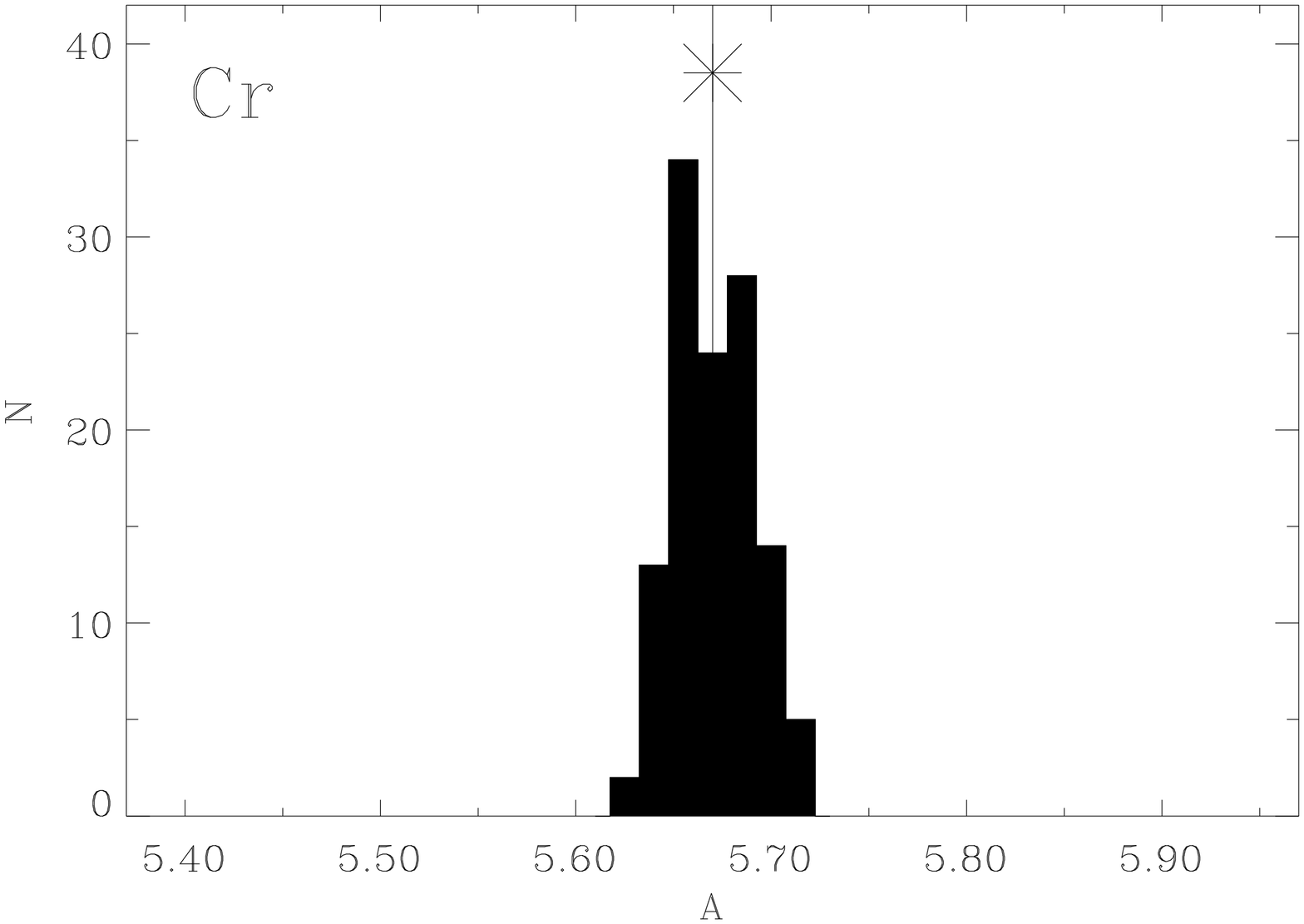}  
\includegraphics[width=5cm,angle=90]{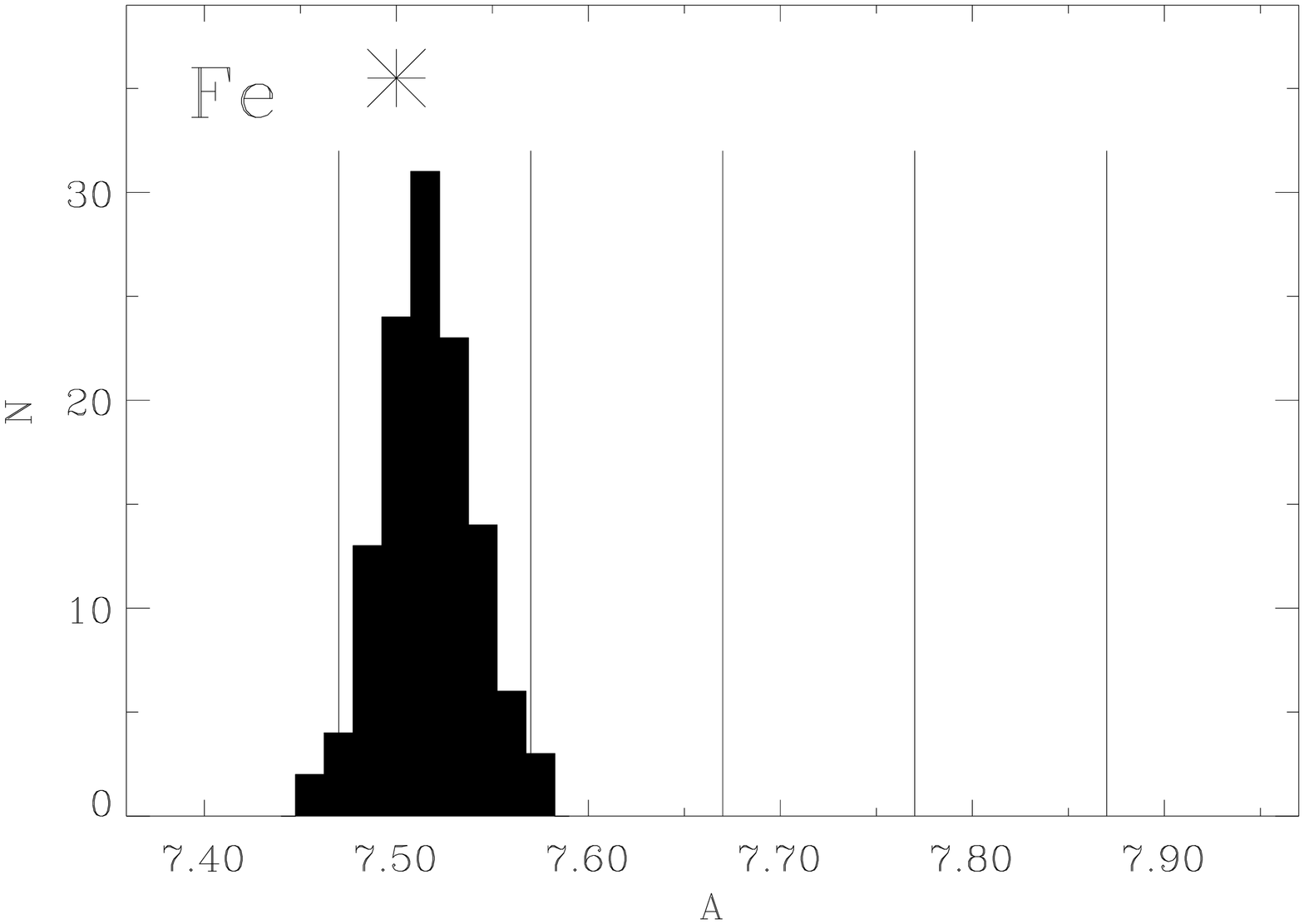}  
\protect\caption[ ]{
Histograms of the initial [spikes at $\Delta$ A $= -0.2, -0.1, 0.0, 
+0.1$ and $+0.2$ dex or simply $\Delta$ A $= 0.0$ from the Anders 
\& Grevesse (1989) photospheric abundances] and
final abundances for the 160  runs with solar data. 
As a reference, the asterisks mark the {\it photospheric}  
abundances published by  Grevesse \& Sauval (1998).
\label{f9}}
\end{figure}

Figure \ref{f9} compares the histograms of the initial (spikes) and final
 (filled bars) abundances for the 160 testing runs. The initial abundances correspond
to the {\it photospheric} abundances of Anders \& Grevesse (1989), which
for Mg, Si, and Fe are altered by 
$\Delta$ A $= 0.0, \pm 0.1$ or $\pm 0.2$ dex, as indicated by
the five spikes in the corresponding panels of the Figure.
The asterisks show the revised
photospheric values of  GS98. The weighted mean
(and standard  deviation) of the abundances recovered for Mg, Si, Ca, Ti, Cr,
and Fe  are listed in Table 3, labeled with "A(Obs)".  
The agreement with GS98,
what we can consider the commonly accepted values,
 is excellent, always within the uncertainties. Although having some
 overlap of the error bars, a slightly larger difference is found for
Ti, whose meteoritic  abundance
 (listed by Grevesse  \& Sauval: $4.94 \pm 0.02$) is in better agreement
with the value we find.

\begin{figure}[!ht]
\centering
\includegraphics[width=6cm,angle=90]{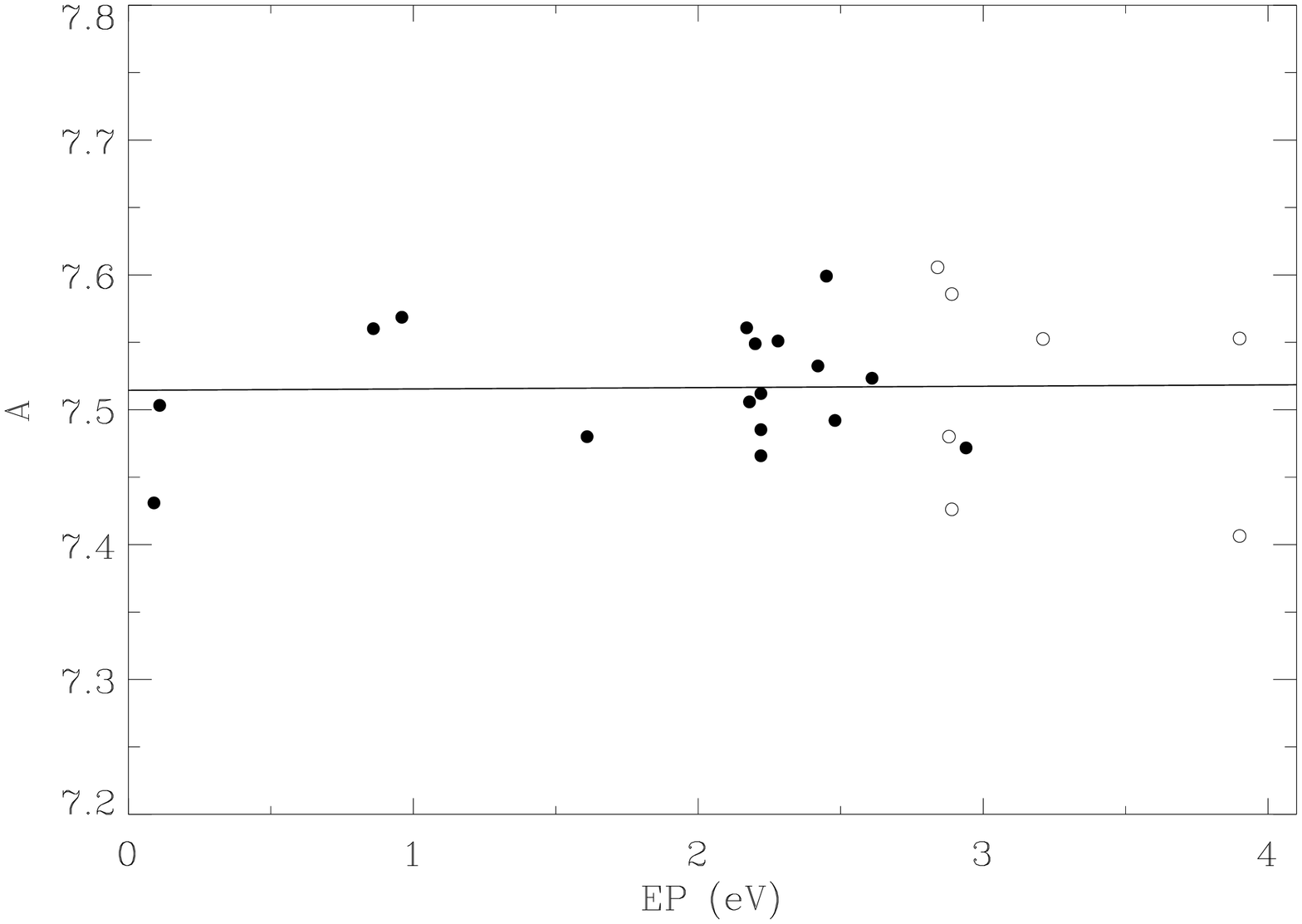}  
\includegraphics[width=6cm,angle=90]{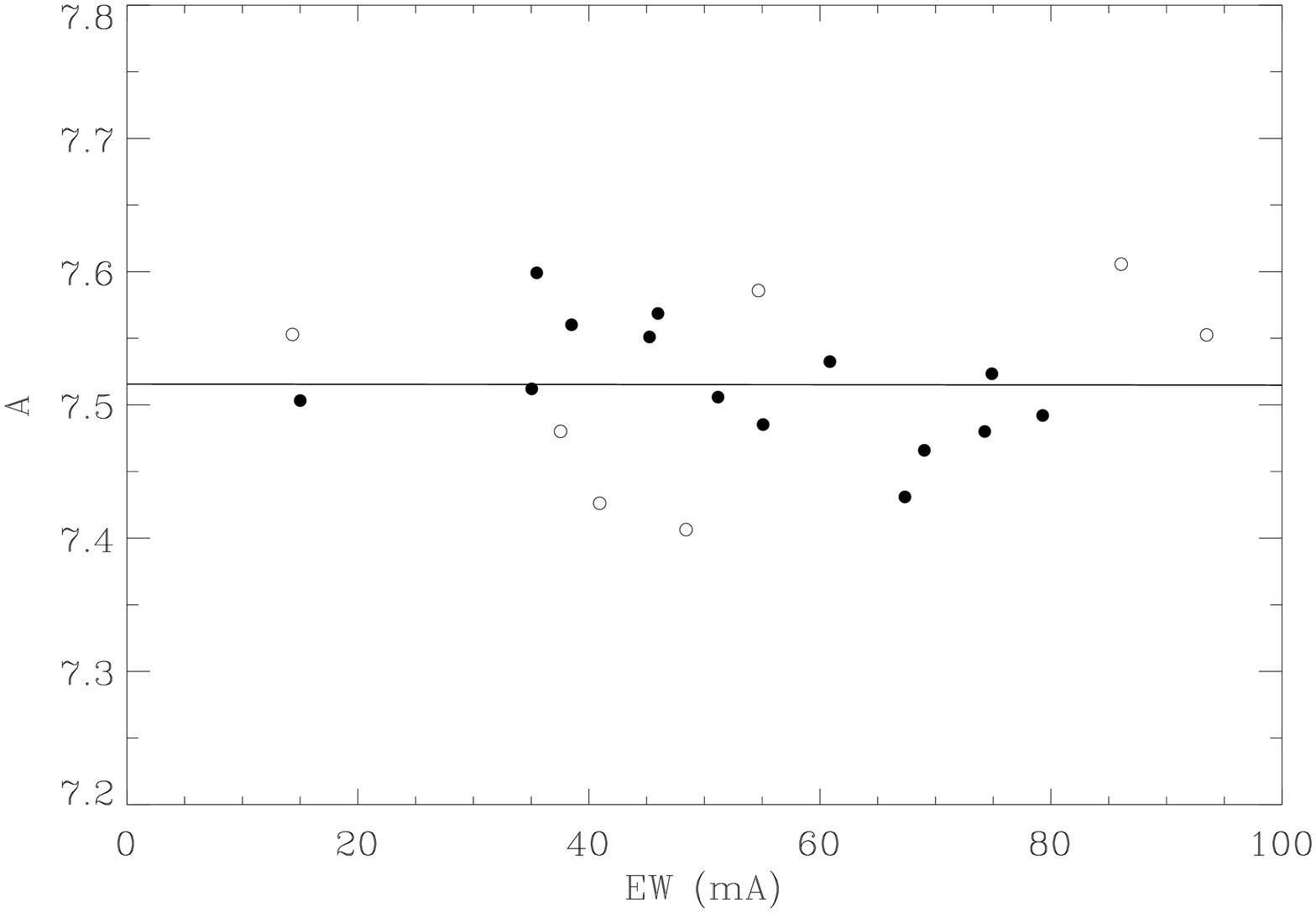}  
\includegraphics[width=6cm,angle=90]{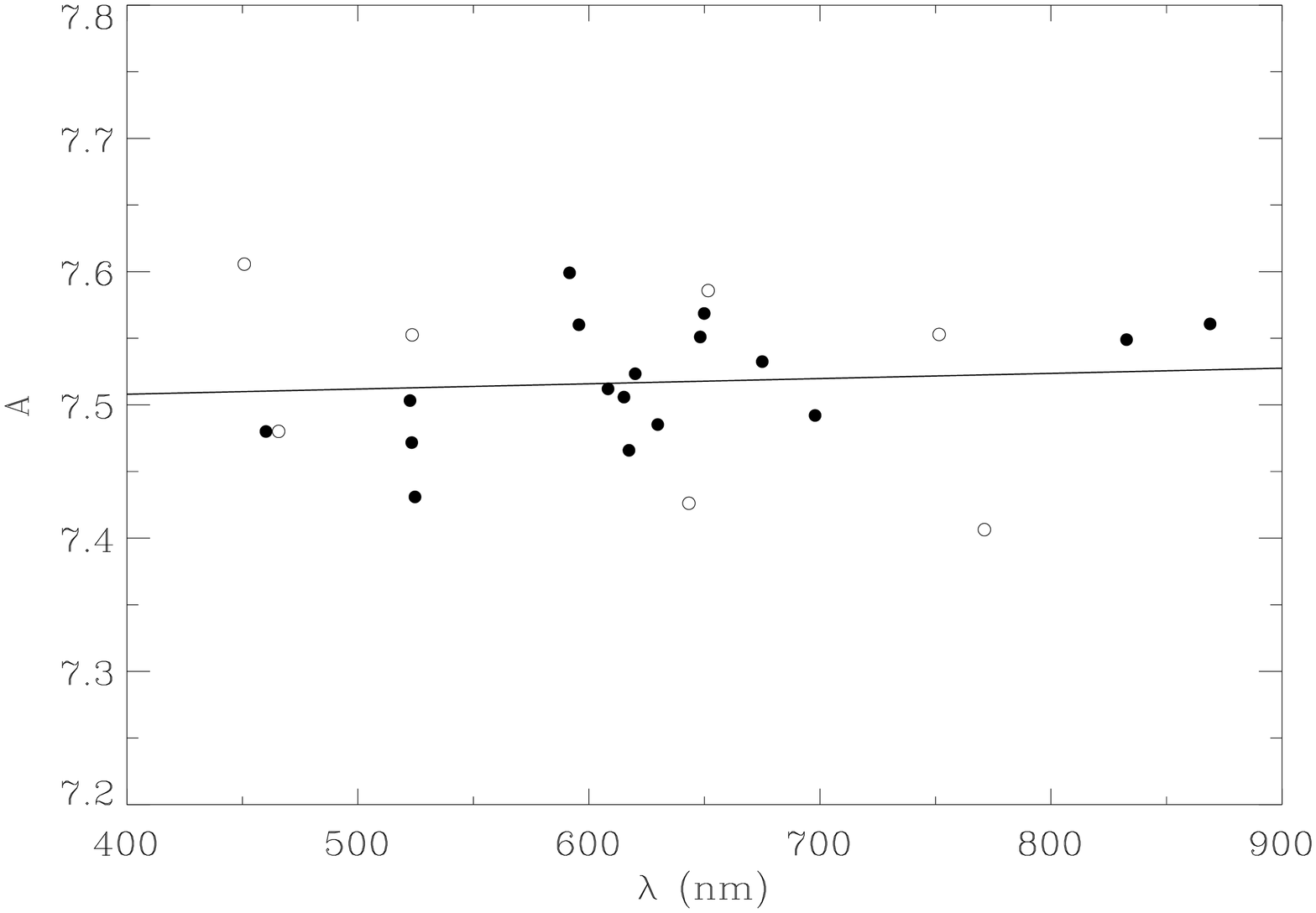}  
\protect\caption[ ]{
The abundances derived from the analysis of individual iron lines 
with the solar model obtained from the inversion are checked for 
trends versus excitation potential (EP), equivalent width (EW) and 
wavelength ($\lambda$). The filled and open symbols identify neutral 
and ionized lines, respectively. The solid lines are linear fits minimizing 
the $\chi^2$ error statistic.
\label{f10}}
\end{figure}

Apart from the quality with which the synthetic profiles fit the 
observations, 
which is limited by the line asymmetries, a rather interesting  way to test 
the new model is to check for  differences between the abundance of a given
element as derived from lines with different atomic parameters. 
Rather than force all the lines of a given element 
to be fitted with a single abundance, we can find the abundance that
best matches each observed profile and then check for trends against
excitation potential (EP), equivalent width (EW), or
 wavelength ($\lambda$). We can fix the model structure 
and invert one line at a time, with the abundance of the 
corresponding element as the only  parameter to be determined. Figure 
\ref{f10} shows that no trend is present when iron lines are considered.
The filled and open symbols  are used to distinguish neutral from ionized 
lines, and there is no difference between those two
groups.  The {\it equivalent width}  in the second panel of the 
figure corresponds to the 
absorption produced by 
the part of the line profile used in the inversion, which is close to, or 
smaller than, the real equivalent width of the whole line profile. As the 
difference between the {\it equivalent width} used in the plot and the real
equivalent width can be very large for the {\it strong} lines, we have limited
the comparison in the middle panel to EW $< 100$ m\AA. 
The solid lines are linear fits to the data, obtained by 
minimizing the $\chi^2$ error statistic, and the slopes of the 
abundance against 
EP, EW and wavelength are $(1 \pm 12) \times 
10^{-3}$, $(-8 \pm 600) \times 10^{-6}$ and $(4 \pm 11) \times 10^{-6}$, 
respectively, and therefore insignificant.

As a test of the continuum opacities, we removed all opacity sources
 except H$^{-}$, finding an atmospheric structure hardly distinguishable 
from the model  in Table 2, and abundances that differed
 from those in Table 3  by less than 0.002 dex. Using our final solar model
 and abundances, we checked the influence of different approximations to 
 estimate the lowering of the ionization potentials by Debye-H\"uckel shielding
 (see, e.g., Drawin \& Felenbok 1965). Although different formulae predict
 very discrepant results, comparison of the most extreme estimates showed that 
 this effect never alters the abundances of the considered species by more 
 than 0.01 dex. The typical abundance shift for most of the lines of 
 neutral species is about $-0.005$ dex, and it is much smaller and has
 the opposite sign for lines of singly-ionized species.

\begin{figure}[!ht]
\centering
\includegraphics[width=10cm,angle=0]{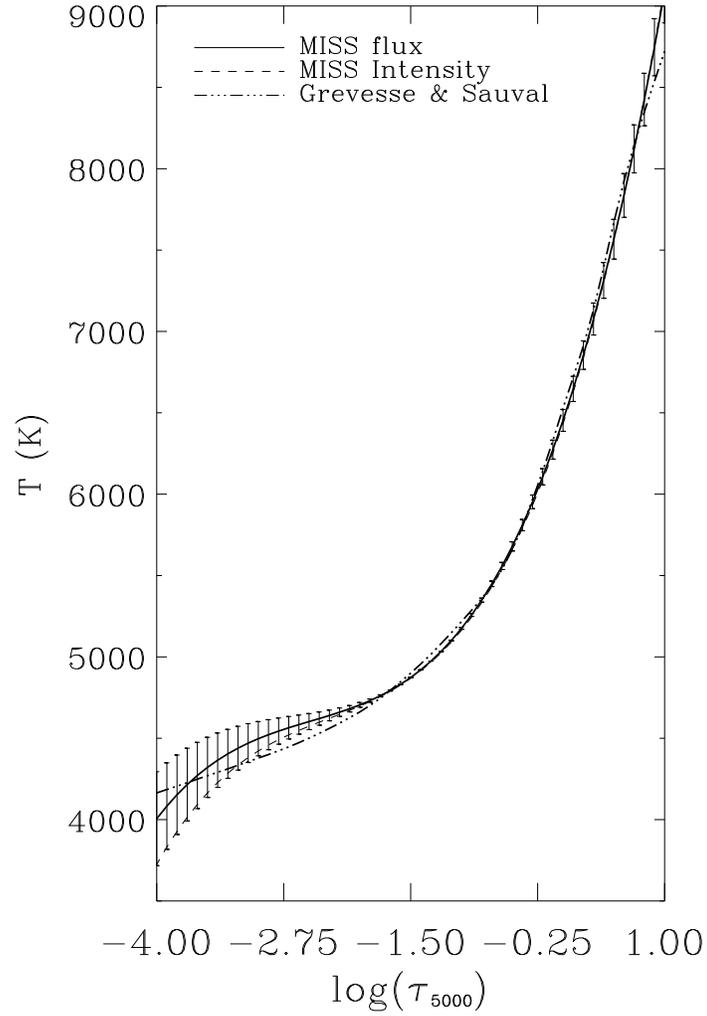}  
\protect\caption[ ]{
Temperature structure derived from  the inversion of the solar flux spectrum (solid line with error bars) and from the inversion of the solar spectrum 
from the center of the disk (dashed curve; error bars omitted for clarity). Also shown is the model of Grevesse \& Sauval (1999), obtained by
modifying 
 the Holweger-M\"uller atmosphere (Holweger \& M\"uller 1974).
\label{f11}}
\end{figure}

The  profiles of the same lines 
we used from the solar atlas of Kurucz et al. (1984), were also extracted from the spectrum of the disk center included in the atlas of Neckel (Brault \& Neckel 1987; Neckel 1999). 
The much smaller region of the disk  that is 
averaged in the aperture, and the vertical view angle, produce
 sharper and much
more asymmetric profiles, which are an even harder challenge for 
the symmetric profiles produced by the static model atmosphere.  
Figure \ref{f11} confirms  that the recovered structure (dashed curve) 
is very similar to that  derived from the  flux spectrum (solid line with error bars), while the micro-turbulence is, understandably, smaller:
 0.8 km s$^{-1}$. For  reference, the empirical model published 
by Grevesse \& Sauval (1999) is plotted with a dash-dotted line.

\begin{figure}[!ht]
\centering
\includegraphics[width=8cm,angle=90]{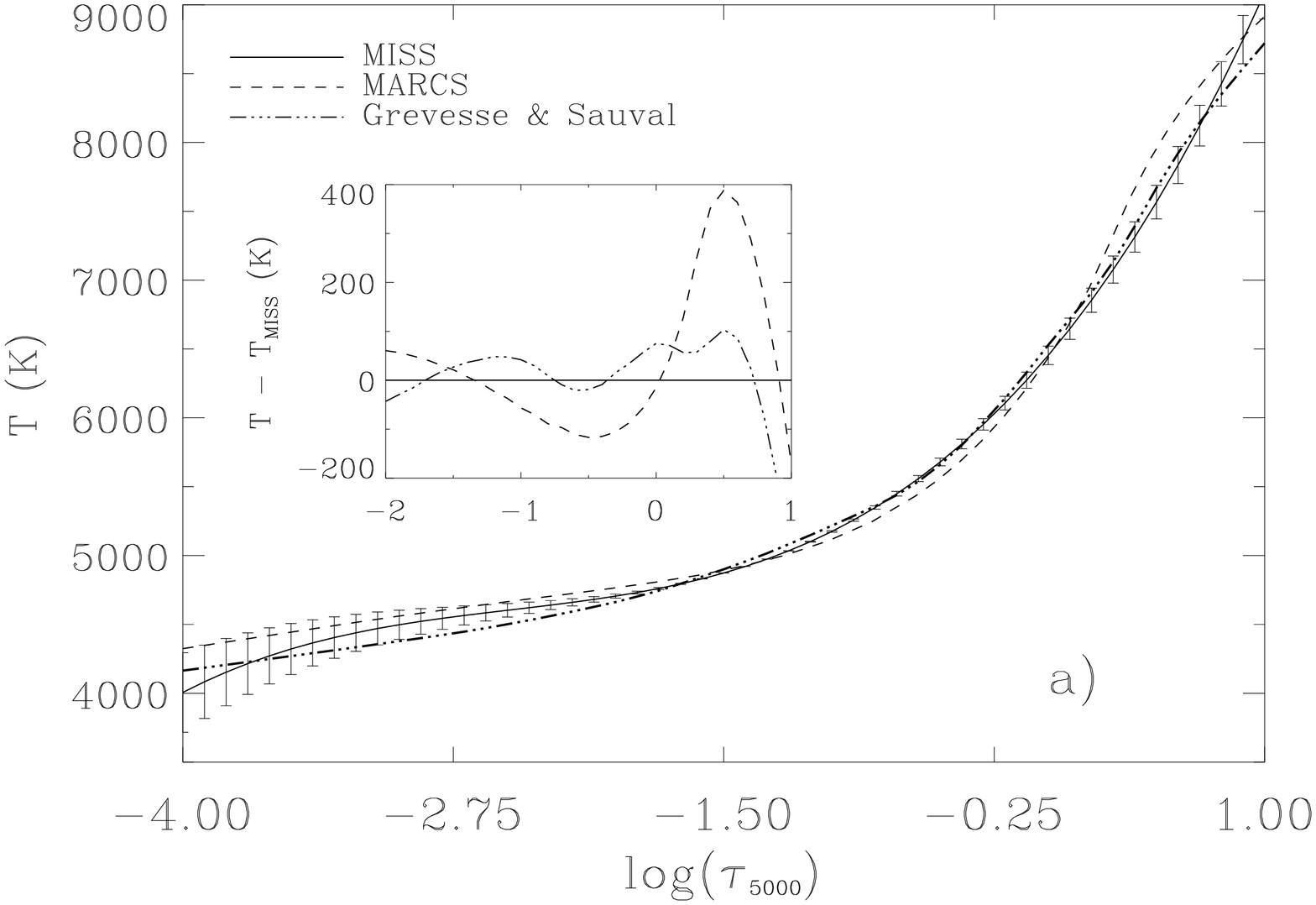}  
\includegraphics[width=8cm,angle=90]{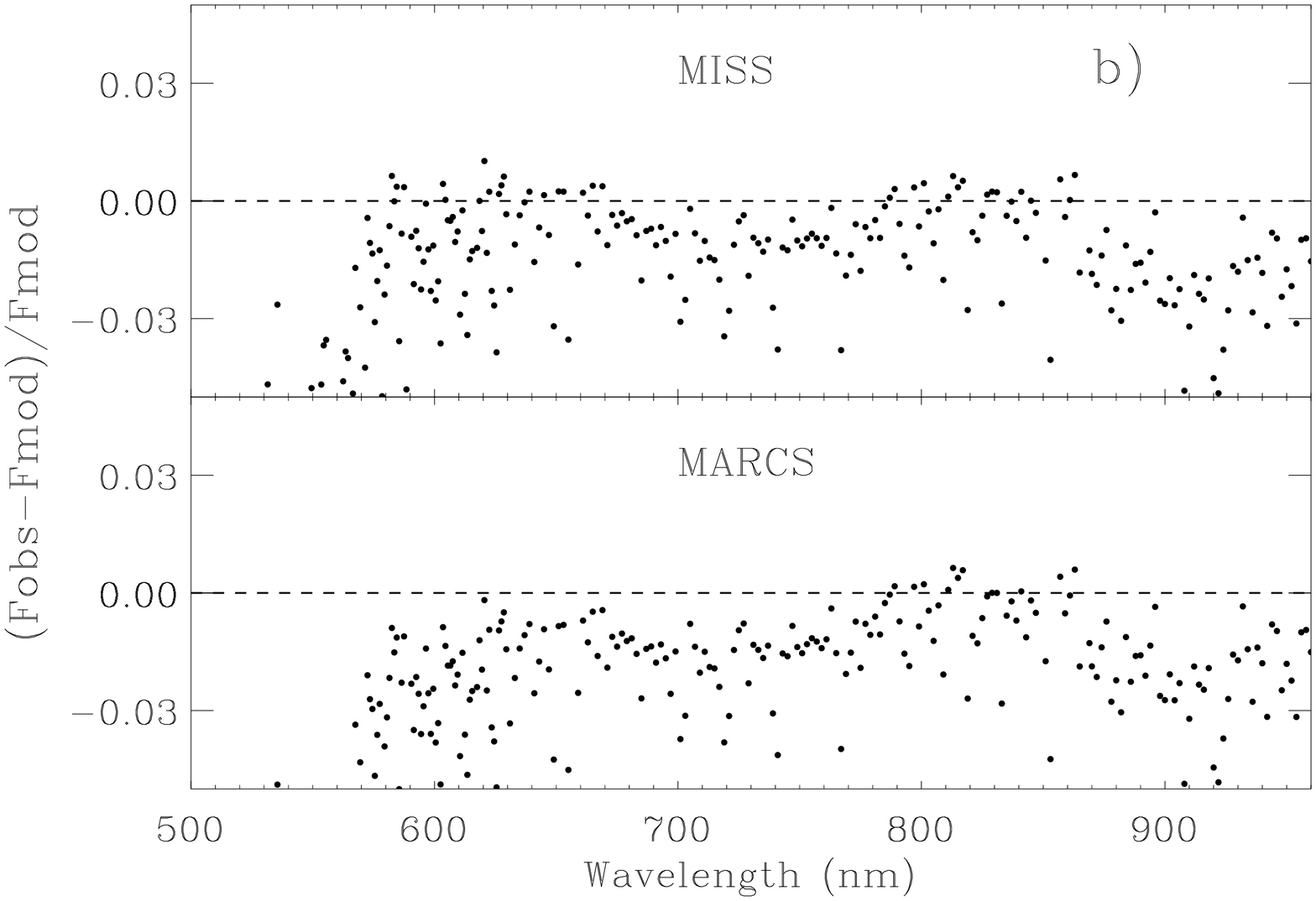}  
\protect\caption[ ]{
a): Comparison between the temperature structure of our semi-empirical model (solid line with error bars), and the flux-constant model computed with the MARCS program
 (dashed line). The semi-empirical model of Grevesse \& Sauval (1999) is also shown as reference (dashed-dotted curve). 
b): Normalized difference between the observed fluxes of the Sun and the predicted 
pure-continuum flux of the semi-empirical {\tt MISS}  and the flux-constant MARCS solar models. The large high frequency scatter is produced by line absorption in the solar atmosphere, whereas the low frequency variations
 are mostly associated with broad absorptions in the Earth's atmosphere. 
 \label{f12}}
 \end{figure}

It is interesting to compare the model we found against
 theoretical models, which are widely used in astrophysical
 applications.  Figure \ref{f12}a compares the temperature structures
derived by {\tt MISS} with a MARCS model (Gustafsson et al. 1975;
Asplund et al. 1997) with $T_{\rm eff} = 5777$ K, $\log g= 4.44$, and
$\xi = 1.00$ km s$^{-1}$, which uses the mixing length
 recipe to account for the convective flux and  Grevesse \& Sauval's
standard {\it photospheric}  composition. The Grevesse \& Sauval (1999)
empirical model, based on a
 modification of the Holweger \& M\"uller (1974) model, is also shown,
 and not surprisingly, is in much closer agreement with our model than
the theoretical structure.  The  temperature of the theoretical model
in the deepest layers
 is  largely dependent on the particular choice for the parameters used
 in the mixing-length formulation.   In particular, the differences
between $-1< \log \tau < 0.2$ are very important for astrophysical
applications.
 The MARCS structure is up to  100 K cooler  than {\tt MISS} between
 $-1< \log \tau < 0$, but   hotter below $\log \tau \simeq 0$.

 Figure \ref{f12}b compares the models' predictions against 
the   flux measured in the optical and
 near-IR, as  critically compiled   by Colina, Bohlin \& Castelli (1996). At
wavelengths shorter than $\sim$ 600 nm the abundance of absorption
lines depresses the observed flux, and makes this comparison
meaningless.  Colina et al. concluded that the observations available
for  wavelengths longer than 960 nm are affected by large
 systematic errors and, therefore, we discarded that region from the
 comparison.   The large high frequency scatter is produced by line
 absorption in the solar atmosphere, whereas the low frequency
variations
 are mostly associated with broad absorptions in the Earth's
 atmosphere.  Surprisingly, the fluxes predicted by the two models are
  closer than expected from the differences in their thermal
 structures. Discrepancies in the continuum opacities and/or the
 equation of state used by {\tt MISS} and MARCS solve the potential problem.
 Such discrepancies should be responsible at least for a share of the
 differences in temperature between the two models, and deserve further
investigation, although they are certainly outside of the scope of this
paper.

\section{Completeness and redundancy in the input data}

From the convergence to very similar structures 
of  inversions with eight different datasets, 
all of them including {\it weak-neutral},  {\it weak-ionized}, 
and {\it strong} lines, it is apparent that  each dataset provides
very similar  information about the solar photosphere.
A practical question immediately arises. Are all the lines  
necessary? We are interested in knowing what features contain 
most of the information and whether  it is possible to use only  a few, 
carefully selected,  lines to recover the photospheric structure.

\begin{figure}[!ht]
\centering
\includegraphics[width=5.2cm,angle=0]{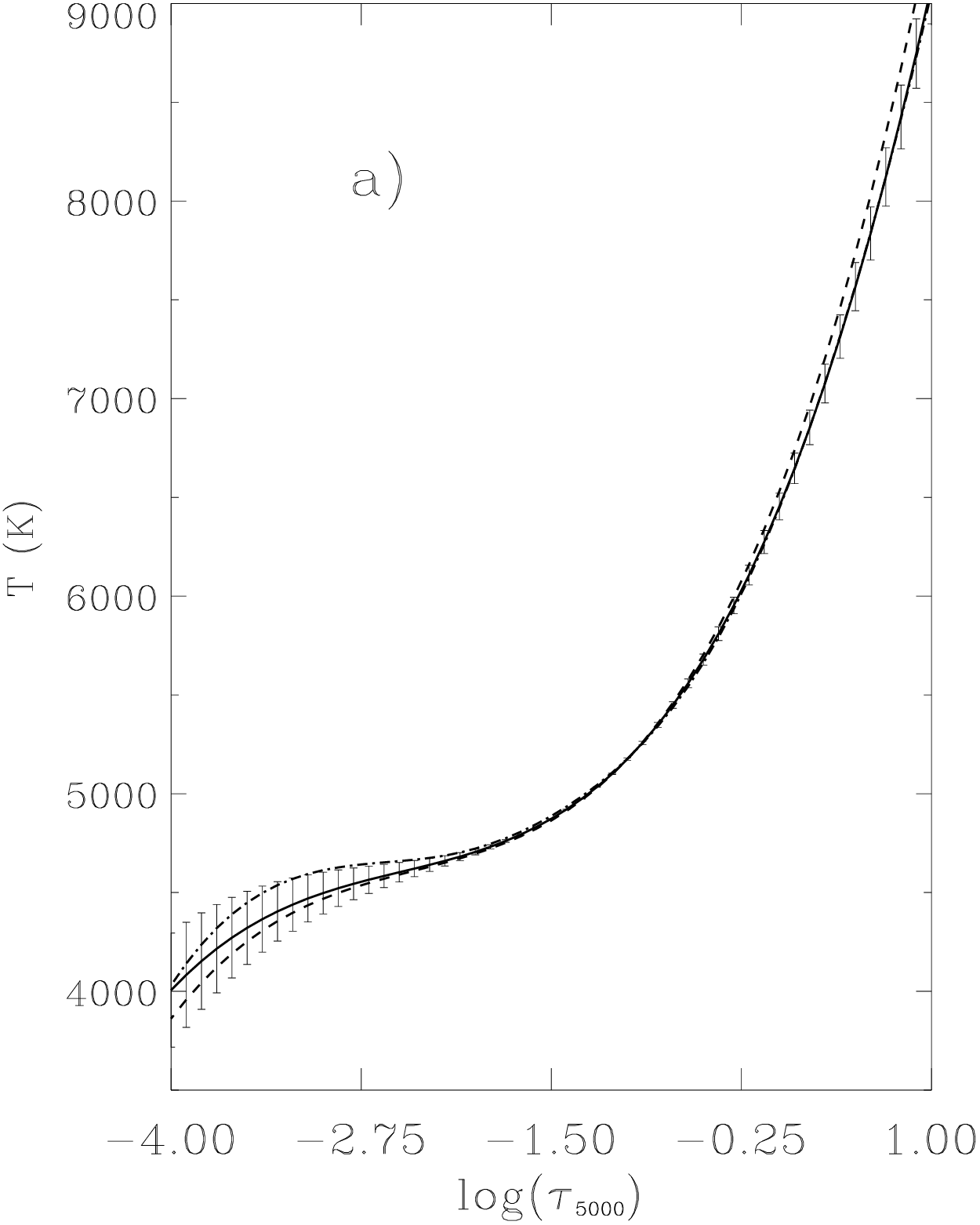}  
\includegraphics[width=5.2cm,angle=0]{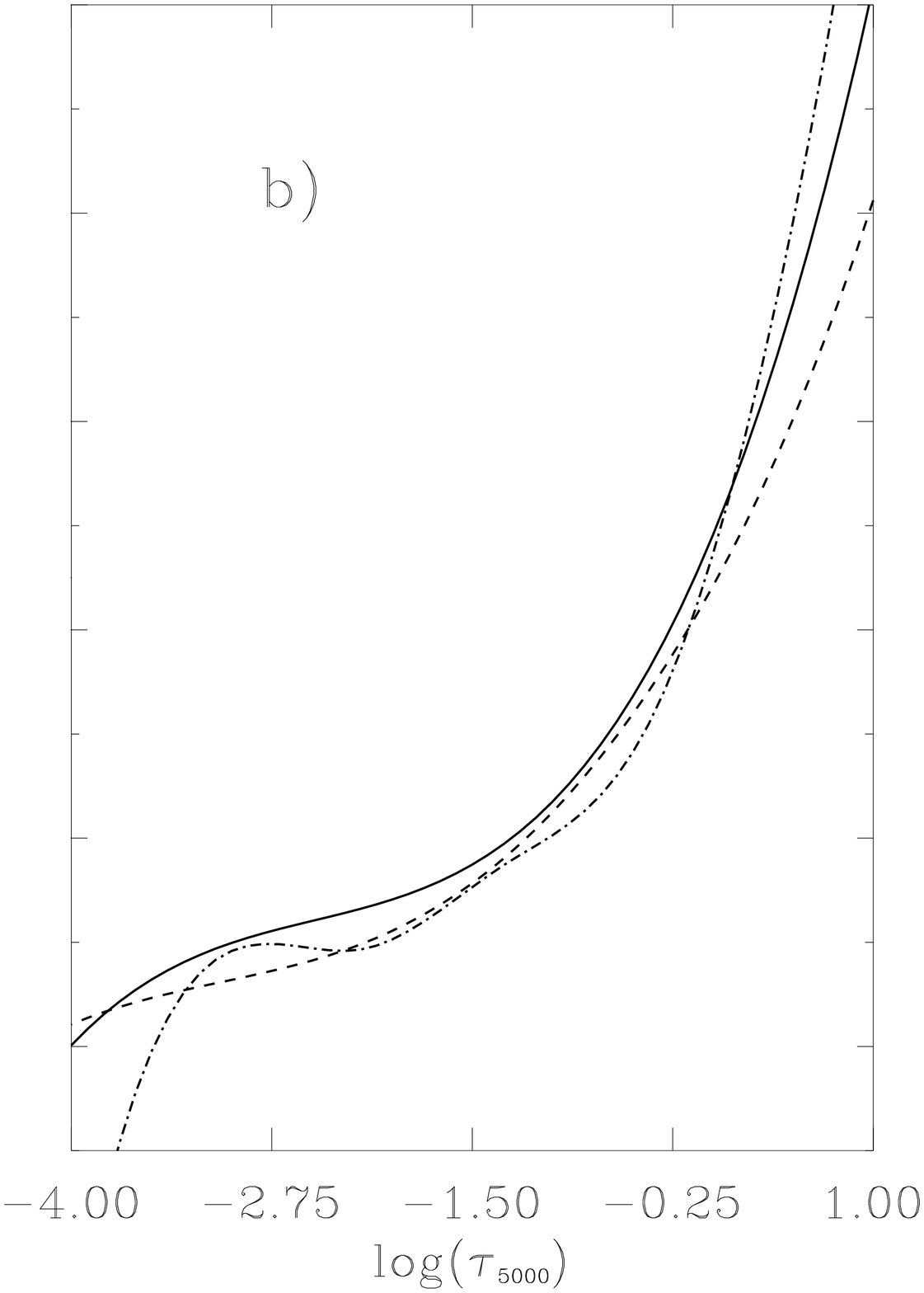} 
\includegraphics[width=5.2cm,angle=0]{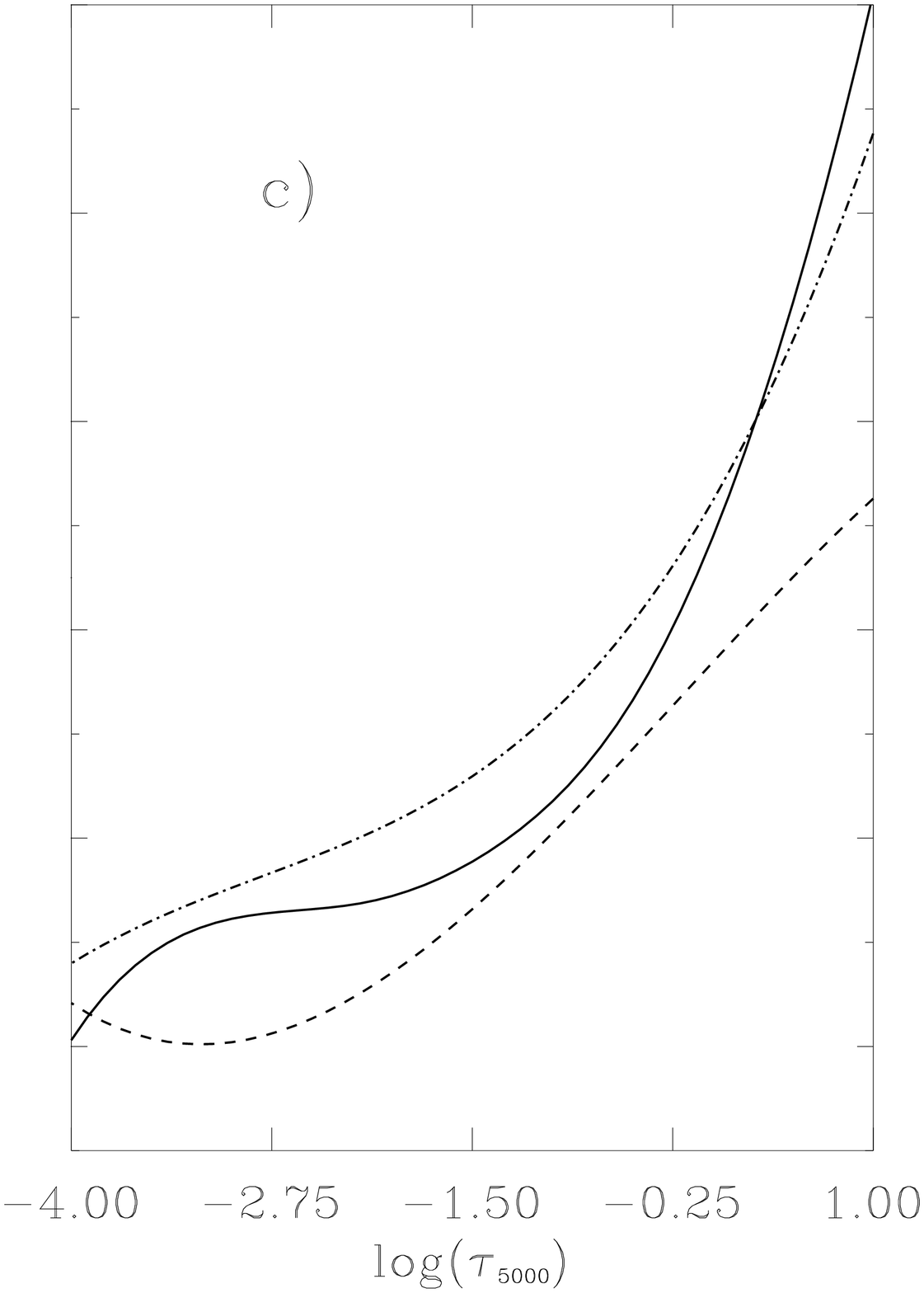}   
\protect\caption[ ]{
a): Atmospheric structures derived from the inversion of all the solar lines in Table 1 (solid line with error bars), only weak lines (dashed curve), and only the seven strongest lines (dashed-dotted). 
The strong lines, which correspond to different elements and ionization stages, contain all the  information that can be 
extracted with our procedure.
b): Structures recovered from the inversion of only neutral (dashed), only ionized (dashed-dotted), or both (solid line) subsets. Lines from 
the two ionization stages, with opposite sensitivity to temperature, 
 are required to constrain the thermal structure. 
c): Structures recovered from the inversion of all 
the strong lines (solid curve),
only strong  lines of neutral species (dashed), and only strong  
lines of ionized species (dashed-dotted). 
\label{f13}}
\end{figure}

Using an inversion procedure like {\tt MISS}, we can only determine the
atmospheric parameters that directly or indirectly affect the formation
of the input spectral lines. Basically, the thermal structure,  the
chemical abundances,  and the micro-turbulence are the most 
important parameters controlling  the overall strength of the
 metal lines. Temperature, gas
 pressure, and (micro/macro) turbulence affect the line width. Any of the
aforementioned parameters has a different impact on a given spectral line profile, but some parameters' influence on that particular line may be
very similar. If our description of the atmosphere and the line formation were perfect, ideal observations of a single profile would suffice 
to disentangle every parameter at play. However, the use of a  
simplified model atmosphere, the
presence of noise in 
observations and atomic data, plus the inclusion
 of a long list of approximations,  
make up  a much more complicated situation. In practice, for any given
line, there is a degeneracy between some parameters, but it can avoided
 by including other lines with different sensitivity:

\begin{itemize}

\item Weak lines cover a limited range of depths, so a set of
different lines is required to constrain tightly the thermal 
structure. Strong lines 
span a larger range of opacity, even just along their wings, and 
consequently, they provide more information. Figure \ref{f13}a
compares the temperature structures recovered from all the lines (solid 
curve with error bars), only  weak lines (dashed) and only 
strong lines (dashed-dotted). We also found that 
reducing the number of weak lines by a factor of two 
 already compromises the depth coverage. Basically, the   
information provided by the  55 weak lines is also contained 
in the set of 7 strong lines. 

When the number of lines used becomes small,  there is little place for
errors in the atomic data. More importantly, the fact that the
atmosphere retrieved from only strong lines is virtually identical to
that from a large number of weak lines highlights that the
pressure-broadening by collisions with hydrogen atoms is properly
described by the theories of Anstee, Barklem \& O'Mara, at least for
the lines used here. The difference between the  abundances derived
from  only-weak and only-strong  lines are 0.03 and 0.00 dex, for
calcium and iron respectively.

\item It is  possible to compensate a change in the chemical
abundance that produces a given line through a change in temperature.
Several lines of the same species  are unlikely to be sufficient to
solve this issue, as their different responses to a fractional change
of the abundance could be compensated by a particular thermal structure. 
A combination of lines of different species will break down the
temperature-abundance degeneracy. Lines produced by different
ionization stages of the same element  will have similar sensitivity to
changes in the abundance, but if one of the species is the dominant,
they will typically show opposite responses to an increase or decrease of the
 temperature. Logically, lines of different elements may have similar
responses to the temperature, but not to any of their abundances.
The temperature-abundance degeneracy can be exemplified by 
an inversion restricted to only neutral or ionized lines.  The dashed line in
Fig. \ref{f13}b corresponds to the thermal structure derived from
only neutral lines, the dashed-dotted line was obtained from only
ionized lines, and the solid curve with error bars was derived with
all, neutral and ionized lines.  The different  thermal structures
bring with them changes in the chemical  abundances of up to 0.1 dex,
not always in the same sense for all the elements. Additionally, the
 micro-turbulence changes slightly  (0.8 and 1.1 km s$^{-1}$ for the
 {\it only-neutral} and {\it only-ionized}
 cases, compared to 1.1 km s$^{-1}$) and so does the macroturbulence
($+0.1$ and $+0.2$ km s$^{-1}$).

\item By definition, the micro-turbulence represents small-scale
velocity fields. If, as in our case, those are assumed isotropic
and Gaussian, it is very difficult to distinguish between thermal and
small-scale turbulent  velocities. This gives place to a second
degeneracy, which involves temperature and  micro-turbulence.
Micro-turbulence cannot, however, mimic the effect that temperature has
on the lines, but if the abundance can change freely to accommodate any
variation of temperature, we are set  for confusion.  Including lines
of the same element with opposite sensitivity to variations in
temperature will automatically constrain that variable.  Several lines of a
given single species could in principle be useful, as  moderately weak and
moderately strong lines are sensitive to micro-turbulence, but the
wings of very strong lines (such as the Ca II lines in Table 1) are not
affected by this parameter. In practice, different ionization stages of
the same element are still necessary, as otherwise it is not possible
to break down the abundance-temperature degeneracy. This is well
illustrated by an inversion of all the strong lines,  only strong lines
of neutral species, and only  strong lines of ionized species.  In Fig.
\ref{f13}c, the solid line was obtained from the inversion of
all the strong lines -- as we have seen,  virtually indistinguishable
from the inversion of all, weak and strong lines. The  dashed curve was
derived from only  strong lines of neutral species, whereas the
dashed-dotted curve from only strong lines of ionized species. The
disparity between the thermal structures corresponds to differences in
the derived abundances that can reach up to  0.3 dex. The {\it key} to
the success of the inversion with only a few strong lines resides in
the presence of two lines of different species of calcium in our list of
{\it strong} lines.

It is interesting to explore whether the group of strong lines suffices
 to constrain the micro-turbulence and, therefore, to confirm that
they are able to retrieve the temperature structure without any
previous knowledge about the micro-turbulence. A set of 20 experiments
corresponding to $\xi_0= 0.5, 1.0, 1.5$, and 2.0 km s$^{-1}$ and
$\Delta$A $= -0.2, -0.1, 0.0 , + 0.1$, and $+0.2$ dex for Mg, Si, and Fe, 
was set, finding that the structures were very similar, and the final 
micro- and macro-turbulence velocities had a mean 
value of $1.18 \pm 0.02$ km
s$^{-1}$ and  $1.86 \pm 0.02$ km s$^{-1}$, respectively. 
As an example of the impact of the scatter on the abundances, 
we found a mean value for the iron abundance of 
$7.500 \pm 0.004$ (SEM) dex, with a standard deviation 
of 0.02 dex. Although the strongest lines in the group, Ca II 
at 8542 and 8662 \AA, are unreactive to 
the micro-turbulence, the other strong, but weaker, lines are 
quite sensitive to this parameter.

\end{itemize}

With only lines of  neutral species, 
Allende Prieto et al. (1998, 2000) 
used the goodness-of-fit between synthetic and observed line
profiles as a criterion to determine the micro-turbulence. 
It has, however,  become 
apparent from  our  experiments that lines of different ionization stages 
are needed to solve properly  the  inversion problem.

Hydrogen lines would be a particularly interesting tool to analyze the
spectra of solar-like 
stars. With H$^{-}$ as the dominant source of  continuum opacity, the
strength of hydrogen lines is a fine sensor for the atmospheric thermal
structure. Spanning a dramatic range in opacity, a single hydrogen line might
be useful to obtain information from the whole photosphere, even extending 
the depth coverage provided by our present database. 
Nonetheless, at this point we have not considered the use of hydrogen lines  
in inversions because their broadening is extremely complex. This 
complexity   is apparent in that
 it is not possible to  reconcile the effective temperatures obtained
from the Balmer lines with other indicators, such as the Infrared Flux Method 
(Fuhrmann et al. 1993; Alonso et al. 1996). We note that 
new theoretical developments  might change this situation in the
near future (Barklem et al. 2000b, c).

\section{Practicalities of the application of inversion methods to other stars}

From the inversion of the hydrodynamical solar simulation, we have
found that systematic errors in the abundances   from single-component
homogeneous inversions of solar-like spectra are likely about 
 10 \% (0.04 dex).  From our inversions of the solar spectrum, we have seen 
that random errors can be a factor of two smaller, provided we 
carry out a careful selection of the input data. This kind of accuracy 
is sufficient for many applications.  

Stellar spectra are typically acquired with  
lower resolution and signal-to-noise ratio than solar spectra, and it is 
necessary to assess the resistance of the inversion procedure to these 
factors. In most cases the stellar gravity is not known with much
precision, but we need this parameter to derive the pressure gradient from 
the thermal structure. Finally, unlike the solar case, the projected rotational velocity of the star is unknown {\it a priori}. 
Unless one of them is significantly
larger than the other, it is likely that there will be
 some confusion between rotation
and macro-turbulence. All these issues are explored below.

\subsection{Lower resolution and signal-to-noise ratio}

A spectral line produced in a one-dimensional static 
atmosphere cannot reproduce the line asymmetries 
observed in real spectra. The  synthetic profiles obtained by our 
inversion procedure will not match  the detailed shape of the observed lines, 
which gives us a hint that we might not need to resolve  such details in the
observed input spectra.  It is also true that  if we try 
to minimize the difference 
between  observed and  synthetic spectra, introducing some   
 noise with a symmetric distribution will not affect the $\chi^2$. In practice,
degrading signal-to-noise and resolution comes with a series of  
 additional problems. First, it becomes harder to identify spectral
features associated to a single transition. Second, it gets 
more difficult to normalize the spectrum, and to determine 
where the center of a line is.
Third, the number of spectral points with independent information diminishes, and so does the available information.

To check the resistance of the inversion method against a degradation in
the signal-to-noise ratio and resolution, we have made use of the  
dataset  from the solar spectrum used in previous sections. We first resampled the spectrum to have two wavelengths per resolution element, then
we  degraded the spectrum by convolving it with a Gaussian profile, and
resampled it to have two wavelengths per resolution element again. 
We generated random numbers following a Poisson distribution (centered
at 1 and scaled to the appropriate width) which were multiplied by the 
spectrum. Finally, every single line was recentered before entering into 
the inversion code. For practical reasons, all the processing was done with 
the  full observed profiles, so  at the end we applied the same mask used
in the solar inversions to exclude   wavelengths apparently 
affected by blends. We tried two different values for the final 
resolving power (R$ = 1 \times 10^5$ and $5 \times 10^4$) 
and  two for the signal-to-noise ratio (SNR $= 100$ and 33). 

\begin{figure}[!ht]
\centering
\includegraphics[width=9cm,angle=90]{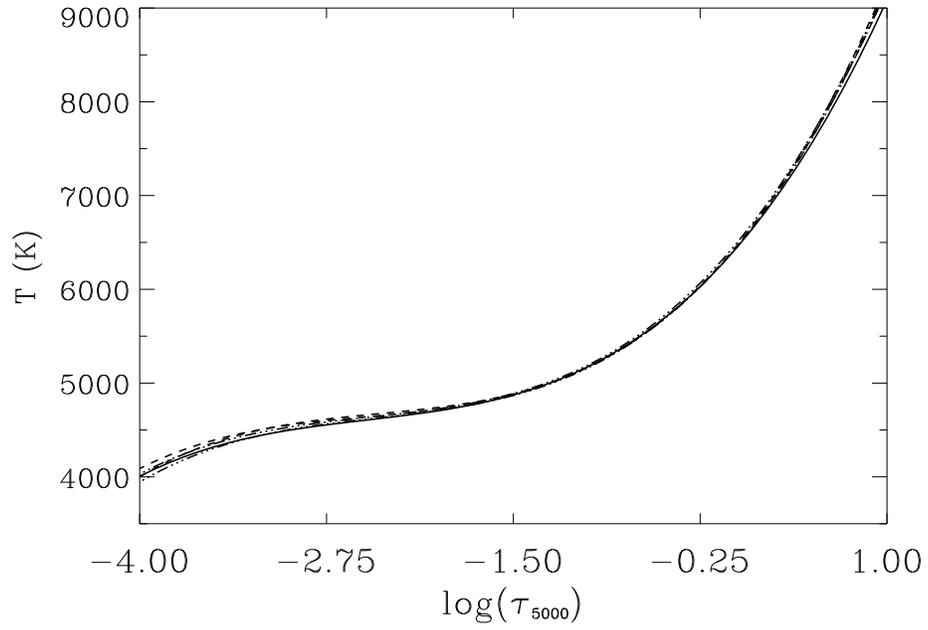}  
\protect\caption[ ]{
Structures derived from the original data from the solar flux atlas (solid line) and from the inversion of the degraded datasets (R$=1 \times 10 ^5, 5 \times10^5$; SNR = 100, 33). 
\label{f14}}
\end{figure}

Figure \ref{f14} shows the structure derived in \S 4 from the
original solar spectrum (solid line) and the new four inversions of the
degraded spectra.  In accordance with the small effect on the thermal
structure, the abundances are affected very little; the iron abundances
obtained for (SNR = 100, R = $10^5$), (SNR = 100, R = $5 \times 10^4$),
(SNR = 33, R = $10^5$), and (SNR = 33, R = $5 \times 10^4$) are 7.50,
7.51, 7.50, and 7.52, respectively. The tests suggests that a correct
 inversion is still possible with a resolution and a signal-to-noise
ratio as low as $5 \times 10^4$ and $33$, respectively. At these
levels, the  weakest lines are simply dissolved in the noise and mapped
with a few  points. However, as we have seen, the strongest lines of
the spectrum already contain all the information, and those features
are very resistant to the decline in  resolution and signal-to-noise
ratio.

\subsection{Uncertainties in gravity}

The surface gravity of the Sun is a well-known parameter, 
$\log g = 4.437368 \pm 0.000087$ (Stix 1991), even if it 
might suffer larger variations than its formal error bar indicates
depending on which definition of {\it radius} is used. The 
overall sensitivity of the spectrum to changes in gravity is relatively small, 
compared to temperature, 
so in this regard an uncertainty in $\log g$ smaller than 0.05 dex is
normally sufficient. For stars closer than about 100 pc, it is possible 
to combine accurate trigonometric parallaxes with calculations of stellar evolution, constraining $\log g$ within roughly 0.06 dex (Allende Prieto \& Lambert 1999). For other  stars, gravities are usually determined from 
spectroscopic analysis or spectroscopic classification, 
rarely achieving a precision better than about 0.2 dex, in some  
cases with much larger  systematic  
errors (Allende Prieto et al. 1999). Definitely, it is important 
to check how sensitive  the results of an inversion are to uncertainties 
in gravity. 

\begin{figure}[!ht]
\centering
\includegraphics[width=6cm,angle=0]{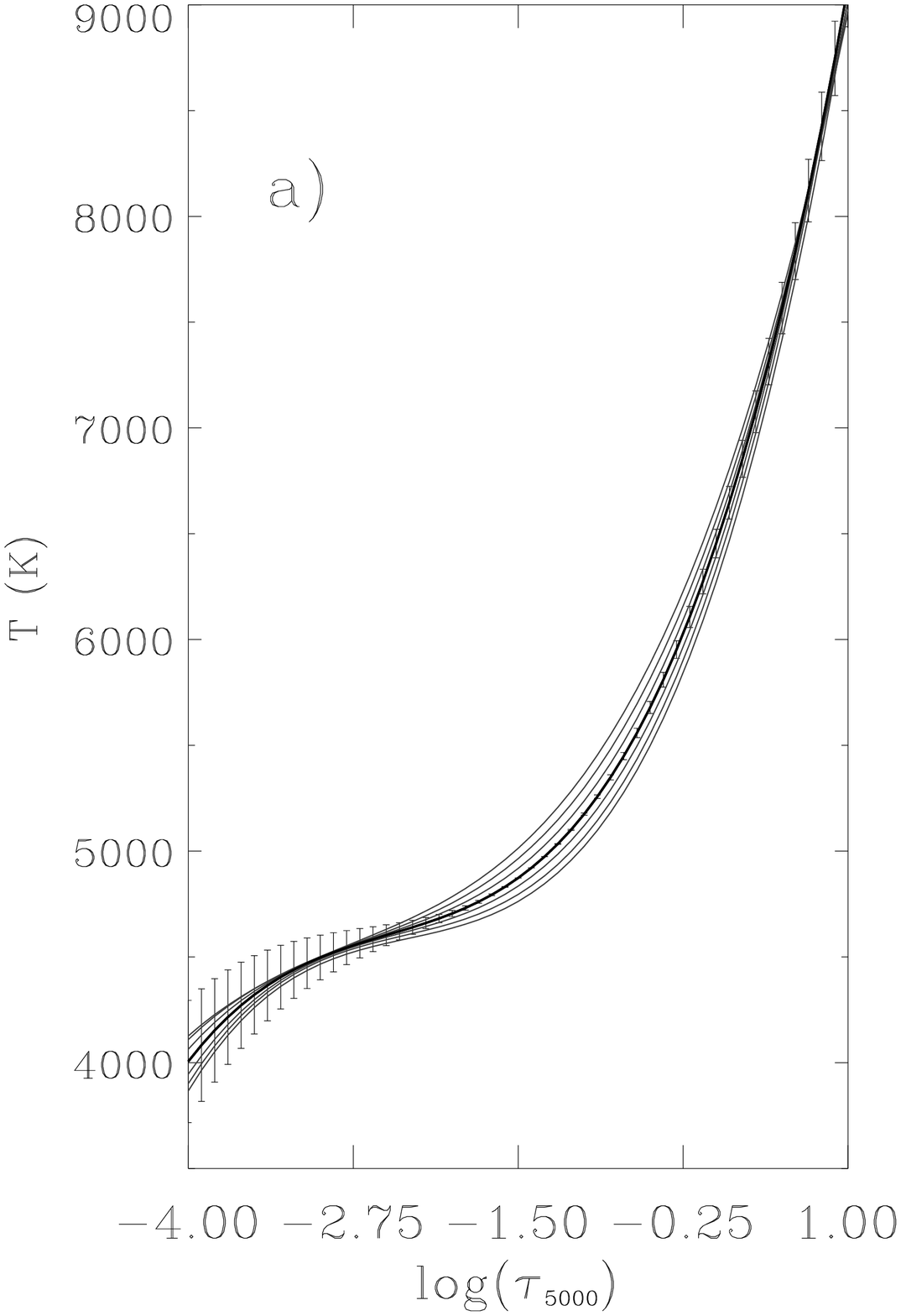}
\includegraphics[width=5cm,angle=90]{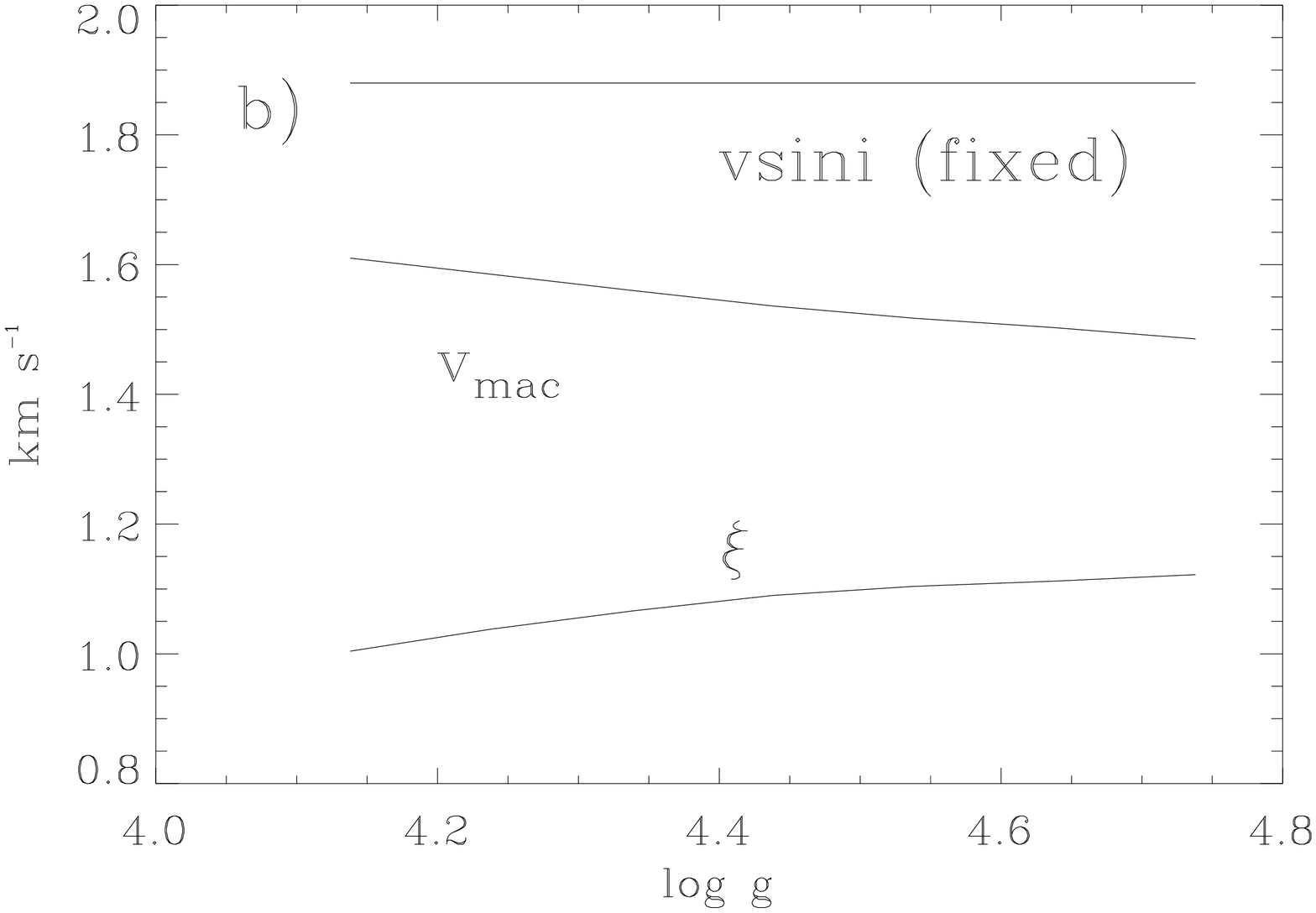}  
\includegraphics[width=5cm,angle=90]{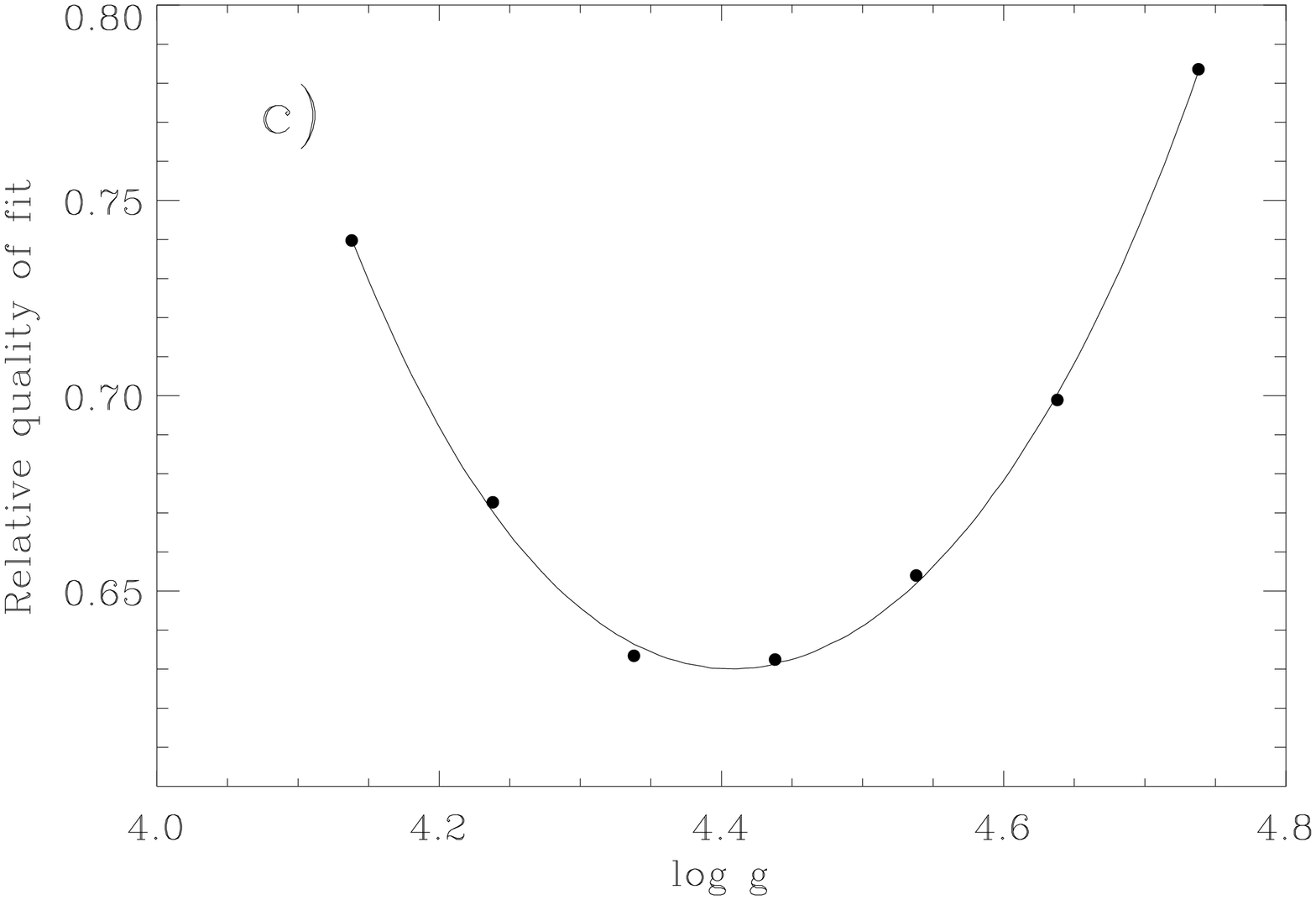}   
\protect\caption[ ]{
a): Thermal structures derived adopting the real solar gravity ($\log g= 4.44$; thick line) and lower and higher values by $\Delta \log g = -0.3, -0.2, -0.1, +0.1, +0.2, +0.3$ dex. A cooler temperature stratification  
tries to compensate for a wrongly set lower gravity.
b): The micro- ($\xi$) and macro-turbulence 
($v_{\rm mac}$) as a function of
the assumed gravity for the solar inversion. These two parameters, 
together with the temperature structure and the abundances, 
are affected  by setting the gravity wrongly. 
c): Quality of the fit to the input line profiles, quantified by the sum of the
square differences between  observed and synthetic normalized
fluxes, as a function of the {\it fixed} surface gravity. The solid line represents
a least-square fit to a second-order polynomial.
\label{f15}}
\end{figure}

We have repeated the inversion of the solar data used in \S 4 but adopting
 a value for the gravity which is different from solar by $-0.3, -0.2, -0.1, +0.1, +0.2$, and $+0.3$ dex. The effect on the recovered thermal structure is 
shown in Figure \ref{f15}a. The thick curve corresponds to the solution
for the right gravity and the thinner lines were derived for the other values.
 Dropping the gravity is compensated by decreasing the temperature and
likewise higher gravities lead to hotter models. Figure \ref{f15}b 
shows that the turbulent velocities derived from the inversion are also
shifted with the gravity (note that $v_{\rm rot} \sin i$ is always held as a known  parameter). 

 In Figure \ref{f15}c we take a close look at the relative quality
of the fits for the different $\log g$ values, quantified by the sum of the
square differences between the observed and synthetic normalized
fluxes. The best fit is found for gravities close to the real value,
revealing that it is possible to extract the gravity from the
inversion. This suggests adding the gravity to the
pool of free parameters to determine in
 the inversion. Frutiger et al. (2000) did so, but they recovered $\log
 g$ values for $\alpha$ Cen B systematically  $0.2-0.4$ dex higher 
than the value derived by Allende Prieto \& Lambert (1999) based on the
 trigonometric parallax. The most recent version of {\tt MISS}  does
 not incorporate gravity as a free parameter, but we will explore this
 promising possibility in the future.

\begin{figure}[!ht]
\centering
\includegraphics[width=10cm,angle=0]{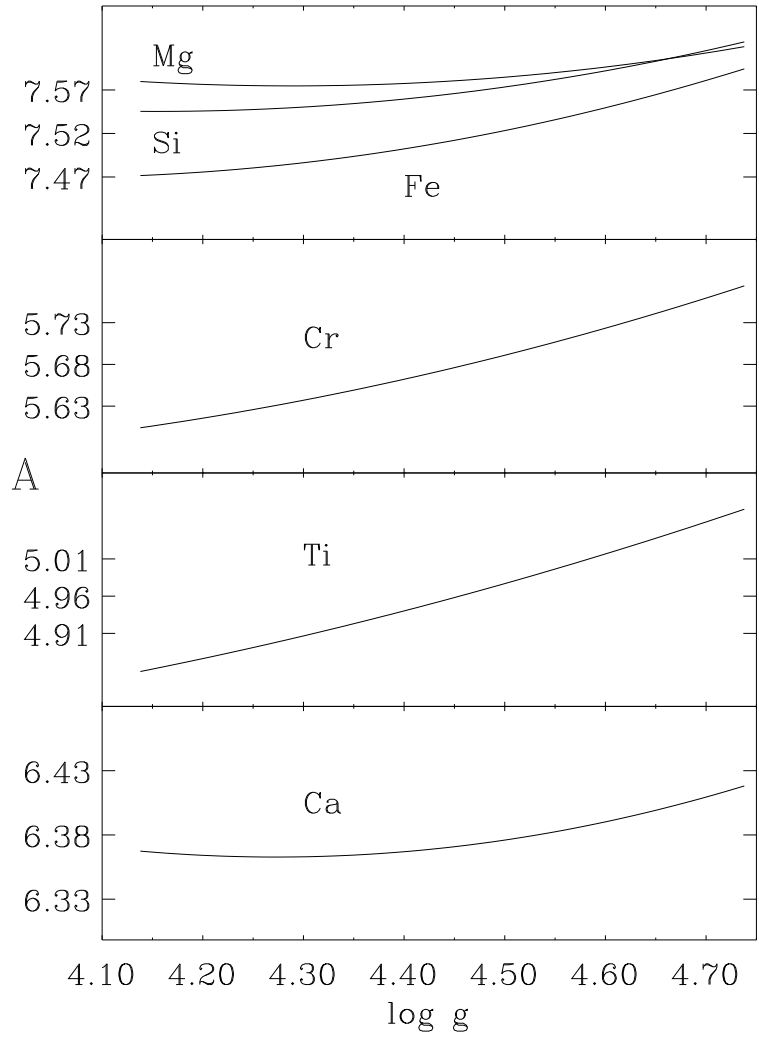}  
\protect\caption[ ]{
Chemical abundances derived as a function of the adopted gravity. 
The separation between vertical tickmarks is always 0.05 dex. 
The original data has been fitted by a second-order polynomial.
\label{f16}}
\end{figure}

As shown in  Figure \ref{f16}, the changes in the atmospheric
structures derived for different gravities are accompanied with
variations in the abundances. The abundances have sensitivities that
reach up to about 0.2 dex per dex (for an uncertainty in the $\log g$
of $\sim 0.2$ dex, some abundances may be wrong by 0.04 dex).  Though
these are not very large errors, they would double the systematic
errors associated with the assumption  of a plane-parallel hydrostatic
atmosphere  studied in \S 3.

\subsection{Rotation and macroturbulence}

If the gravity is usually uncertain, the rotational velocity is completely 
unknown. For a spherically symmetric star without surface features,
 all we can aspire to deduce from the spectral lines is the 
product $v_{\rm rot} \sin i$. In many cases, not even that, as 
disentangling rotation  from other competing broadening mechanisms is
not trivial. The projected rotational velocity can be combined with other
sources of information to extract the angular velocity, which is by itself
an important parameter as it affects the star's structure and evolution.

\begin{figure}[!ht]
\centering
\includegraphics[width=7cm,angle=90]{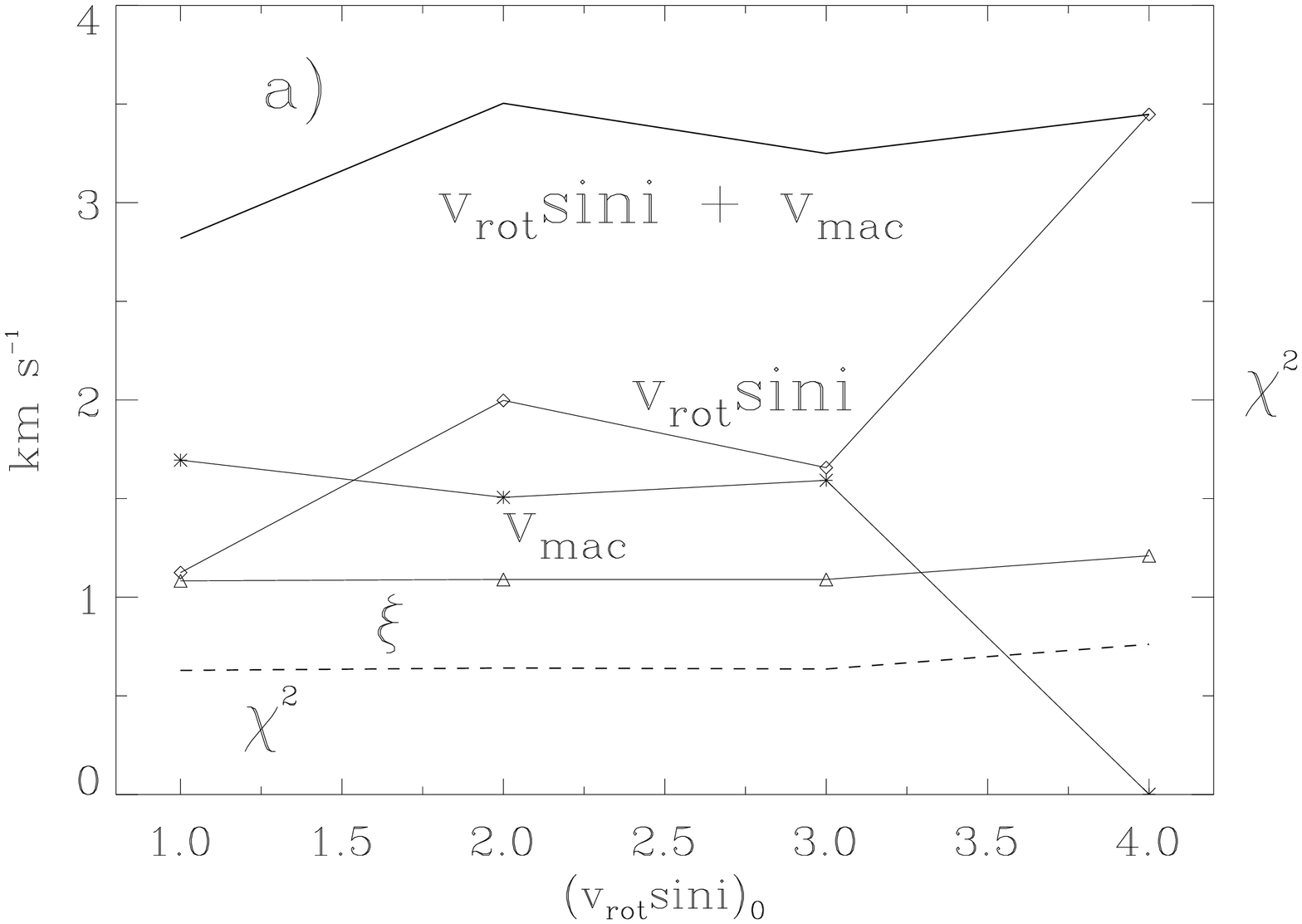} 
\includegraphics[width=7cm,angle=0]{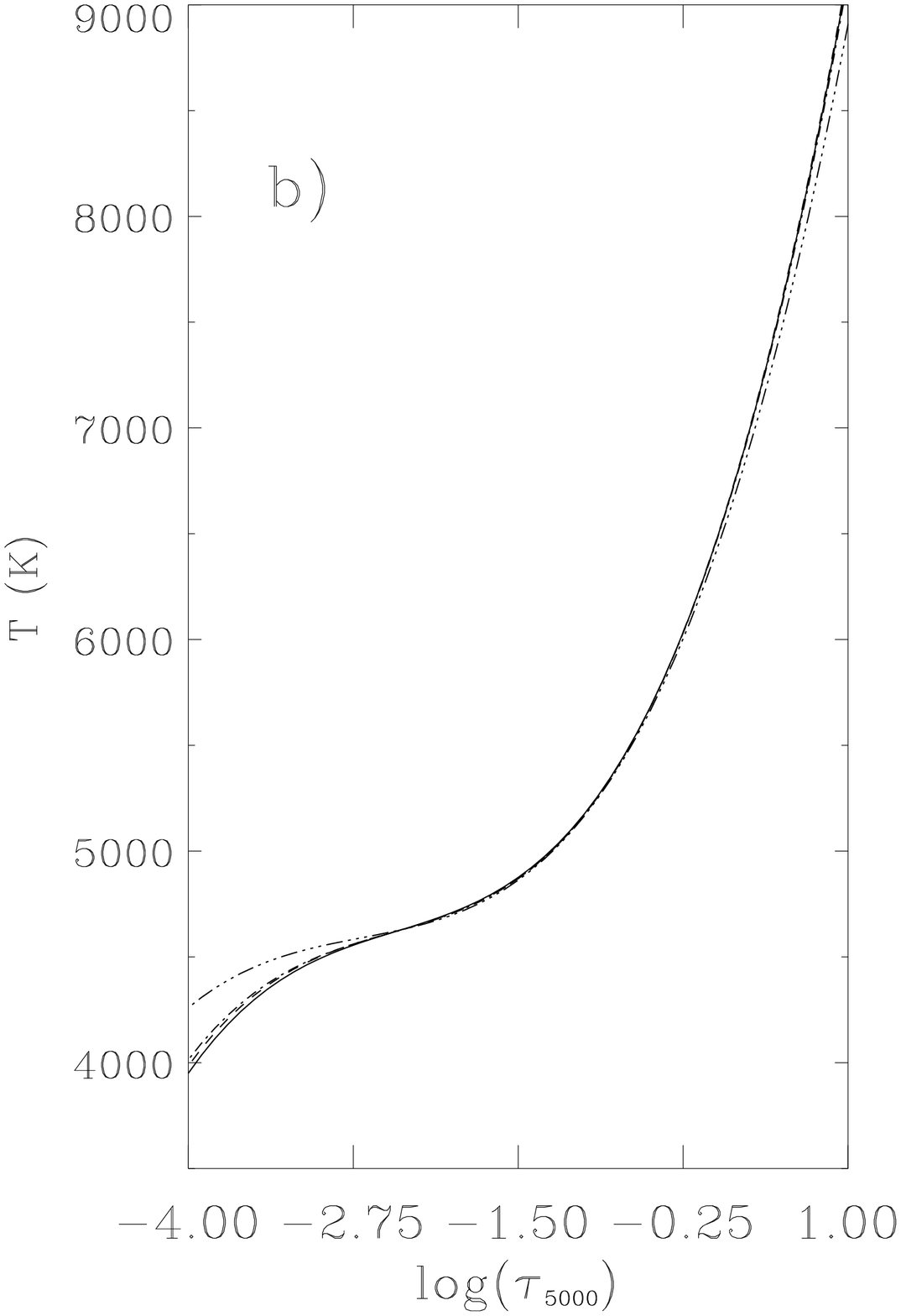}   
\protect\caption[ ]{
a): The projected rotational velocity of the Sun is now a free parameter. This figure shows the final values for the micro- ($\xi$) and macro-turbulence ($v_{\rm mac}$), $v_{\rm rot} \sin i$, and the quality of the fit (reduced $\chi^2$), as a function of the initial guess.
b): Recovered structures when different values are initially adopted for the rotational velocity of the Sun. The styles of the lines, solid, dashed, dashed-dotted, and three-dotted-dashed, correspond to $(v_{\rm rot} \sin i)_0 = 1, 2, 3$, and 4, respectively. 
\label{f17}}
\end{figure}

To learn what  the prospects are for the inversion  to obtain
accurate projected rotational velocities, whether rotation can be 
confused with other velocity fields with smaller characteristic length-scales
(macro-turbulence), and the possible influence that such confusion
would have on the retrieved abundances, we have introduced rotation 
as a free parameter in dealing with the solar spectrum. We have carried out  inversions with an initial guess $(v_{\rm rot} \sin i)_0$ of 1, 2, 3, and
4 km s$^{-1}$. The solid lines in Figure \ref{f17}a show the 
recovered $v_{\rm rot} \sin i$, $v_{\rm mac}$, $\xi$, and the sum of $v_{\rm rot} \sin i + v_{\rm mac}$, whereas the dashed line shows the reduced $\chi^2$, adopting $\sigma_i = 1$, for each final solution (the particular value of $\sigma_i$ does not matter, as we are interested in relative values).
 The sum of the macro-turbulence and the projected 
rotational velocity is roughly conserved, but that does not apply for each of
the two line broadening components. 

We are not very surprised by this result. A Gaussian isotropic
macro-turbulence is not a very good description of the atmospheric
large-scale convective motions (Asplund et al. 2000a), and in the solar case macro-turbulence and
projected rotational velocity happen to be of similar size,
further complicating their separation.  A homogeneous model atmosphere cannot
reproduce the asymmetric spectral lines produced by convective
turbulence, limiting the information we are able to extract from the
detailed shapes  of the line profiles. It is possible that a 
different description of turbulence, such as the radial-tangential recipe
(Gray 1975) could be useful to improve these results. It might be
possible as well that homogeneous models with two or more components
and depth-dependent velocities could be  useful to this end. At
this time we do not have an answer, but it is worthwhile to explore
this issue more deeply.

Figure \ref{f17}b shows that the shape of the atmospheric structures 
has almost no dependence on 
 the final rotational velocity. Consequently, the retrieved
abundances are also very weakly affected. The standard deviations of
the abundances of Si, Ca, Ti, Cr and Fe found for the four different values of
$(v_{\rm rot} \sin i)_0$ are less than or about 0.01 dex.
As our data includes a single Mg line, the abundance of this element 
is much more sensitive, showing a standard deviation of 0.04 dex.

\section{Conclusions and prospects}

There are two major difficulties to describe accurately the energy
 balance in the atmospheres of  late-type stars: line blanketing and 
convection. Line blanketing is a problem because it 
 requires including a very large number
of frequencies, and the atomic data for many lines is uncertain or
unavailable. Convection accounts for a significant part of the energy
transferred outwards in the deepest photospheric layers, and the simplistic
approaches to take it into account within the
framework of static and homogeneous models have  many free 
parameters. A proper description of convection has been achieved through
numerical hydrodynamical modeling (Stein \& Nordlund 1998, Asplund
et al. 2000a), but this is  computationally very
involved, and currently impractical for carrying out chemical analyses of 
large samples of stars. 

A possible way to bypass the theoretical annoyance is to use observations 
to constrain as many parameters as possible, rather than just a few. 
Inversion methods can be applied to stellar spectra  and so relax some
theoretical assumptions. In the interleaved measuring-modeling process 
that is used to determine chemical abundances from stellar spectra,  
these methods try to effectively shift the balance  towards a pure measurement.
Among these codes, {\tt MISS} (Allende Prieto et al. 1998) 
adopts a one-dimensional single-component  model atmosphere in LTE
and hydrostatic equilibrium. 

Recent calculations show that departures from LTE in the photospheres
of solar-like stars are, on average, relatively small
 both in the line formation (Shchukina \& Trujillo Bueno 2001, Gehren 
 et al. 2001) and in
the photospheric structure (Hauschildt, Allard, \& Baron 1999).
Shchukina \& Trujillo Bueno (2001) find that assuming LTE can artificially
decrease the mean iron abundance obtained from the analysis of Fe I
intensity profiles with a three-dimensional model atmosphere by $\sim
0.07$ dex, and with the Holweger \& M\"uller solar model by $\sim$ 0.06
dex. If this is right, a non-LTE line-formation analysis of iron 
in 1D would provide 
accurate corrections to $\sim 0.01$ dex. However, a  previous
calculation by Th\'evenin \& Idiart (1999) found that typical errors
in the LTE abundances from disk-integrated  observations analyzed with
  homogeneous solar model atmospheres were essentially zero. The
 differences between these two works might reside in the fact that
Shchukina \& Trujillo Bueno used intensity, whereas Th\'evenin \&
Idiart employed flux, the particular lines employed, or  
the atomic data adopted.  We should emphasize that depending on which 
element is studied, the specifics of the atmosphere, 
and the particular set of spectral lines  employed,  non-LTE corrections 
may be large, and also may differ significantly for 3D and 1D 
calculations (see, e.g.,  lithium in metal-poor dwarfs; Asplund et al. 1999).

We have tried to fill an important gap, as there was no quantitative
assessment of the systematic errors incurred by the use of a static and
homogeneous model atmosphere as part of an inversion method. In our
experiment, we removed  non-LTE effects
 by adopting LTE in both the synthesis of the profiles emerging from
 the simulation and in the inversion. We have employed a detailed
three-dimensional time-dependent hydrodynamical simulation of the solar
surface to calculate a spatial-, angular- and time-averaged spectrum,
which has been fed to {\tt MISS}.  The rms difference between the
abundances recovered by the inversion code and those used in the
three-dimensional simulation was only 10 \% (0.04 dex), and the mean
difference was $-0.01 \pm 0.02$ (SEM) dex.
Speaking generally, it is apparent that even with the worst 
expectations, departures from LTE will
 not make this error much larger than 0.07 dex. This kind of accuracy
is sufficient for many applications.  From our inversions of the solar
spectrum, we have seen than internal random errors can be reduced to 
about 0.02 dex, provided we carry out a careful selection of the input data.

We have compiled a  careful selection of spectral features of 
Mg, Si, Ca, Ti, Cr, and Fe,  including strong lines for which 
calculations of the 
collisional damping due to neutral hydrogen are available.
Normalized line profiles are used to carry out an inversion of the
solar spectrum, searching for the chemical abundances of the 
involved elements, a depth-independent micro-turbulence, a Gaussian
macro-turbulence, and the atmospheric thermal structure,  that 
reproduce the observations best. There are no surprises, 
the abundances recovered are similar to
the commonly accepted values listed by Grevesse \& Sauval (1998).

We realize that a very limited number of strong lines from different 
ionization stages suffices to constrain the atmospheric structure 
and the abundances. In fact, we tested our inversion procedure 
against a degradation of the
spectral resolution and signal-to-noise ratio in the data, finding 
 that reliable inversions are possible at R $\equiv \lambda/\delta\lambda \sim 5 \times 10^4$ and SNR $\sim 30$. The
stellar gravity, usually poorly determined, can also be 
recovered by the inversion. However, the projected rotational velocity and 
the Gaussian macro-turbulence are easily confused when their size is similar,
but this has a very small effect on the abundances, at least for the 
solar case.

With these new results, we foresee no major difficulty to make use of
inversions of stellar spectra to perform extensive analysis of chemical
abundances in  stars with solar or similar temperatures and
 metallicities. Further and in progress development of our code for
this task includes a more detailed treatment of the macro-turbulence
(expected to be useful in separating macro-turbulence and rotation), 
the addition
of a second one-dimensional component, and vertical velocities,
 to the model atmosphere   (aimed at extracting information on the
granulation and achieving a closer match of the observations), and the
addition of  gravity to the pool of parameters to be determined
 automatically. 

Further extension of tests like the one described in
 this paper, and on departures from LTE,  to other domains in the HR
diagram are also in progress and will be the subject of  forthcoming
papers. So far, we should mention that there is evidence of much larger
systematic effects for stars slightly  warmer than the Sun. For
example, in a three-dimensional simulation for Procyon ($T_{\rm eff}
\simeq
 6500$ K) the layers with the largest temperature inhomogeneities
 overlapped with the continuum-forming region, whereas in the solar
 simulation such layers are  below the continuum-forming region
(Nordlund \& Dravins 1990, Allende Prieto et al. 2001). Hauschildt et
al. (1999) detected large departures from LTE, even affecting the
calculation of the (horizontally-homogeneous) atmospheric structure,
for a star like Vega ($T_{\rm eff} \simeq 10,000$ K). Evidence is also
present for larger-than-solar inhomogeneities in metal-deficient stars
(Allende Prieto et al. 1999, Asplund et al. 1999), as it is for
larger-than-solar departures from LTE (Th\'evenin \&  Idiart 1999).

\acknowledgments

We are indebted to Luis Colina for  the solar reference spectrum, to
Ram\'on  Garc\'{\i}a L\'opez, David Lambert, and Gajendra Pandey for
fruitful discussions, to Thomas Meylan for his line list, to Heinz
Neckel for  the solar spectrum at the center of the disk and helpful
explanations,  and to Ward Whaling for providing Fe II $f-$values. It is 
a pleasure to thank the referee, Phillip Judge, for his detailed 
comments.

NSO/Kitt Peak FTS data used here were produced by NSF/NOAO.  We have
taken advantage of the Vienna Atomic Line Database (VALD) and  NASA's
Astrophysics Data System Abstract Service. This work has been partially 
supported by NSF through  grant AST-0086321.

\clearpage

\begin{deluxetable}{cccccccc}
\tablecolumns{8}
\tablewidth{0pc}
\tablecaption{Spectral lines and atomic data used in the inversion.  Cross-sections $\sigma$ for the collisional broadening by hydrogen are given for $v=10^4$ m s$^{-1}$, and $\alpha$ describes the velocity dependence assuming $\sigma(v) \propto v^{-\alpha}$ \label{tab:atomic_data}}
\tablehead{ \colhead{Species}  & 
  \colhead{Wavelength} &  \colhead{$E_{\mathrm{low}}$}  & 
  \colhead{$\log(gf)$} & 
  \colhead{$\sigma$} & \colhead{$\alpha$} & 
  \colhead{$\log\Gamma_{\mathrm{rad}}$} & 
  \colhead{$\log(gf)$ source}  \\
  \colhead{}    & 
  \colhead{(\AA)}      	      &  \colhead{(eV)}&
  \colhead{}                  &  \colhead{($a_0^2$)}  &
  \colhead{}        	      &  \colhead{(rad s$^{-1}$)} & 
  \colhead{} 	    }
\startdata
\cutinhead{Weak--Neutral}
 Si   I &  5665.557 &  4.92 &  $-$1.940  & 1772 &  0.222 & 8.290  & BZH\tablenotemark{a}                  \\
 Si   I &  5684.490 &  4.95 &  $-$1.550  & 1798 &  0.221 & 8.250  & BZH\tablenotemark{a}                  \\
 Si   I &  5690.425 &  4.93 &  $-$1.770  & 1772 &  0.222 & 8.300  & BZH\tablenotemark{a}                  \\
 Si   I &  5701.106 &  4.93 &  $-$1.950  & 1768 &  0.222 & 8.310  & BZH\tablenotemark{a}                  \\
 Si   I &  5708.402 &  4.95 &  $-$1.370  & 1787 &  0.222 & 8.270  & BZH\tablenotemark{a}                  \\
 Si   I &  5948.545 &  5.08 &  $-$1.130  & 1845 &  0.222 & 8.330  & BZH\tablenotemark{a}                  \\
 Si   I &  7680.271 &  5.86 &  $-$0.590  & 2107 &  0.495 & 7.660  & BZH\tablenotemark{a}                 \\
 Si   I &  7918.387 &  5.95 &  $-$0.510  & 2934 &  0.232 & 7.530  & BZH\tablenotemark{a}             \\
 Si   I &  7932.356 &  5.96 &  $-$0.370  & 2985 &  0.235 & 7.570  & BZH\tablenotemark{a}            \\
 Ca   I &  6161.297 &  2.52 &  $-$1.266  &  978 &  0.257 & 7.274  & Oxford                                  \\
 Ca   I &  6166.441 &  2.52 &  $-$1.142  &  976 &  0.257 & 7.269  & Oxford                                   \\
 Ca   I &  6455.604 &  2.52 &  $-$1.290  &  365 &  0.241 & 7.667  & Oxford                                   \\
 Ca   I &  6499.656 &  2.52 &  $-$0.818  &  364 &  0.239 & 7.640  & Oxford                                   \\
 Ti   I &  4758.122 &  2.25 &     0.481  &  326 &  0.246 & 8.029  & GBP                                   \\
 Ti   I &  4759.274 &  2.25 &     0.570  &  327 &  0.246 & 8.029  & GBP                                   \\
 Ti   I &  5113.445 &  1.44 &  $-$0.727  &  298 &  0.243 & 7.332  & GBP                                   \\
 Ti   I &  5295.781 &  1.05 &  $-$1.577  &  278 &  0.253 & 6.687  & GBP                                   \\
 Ti   I &  5490.154 &  1.46 &  $-$0.877  &  374 &  0.262 & 8.155  & GBP                                   \\
 Ti   I &  5866.457 &  1.07 &  $-$0.784  &  259 &  0.262 & 7.643  & GBP                                   \\
 Ti   I &  5922.115 &  1.05 &  $-$1.410  &  313 &  0.242 & 7.853  & GBP                                   \\
 Ti   I &  6092.799 &  1.89 &  $-$1.323  &  398 &  0.239 & 8.104  & GBP                                   \\
 Ti   I &  6258.109 &  1.44 &  $-$0.299  &  355 &  0.237 & 8.230  & GBP                                   \\
 Ti   I &  7357.735 &  1.44 &  $-$1.066  &  329 &  0.244 & 7.846  & GBP                                   \\
 Cr   I &  4801.028 &  3.12 &  $-$0.131  &  348 &  0.240 & 7.857  & Oxford                                   \\
 Cr   I &  4964.931 &  0.94 &  $-$2.527  &  262 &  0.261 & 8.236  & Oxford                                   \\
 Cr   I &  5272.002 &  3.45 &  $-$0.422  &  757 &  0.238 & 8.401  & Oxford                                   \\
 Cr   I &  5300.751 &  0.98 &  $-$2.129  &  329 &  0.263 & 7.716  & Oxford                                   \\
 Cr   I &  5312.859 &  3.45 &  $-$0.562  &  751 &  0.238 & 8.402  & Oxford                                   \\
 Cr   I &  5787.922 &  3.32 &  $-$0.083  & 1097 &  0.291 & 8.002  & Oxford                                   \\
 Cr   I &  7355.899 &  2.89 &  $-$0.285  &  902 &  0.246 & 7.822  & Oxford                                   \\
\break
 Fe   I &  4602.006 &  1.61 &  $-$3.150  &  296 &  0.260 & 8.083  & Oxford                                   \\
 Fe   I &  5225.533 &  0.11 &  $-$4.790  &  207 &  0.253 & 3.643  & Oxford                                   \\
 Fe   I &  5247.057 &  0.09 &  $-$4.950  &  206 &  0.253 & 3.894  & Oxford                                   \\
 Fe   I &  5916.254 &  2.45 &  $-$2.990  &  341 &  0.238 & 8.009  & Oxford                                   \\
 Fe   I &  5956.700 &  0.86 &  $-$4.610  &  227 &  0.252 & 4.433  & Oxford                                   \\
 Fe   I &  6082.715 &  2.22 &  $-$3.570  &  306 &  0.271 & 6.886  & Oxford                                   \\
 Fe   I &  6151.623 &  2.18 &  $-$3.300  &  277 &  0.263 & 8.190  & Oxford                                   \\
 Fe   I &  6173.342 &  2.22 &  $-$2.880  &  281 &  0.266 & 8.223  & Oxford                                   \\
 Fe   I &  6200.321 &  2.61 &  $-$2.440  &  350 &  0.235 & 8.013  & Oxford                                   \\
 Fe   I &  6297.801 &  2.22 &  $-$2.740  &  278 &  0.264 & 8.190  & Oxford                                   \\
 Fe   I &  6481.878 &  2.28 &  $-$2.980  &  308 &  0.243 & 8.190  & Oxford                                   \\
 Fe   I &  6498.945 &  0.96 &  $-$4.700  &  226 &  0.253 & 4.638  & Oxford                                   \\
 Fe   I &  6750.161 &  2.42 &  $-$2.620  &  335 &  0.241 & 6.886  & Oxford                                   \\
 Fe   I &  6978.861 &  2.48 &  $-$2.500  &  337 &  0.241 & 6.886  & Oxford                                   \\
\cutinhead{Weak--Ionized}
 Si  II &  6371.361  &  8.12 & $-$0.000  &  389 &  0.189 & 9.080 &  BZH\tablenotemark{a}       \\
 Ti  II &  4798.535  &  1.08 & $-$2.670  &  211 &  0.209 & 8.283 &  BHN                                     \\
 Ti  II &  5336.793  &  1.58 & $-$1.630  &  272 &  0.314 & 8.207 &  BHN                                      \\
 Ti  II &  5418.773  &  1.58 & $-$2.110  &  270 &  0.315 & 8.199 &  BHN                                      \\
 Fe  II &  4508.287  &  2.84 & $-$2.520  &  188 &  0.267 & 8.617 &  PGH Resc.                                \\
 Fe  II &  4656.979  &  2.88 & $-$3.580  &  190 &  0.330 & 8.612 &  HK Resc.\tablenotemark{b}         \\
 Fe  II &  5234.632  &  3.21 & $-$2.230  &  188 &  0.268 & 8.487 &  HK Resc.\tablenotemark{c}            \\
 Fe  II &  6432.684  &  2.89 & $-$3.510  &  174 &  0.270 & 8.462 &  SKH\tablenotemark{d}         \\
 Fe  II &  6516.086  &  2.89 & $-$3.380  &  174 &  0.270 & 8.464 &  SKH                 \tablenotemark{c}            \\
 Fe  II &  7515.836  &  3.90 & $-$3.450  &  187 &  0.271 & 8.612 &  PGH, HK Aver. Resc.\tablenotemark{e}         \\
 Fe  II &  7711.730  &  3.90 & $-$2.450  &  186 &  0.264 & 8.615 &  PGH, HK Aver. Resc. \tablenotemark{f}         \\
\cutinhead{Strong}
 Mg   I &  8806.778  & 4.33  & $-$0.120  &  531 &  0.292 & 8.690 & RS              \tablenotemark{g}     \\
 Ca   I &  6162.183  & 1.89  & $-$0.097  &  878 &  0.236 & 7.860 & Oxford                                  \\    
 Ca  II &  8542.120  & 1.70  & $-$0.463  &  291 &  0.275 & 8.164 & THE            \tablenotemark{h}                  \\
 Ca  II &  8662.169  & 1.69  & $-$0.723  &  291 &  0.275 & 8.152 & THE             \tablenotemark{h}                 \\ 
 Fe   I &  5232.952  & 2.94  & $-$0.058  &  713 &  0.238 & 8.009 & OWL                                      \\
 Fe   I &  8327.067  & 2.20  & $-$1.525  &  258 &  0.247 & 7.303 & Oxford                                   \\ 
 Fe   I &  8688.643  & 2.17  & $-$1.212  &  253 &  0.245 & 7.276 & Oxford                                  \\
\enddata
\tablenotetext{a}{Rescaled GARZ}
\tablenotetext{b}{HLG=$-$3.59}
\tablenotetext{c}{HLG same}
\tablenotetext{d}{HLG=$-$3.50}
\tablenotetext{e}{HLG=$-$3.44}
\tablenotetext{f}{HLG=$-$2.47}
\tablenotetext{g}{Critically averaged}
\tablenotetext{h}{Theory}
\tablerefs{
Sources of $f$-values: BZH = Becker, Zimmermann, \& Holweger (1980); 
BHN = Bizzarri et al. (1993); GARZ = Garz~(1973); GBP = Grevesse, Blackwell \& Petford 1989; HLG = Hannaford et al. (1992); HK = Heise \& Kock (1990); OWL = O'Brian et al. (1991); Oxford = Blackwell \& Shallis (1979) and following papers; Pauls, Grevesse, \& Huber (1990); RS = Ruck \& Smith (1993); SKH = Schnabel, Kock, \& Holweger (1999); THE = Theodosiou (1989)
}
\tablecomments{
Aver. = averaged ; Resc. = Rescaled to averaged lifetime of Schnabel et al. (1999) and Guo et al. (1992)
}
\end{deluxetable}

\begin{deluxetable}{cccccc}
\tablecolumns{6}
\tablewidth{0pc}
\tablecaption{Solar model atmosphere derived from the inversion}
\tablehead{ \colhead{$\log \tau_{5000}$} &  \colhead{T} &  \colhead{Pe}&  \colhead{Pg}&  \colhead{$\rho$}&  \colhead{column mass} \\
  \colhead{} &  \colhead{(K)} &  \colhead{(dyn cm$^{-2}$)}&  \colhead{(dyn cm$^{-2}$)}&  \colhead{(g cm$^{-3}$)}&  \colhead{(g cm$^{-2}$)}}
\startdata
$-$4.0 & 4003.8 & 0.2636$\times 10^{-01}$ & 0.2903$\times 10^{+03}$ & 0.1132$\times 10^{-08}$ & 0.0000$\times 10^{+00}$ \\
$-$3.9 & 4081.5 & 0.4348$\times 10^{-01}$ & 0.4798$\times 10^{+03}$ & 0.1835$\times 10^{-08}$ & 0.9047$\times 10^{-02}$ \\
$-$3.8 & 4151.3 & 0.6163$\times 10^{-01}$ & 0.6705$\times 10^{+03}$ & 0.2521$\times 10^{-08}$ & 0.1714$\times 10^{-01}$ \\
$-$3.7 & 4214.0 & 0.8146$\times 10^{-01}$ & 0.8715$\times 10^{+03}$ & 0.3228$\times 10^{-08}$ & 0.2524$\times 10^{-01}$ \\
$-$3.6 & 4270.1 & 0.1033$\times 10^{+00}$ & 0.1089$\times 10^{+04}$ & 0.3979$\times 10^{-08}$ & 0.3374$\times 10^{-01}$ \\
$-$3.5 & 4320.0 & 0.1274$\times 10^{+00}$ & 0.1326$\times 10^{+04}$ & 0.4791$\times 10^{-08}$ & 0.4288$\times 10^{-01}$ \\
$-$3.4 & 4364.3 & 0.1540$\times 10^{+00}$ & 0.1588$\times 10^{+04}$ & 0.5678$\times 10^{-08}$ & 0.5285$\times 10^{-01}$ \\
$-$3.3 & 4403.5 & 0.1834$\times 10^{+00}$ & 0.1878$\times 10^{+04}$ & 0.6658$\times 10^{-08}$ & 0.6385$\times 10^{-01}$ \\
$-$3.2 & 4438.2 & 0.2159$\times 10^{+00}$ & 0.2202$\times 10^{+04}$ & 0.7745$\times 10^{-08}$ & 0.7605$\times 10^{-01}$ \\
$-$3.1 & 4469.2 & 0.2520$\times 10^{+00}$ & 0.2565$\times 10^{+04}$ & 0.8959$\times 10^{-08}$ & 0.8966$\times 10^{-01}$ \\
$-$3.0 & 4496.9 & 0.2919$\times 10^{+00}$ & 0.2971$\times 10^{+04}$ & 0.1031$\times 10^{-07}$ & 0.1049$\times 10^{+00}$ \\
$-$2.9 & 4521.2 & 0.3362$\times 10^{+00}$ & 0.3429$\times 10^{+04}$ & 0.1184$\times 10^{-07}$ & 0.1220$\times 10^{+00}$ \\
$-$2.8 & 4543.6 & 0.3853$\times 10^{+00}$ & 0.3942$\times 10^{+04}$ & 0.1354$\times 10^{-07}$ & 0.1412$\times 10^{+00}$ \\
$-$2.7 & 4564.2 & 0.4400$\times 10^{+00}$ & 0.4521$\times 10^{+04}$ & 0.1546$\times 10^{-07}$ & 0.1629$\times 10^{+00}$ \\
$-$2.6 & 4583.6 & 0.5008$\times 10^{+00}$ & 0.5172$\times 10^{+04}$ & 0.1761$\times 10^{-07}$ & 0.1874$\times 10^{+00}$ \\
$-$2.5 & 4602.3 & 0.5688$\times 10^{+00}$ & 0.5906$\times 10^{+04}$ & 0.2003$\times 10^{-07}$ & 0.2150$\times 10^{+00}$ \\
$-$2.4 & 4620.8 & 0.6451$\times 10^{+00}$ & 0.6737$\times 10^{+04}$ & 0.2275$\times 10^{-07}$ & 0.2463$\times 10^{+00}$ \\
$-$2.3 & 4639.9 & 0.7309$\times 10^{+00}$ & 0.7672$\times 10^{+04}$ & 0.2581$\times 10^{-07}$ & 0.2817$\times 10^{+00}$ \\
$-$2.2 & 4659.7 & 0.8277$\times 10^{+00}$ & 0.8729$\times 10^{+04}$ & 0.2924$\times 10^{-07}$ & 0.3219$\times 10^{+00}$ \\
$-$2.1 & 4681.6 & 0.9380$\times 10^{+00}$ & 0.9923$\times 10^{+04}$ & 0.3308$\times 10^{-07}$ & 0.3675$\times 10^{+00}$ \\
$-$2.0 & 4705.1 & 0.1064$\times 10^{+01}$ & 0.1128$\times 10^{+05}$ & 0.3741$\times 10^{-07}$ & 0.4193$\times 10^{+00}$ \\
$-$1.9 & 4731.3 & 0.1209$\times 10^{+01}$ & 0.1281$\times 10^{+05}$ & 0.4227$\times 10^{-07}$ & 0.4781$\times 10^{+00}$ \\
$-$1.8 & 4760.9 & 0.1376$\times 10^{+01}$ & 0.1455$\times 10^{+05}$ & 0.4768$\times 10^{-07}$ & 0.5449$\times 10^{+00}$ \\
$-$1.7 & 4793.8 & 0.1571$\times 10^{+01}$ & 0.1652$\times 10^{+05}$ & 0.5380$\times 10^{-07}$ & 0.6209$\times 10^{+00}$ \\
$-$1.6 & 4831.2 & 0.1800$\times 10^{+01}$ & 0.1878$\times 10^{+05}$ & 0.6067$\times 10^{-07}$ & 0.7072$\times 10^{+00}$ \\
$-$1.5 & 4873.4 & 0.2069$\times 10^{+01}$ & 0.2134$\times 10^{+05}$ & 0.6833$\times 10^{-07}$ & 0.8055$\times 10^{+00}$ \\
$-$1.4 & 4920.8 & 0.2388$\times 10^{+01}$ & 0.2425$\times 10^{+05}$ & 0.7693$\times 10^{-07}$ & 0.9171$\times 10^{+00}$ \\
$-$1.3 & 4974.3 & 0.2769$\times 10^{+01}$ & 0.2758$\times 10^{+05}$ & 0.8654$\times 10^{-07}$ & 0.1044$\times 10^{+01}$ \\
$-$1.2 & 5034.2 & 0.3227$\times 10^{+01}$ & 0.3138$\times 10^{+05}$ & 0.9727$\times 10^{-07}$ & 0.1188$\times 10^{+01}$ \\
$-$1.1 & 5101.0 & 0.3784$\times 10^{+01}$ & 0.3572$\times 10^{+05}$ & 0.1093$\times 10^{-06}$ & 0.1352$\times 10^{+01}$ \\
$-$1.0 & 5174.9 & 0.4469$\times 10^{+01}$ & 0.4071$\times 10^{+05}$ & 0.1228$\times 10^{-06}$ & 0.1538$\times 10^{+01}$ \\
$-$0.9 & 5257.8 & 0.5328$\times 10^{+01}$ & 0.4637$\times 10^{+05}$ & 0.1377$\times 10^{-06}$ & 0.1747$\times 10^{+01}$ \\
$-$0.8 & 5348.7 & 0.6427$\times 10^{+01}$ & 0.5284$\times 10^{+05}$ & 0.1542$\times 10^{-06}$ & 0.1982$\times 10^{+01}$ \\
$-$0.7 & 5449.1 & 0.7879$\times 10^{+01}$ & 0.6018$\times 10^{+05}$ & 0.1724$\times 10^{-06}$ & 0.2243$\times 10^{+01}$ \\
$-$0.6 & 5559.0 & 0.9866$\times 10^{+01}$ & 0.6852$\times 10^{+05}$ & 0.1924$\times 10^{-06}$ & 0.2528$\times 10^{+01}$ \\
$-$0.5 & 5679.5 & 0.1268$\times 10^{+02}$ & 0.7790$\times 10^{+05}$ & 0.2141$\times 10^{-06}$ & 0.2835$\times 10^{+01}$ \\
$-$0.4 & 5810.2 & 0.1676$\times 10^{+02}$ & 0.8831$\times 10^{+05}$ & 0.2372$\times 10^{-06}$ & 0.3157$\times 10^{+01}$ \\
$-$0.3 & 5952.6 & 0.2286$\times 10^{+02}$ & 0.9989$\times 10^{+05}$ & 0.2619$\times 10^{-06}$ & 0.3486$\times 10^{+01}$ \\
$-$0.2 & 6107.1 & 0.3209$\times 10^{+02}$ & 0.1124$\times 10^{+06}$ & 0.2872$\times 10^{-06}$ & 0.3814$\times 10^{+01}$ \\
$-$0.1 & 6273.6 & 0.4617$\times 10^{+02}$ & 0.1257$\times 10^{+06}$ & 0.3127$\times 10^{-06}$ & 0.4133$\times 10^{+01}$ \\
0.0 & 6453.3 & 0.6784$\times 10^{+02}$ & 0.1397$\times 10^{+06}$ & 0.3376$\times 10^{-06}$ & 0.4438$\times 10^{+01}$ \\
0.1 & 6646.8 & 0.1013$\times 10^{+03}$ & 0.1539$\times 10^{+06}$ & 0.3611$\times 10^{-06}$ & 0.4724$\times 10^{+01}$ \\
0.2 & 6854.2 & 0.1529$\times 10^{+03}$ & 0.1681$\times 10^{+06}$ & 0.3824$\times 10^{-06}$ & 0.4990$\times 10^{+01}$ \\
0.3 & 7076.1 & 0.2325$\times 10^{+03}$ & 0.1822$\times 10^{+06}$ & 0.4015$\times 10^{-06}$ & 0.5235$\times 10^{+01}$ \\
0.4 & 7313.4 & 0.3550$\times 10^{+03}$ & 0.1961$\times 10^{+06}$ & 0.4177$\times 10^{-06}$ & 0.5460$\times 10^{+01}$ \\
0.5 & 7566.4 & 0.5424$\times 10^{+03}$ & 0.2094$\times 10^{+06}$ & 0.4309$\times 10^{-06}$ & 0.5665$\times 10^{+01}$ \\
0.6 & 7835.6 & 0.8272$\times 10^{+03}$ & 0.2221$\times 10^{+06}$ & 0.4409$\times 10^{-06}$ & 0.5853$\times 10^{+01}$ \\
0.7 & 8121.4 & 0.1256$\times 10^{+04}$ & 0.2341$\times 10^{+06}$ & 0.4475$\times 10^{-06}$ & 0.6023$\times 10^{+01}$ \\
0.8 & 8425.0 & 0.1897$\times 10^{+04}$ & 0.2452$\times 10^{+06}$ & 0.4508$\times 10^{-06}$ & 0.6176$\times 10^{+01}$ \\
0.9 & 8746.2 & 0.2840$\times 10^{+04}$ & 0.2550$\times 10^{+06}$ & 0.4500$\times 10^{-06}$ & 0.6315$\times 10^{+01}$ \\
1.0 & 9086.0 & 0.4209$\times 10^{+04}$ & 0.2633$\times 10^{+06}$ & 0.4450$\times 10^{-06}$ & 0.6438$\times 10^{+01}$ \\
1.1 & 9444.8 & 0.6165$\times 10^{+04}$ & 0.2700$\times 10^{+06}$ & 0.4358$\times 10^{-06}$ & 0.6548$\times 10^{+01}$ \\
1.2 & 9823.2 & 0.8904$\times 10^{+04}$ & 0.2748$\times 10^{+06}$ & 0.4219$\times 10^{-06}$ & 0.6645$\times 10^{+01}$ \\
1.3 & 10221. & 0.1265$\times 10^{+05}$ & 0.2773$\times 10^{+06}$ & 0.4030$\times 10^{-06}$ & 0.6730$\times 10^{+01}$ \\
1.4 & 10640. & 0.1763$\times 10^{+05}$ & 0.2772$\times 10^{+06}$ & 0.3788$\times 10^{-06}$ & 0.6804$\times 10^{+01}$ \\
\enddata
\end{deluxetable}

\begin{deluxetable}{ccccccc}
\tablecolumns{7}
\tablewidth{0pc}
\tablecaption{Abundances used in the numerical simulation (Grevesse \& Sauval 1998 $\equiv$ GS98) and those obtained from the inversion of the simulation (Sim) and solar observations (Obs)}
\tablehead{ \colhead{Element} & \colhead{A (GS98)} &  \colhead{$\sigma$\tablenotemark{a}} & \colhead{A (Sim)} &  \colhead{$\sigma$}&  \colhead{A (Obs)} &  \colhead{$\sigma$}}
\startdata
Mg & 7.58 & 0.05 & 7.65 & 0.02 & 7.58 & 0.02 \\
Si & 7.55 & 0.05 & 7.52 & 0.02 & 7.57 & 0.02 \\
Ca & 6.36 & 0.02 & 6.36 & 0.03 & 6.37 & 0.03 \\
Ti & 5.02 & 0.06 & 5.07 & 0.02 & 4.95 & 0.03 \\
Cr & 5.67 & 0.03 & 5.68 & 0.02 & 5.67 & 0.02 \\
Fe & 7.50 & 0.05 & 7.48 & 0.03 & 7.51 & 0.02 \\
\enddata
\tablenotetext{a}{Note that this $\sigma$ is the uncertainty listed by GS98, and it is not clear whether it can be directly compared to the $\sigma$s derived from our set of experiments with the inversion code.}
\end{deluxetable}

\clearpage

\clearpage
\clearpage

\end{document}